\documentclass[iop,twocolappendix,numberedappendix]{emulateapj}

\usepackage{graphicx}
\usepackage{float}
\usepackage{amsmath}
\usepackage{xcolor}
\usepackage{hyperref}
\usepackage{enumitem}
\setlist{parsep=0pt,listparindent=\parindent}
    {\pagebreak[4]\global\pdfpageattr\expandafter{\the\pdfpageattr/Rotate 90}}%
    {\pagebreak[4]\global\pdfpageattr\expandafter{\the\pdfpageattr/Rotate 0}}%

\newcommand{\err}[2]{\ensuremath{^{+#1}_{-#2}}}

\newcommand{\JHU}{Department of Physics and Astronomy, The Johns Hopkins University, Baltimore, MD 21218.}
\newcommand{\STScI}{Space Telescope Science Institute, Baltimore, MD 21218.}
\newcommand{\Kavli}{University of Chicago, Kavli Institute for Cosmological Physics, Chicago, IL, USA.}

\altaffiltext{1}{\JHU}
\altaffiltext{2}{\Kavli}
\altaffiltext{3}{Hubble, KICP Fellow}
\altaffiltext{4}{\STScI}
\altaffiltext{5}{Harvard-Smithsonian Center for Astrophysics, 60 Garden Street, Cambridge, MA 02138, USA}
\altaffiltext{6}{Department of Physics, Harvard University, Cambridge, MA 02138, USA}
\altaffiltext{7}{Department of Astronomy and Astrophysics, University of California, Santa Cruz, CA 92064, USA}
\altaffiltext{8}{Astrophysical Institute, Department of Physics and Astronomy, 
251B Clippinger Lab, Ohio University, Athens, OH 45701, USA}
\altaffiltext{9}{Institute for Astronomy, University of Hawaii at Manoa, Honolulu, HI 96822, USA}
\altaffiltext{10}{Department of Physics, Durham University, South Road, Durham DH1 3LE, UK}

\def\agn{86}

\def\speccontam{$1.4\% \pm 1.3$\%}

\def\totalcontam{$2.6\% \pm 1.4$\%}

\def\numspec{3,930}
\def\numspecgood{3,147}
\def\numspecfinal{1,020}
\def\numsnetotal{1,145}

\def\fp{\textit{Fitprob}}

\def\wbias{-0.005$\pm$0.004}
\def\wbiasper{10\%}

\def\simsysp{0.014$\pm$0.007}
\def\simsyspper{30\%}

\begin{document}

\title{Measuring the Properties of Dark Energy with Photometrically Classified Pan-STARRS Supernovae. I. Systematic Uncertainty from Core-Collapse Supernova Contamination}
\author{D. O. Jones\altaffilmark{1}, D. M. Scolnic\altaffilmark{2,3}, A. G. Riess\altaffilmark{1,4}, 
R. Kessler\altaffilmark{2}, A. Rest\altaffilmark{4}, R. P. Kirshner\altaffilmark{5,6}, E.
  Berger\altaffilmark{5}, C. A. Ortega\altaffilmark{1}, R. J. Foley\altaffilmark{7},
R. Chornock\altaffilmark{8}, P. J. Challis\altaffilmark{5}, 
W. S. Burgett\altaffilmark{9},
K. C. Chambers\altaffilmark{9}, P. W. Draper\altaffilmark{10}, 
H. Flewelling\altaffilmark{9},
M. E. Huber\altaffilmark{9}, N. Kaiser\altaffilmark{9}, 
R.-P. Kudritzki\altaffilmark{9}, N. Metcalfe\altaffilmark{10}, 
R. J. Wainscoat\altaffilmark{9}, C. Waters\altaffilmark{9}}

\begin{abstract} 

The Pan-STARRS (PS1) Medium Deep Survey discovered over
5,000 likely supernovae (SNe) but obtained spectral classifications 
for just 10\% of its SN candidates.  We measured spectroscopic 
host galaxy redshifts for 3,147 of these likely
SNe and estimate that $\sim$1,000 are Type Ia SNe (SNe\,Ia)
with light-curve quality sufficient for a cosmological analysis.
We use these data with simulations to determine the impact of
core-collapse SN (CC\,SN) contamination on measurements of the dark energy 
equation of state parameter, $w$.  Using the method of Bayesian Estimation Applied to 
Multiple Species (BEAMS), distances to SNe\,Ia and the contaminating
CC\,SN distribution are simultaneously determined.  We test
light-curve based SN classification priors for BEAMS 
as well as a new classification method that relies upon host galaxy
spectra and the association of SN type with host type.  
By testing several SN classification methods and 
CC\,SN parameterizations on large SN simulations, we
estimate that CC\,SN contamination gives
a systematic error on $w$ ($\sigma_w^{CC}$) of 0.014,
29\% of the statistical uncertainty.  Our best method
gives $\sigma_w^{CC} = 0.004$, just 8\% of the statistical uncertainty,
but could be affected by incomplete knowledge of the CC\,SN distribution.
This method determines the SALT2 color and shape coefficients, 
$\alpha$ and $\beta$, with $\sim$3\% bias.  However, we find that
some variants require $\alpha$ and $\beta$ to be fixed
to known values for
BEAMS to yield accurate measurements of $w$.
Finally, the inferred abundance of bright CC\,SNe in our sample is greater than
expected based on measured CC\,SN rates and luminosity functions.


\end{abstract}

\section{Introduction}
\label{sec:intro}

Since the discovery of cosmic acceleration 
\citep{Riess98,Perlmutter99}, measuring the properties of dark energy with Type Ia 
supernovae (SNe\,Ia) has been predicated on the 
spectroscopic confirmation of SN\,Ia candidates.
However, as the size of individual SN\,Ia samples surpasses 1,000 SNe, 
obtaining spectra for each Type Ia candidate is 
becoming prohibitively expensive.  Only a small fraction 
of SNe\,Ia from current and future surveys such as the 
Dark Energy Survey (DES) and the Large Synoptic Survey 
Telescope (LSST) will have spectroscopic classification.
Without spectroscopic classification, core-collapse SN (CC\,SN) 
contamination can bias our estimates of cosmological parameters 
(\citealp{Falck10}, \citealp*{Kunz07}).

Without SN spectroscopy, the shape and color of a
photometric SN light curve can be used as a less precise
diagnostic of the type.  \citet{Campbell13} used SDSS $ugriz$ light curves 
to classify 752 SNe as likely Type Ia, enough to measure the dark 
energy equation of state parameter, $w$, with $\sim$10\% 
statistical uncertainty.  Their sample was selected from light curve
properties and a classifier 
that compares each observed light curve to SN\,Ia and CC\,SN 
templates (PSNID; \citealp{Sako11}).  Their final sample comprised 
just 3.9\% CC\,SNe.  While \citet{Campbell13}
is the \textit{only} SN\,Ia-based measurement of $w$ to date that does not use
spectroscopic classification for its SNe, the measurement did 
not include systematic uncertainties.  In addition, contaminating CC\,SNe bias
their measurements of SN\,Ia dispersion and the correlation between
SN luminosity and light curve rise/decline rate by $\sim$60\%.

Many light curve classifiers use the ``na\"{\i}ve Bayes'' approximation, 
which assumes that all observables that indicate SN type are uncorrelated.
Machine learning techniques can often outperform these classifiers, 
yielding higher SN\,Ia classification efficiency (the fraction of SNe\,Ia classified 
correctly) and lower CC\,SN contamination \citep{Lochner16,Moller16}.
On SDSS SN data, the \citet{Sako14} kd-tree nearest neighbor (NN) 
method has a purity comparable to that of \citet{Campbell13} but accurately classifies
$\sim$1.4 times as many real SNe\,Ia in a given sample.

An important caveat is that nearly all classifiers are optimized on 
simulations with little evaluation on real data.  Simulations, in turn, 
depend on CC\,SN templates and knowledge of the CC\,SN luminosity functions 
(LFs) and rates.  CC\,SNe are diverse, far more so than SNe\,Ia, and only a 
limited number of high-quality templates are publicly available.  
Training a classifier directly on survey data is possible but can be
sub-optimal due to limited numbers of CC\,SNe observed and the dependence
of classifier results on the specific survey characteristics 
(e.g. observing cadences, filters, and signal-to-noise ratios).

We can make SN classification less dependent on 
CC\,SN templates, LFs, and rates by incorporating 
host galaxy data.  Because many SNe\,Ia have a $\gtrsim$1 Gyr
delay time between progenitor formation and explosion \citep{Rodney14},
they are the only type of SNe found in early-type galaxies 
(with very few known 
exceptions; \citealp{Suh11}).  \citet{Foley13} found that it was 
possible to accurately classify the $\sim$20\% of SNe\,Ia found 
in elliptical galaxies if the morphology of their host galaxy is known.

Though these results are encouraging, 
light curve and host galaxy classification alone may not be enough to 
enable a measurement of $w$ as precise as 
measurements using spectroscopically classified SNe 
(e.g. \citealp{Betoule14}, $w = -1.027 \pm 0.055$).  A difference 
in $w$ of 5\% corresponds to a change of 0.02 mag from $z = 0$ 
to $z = 0.5$; if CC\,SNe are 1 mag fainter than SNe\,Ia on 
average, a bias of 0.02 mag can be induced by just 2\% CC\,SN contamination 
in a high-$z$ sample such as PS1.  If the contaminating distribution
of CC\,SNe is more than 1 mag fainter (this depends on survey
Malmquist bias), it takes even fewer CC\,SNe to bias $w$ by an
equivalent amount.

A Bayesian method, however, could use the probabilities that SNe are of Type 
Ia as priors to simultaneously 
determine distances to Ia and CC\,SNe without bias.  We refer 
to this method as Bayesian Estimation Applied to Multiple 
Species (BEAMS) following \citet*{Kunz07} (hereafter KBH07; see also 
\citealp{Press97} and \citealp{Rubin15}).  KBH07 test BEAMS 
on a simplistic SN simulation and 
find that it gives near-optimal accuracy and uncertainties 
on SN\,Ia distances.

\citet{Hlozek12} test BEAMS further with Monte Carlo simulations of
the Sloan Digital Sky Survey SN survey (SDSS-SN; \citealp{Frieman08}; 
\citealp{Kessler09}).  BEAMS biases measurements of 
the cosmic matter and dark energy densities, $\Omega_M$ and 
$\Omega_{\Lambda}$, by less than the statistical uncertainties 
measured from their simulations.  Their results demonstrated
that SDSS SNe without spectroscopic classification 
can significantly improve cosmological constraints 
relative to the SDSS spectroscopic sample \citep{Kessler09}.  
\citet{Hlozek12} did not measure
the systematic uncertainties from their method.

As with SDSS, Pan-STARRS (PS1) discovered far more SNe\,Ia 
than could be observed spectroscopically.  
Spectroscopically confirmed SNe\,Ia from the first $\sim$1/3 of 
PS1 have been used to measure cosmological parameters but 
constitute only a small fraction of the available data 
(\citealp{Rest14}, hereafter R14; \citealp{Scolnic14}).  In 
this study, we use PS1 SNe with and without spectroscopic classification 
as a tool for testing SN classifiers, understanding CC\,SN contaminants, and measuring 
the systematic error due to CC\,SN contamination.  
In total, PS1 has \numsnetotal\ SNe with high-quality
light curves and spectroscopic redshifts $-$ both host
galaxy and SN redshifts $-$ that can be
used to measure cosmological parameters (including a $\sim$few
percent CC\,SN contamination).  Here, we focus on the \numspecfinal\
likely SNe\,Ia with spectroscopic host galaxy redshifts, 143 of which are
spectroscopically confirmed, in order to study a sample with fewer
selection biases (\S\ref{sec:hostz}).

The goal of this study is to develop the methods necessary to measure
cosmological parameters robustly using PS1 SNe without spectroscopic
classifications (hereafter referred to as photometric SNe).
Our full cosmological results from these data will be presented 
in a future analysis.

In \S\ref{sec:ps1}, we present the sample and our host galaxy redshift follow-up
survey.  \S\ref{sec:sim} discusses our SNANA simulations of the PS1
sample and our assumptions about the CC\,SN population.  \S\ref{sec:bayes}
describes our Bayesian parameter estimation
methodology.  In \S\ref{sec:results} 
we test BEAMS on simulations and subsamples of PS1 photometric SNe.  
In \S\ref{sec:varresults} we test the robustness
of these results by exploring several variants of the method.
The uncertainties in our 
simulations and methodology are discussed in \S\ref{sec:discussion} and 
our conclusions are in \S\ref{sec:conclusions}.

\section{The Pan-STARRS Photometric Supernova Sample}
\label{sec:ps1}

The Pan-STARRS medium deep survey covers 10 7-square degree fields in 
five broadband filters, with typical $griz_{P1}$ observational cadences 
of 6 images per 10 days and a 5 day gap during bright time during 
which $y_{P1}$ images are taken.  Typical 5$\sigma$ detection limits are 
$\sim$23 AB mag for $griz_{P1}$, albeit with significant variation.  For 
a complete description of the PS1 survey, see \citet{Kaiser10} and 
R14.

PS1 images are processed using an image subtraction pipeline
that is described in detail in \citet{Rest05} and R14.
To measure final light curves for the PS1 photometric
sample (and the full spectroscopic sample; Scolnic et al. in prep), we
made several improvements to that pipeline.
We more than doubled the typical number of images that are
combined to create a deep template for subtraction, we
refined our method of selecting stars to build the
point spread function (PSF) model, and we improved the
zeropoint calibration.  These
improvements will be described in detail in Scolnic et al. (in prep.).

Pan-STARRS discovered 5,235 likely SNe during its four years of 
operation and obtained spectra for 520 SNe.  We collected \numspecgood\
spectroscopic host galaxy redshifts of these likely SNe (\S\ref{sec:hostz}).
In addition to SN candidates, we observed spectra for thousands of 
variable stars, AGN, flaring M dwarfs, and other transients that 
will be published in future work.

\subsection{Host Galaxy Redshift Survey}
\label{sec:hostz}

\begin{deluxetable*}{lcccccc}
\tabletypesize{\scriptsize}
\tablewidth{0pt}
\tablecaption{Redshift Follow-up Summary}
\tablehead{
\colhead{Telescope} & \colhead{Instrument} & \colhead{SN Redshifts\tablenotemark{a}} &
\colhead{$\lambda_{min} - \lambda_{max}$} & \colhead{Avg. Exp. Time} &
\colhead{Approx. Resolution} & \colhead{$z_{\textrm{median}}$}\\*[2pt]
&&&\AA&min.&\AA\ pix$^{-1}$&}
\startdata
AAT&AAOmega&512&3700 $-$ 8500&180&6&0.15\\
APO&DIS&10&3500 $-$ 9800&60&2.5&0.24\\
MMT&Hectospec&2348&3700 $-$ 9200&90&5&0.33\\
SDSS&BOSS&250&3800 $-$ 9200&45&2.5&0.20\\
WIYN&Hydra&45&3700 $-$ 6500&180&4.5&0.34\\
Other\tablenotemark{b}&\nodata&361&\nodata&\nodata&\nodata&0.19\\
\tableline\\
Total&\nodata&\numspecgood&\nodata&\nodata&\nodata&0.30\\
\enddata
\label{table:z}
\tablecomments{Some transient hosts were observed with multiple
  telescopes.  Numbers include host galaxy observations of
  both spectroscopically confirmed and unconfirmed SN candidates.}
\tablenotetext{a}{Number of SN candidates with reliable redshifts.}
\tablenotetext{b}{Includes redshifts from 2dFGRS \citep{Colless03},
  6dFGS \citep{Jones09}, DEEP2 \citep{Newman13}, VIPERS
  \citep{Scodeggio16}, VVDS \citep{LeFevre05},
  WiggleZ \citep{Blake08}, and zCOSMOS \citep{Lilly07}.}
\end{deluxetable*}

\begin{figure}
\centering
\includegraphics[width=3.5in]{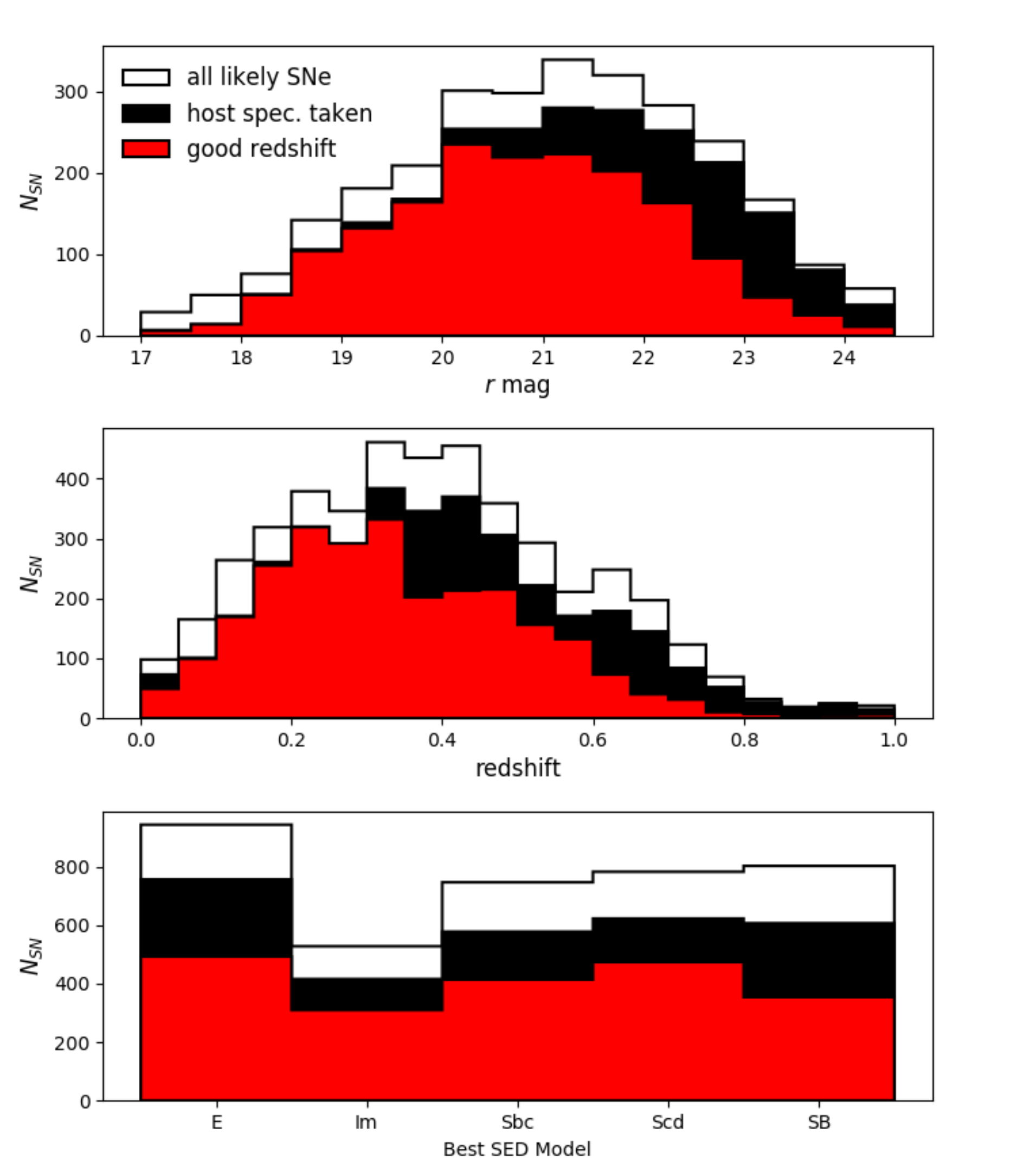}
\caption{Host properties from PS1 as a function of $r$ mag, redshift,
   and best-fit SED model.  Out of the full 
  sample of 5,235 PS1 SNe (white; host galaxy photo-$z$), we observed \numspec\ 
  hosts (black; photo-$z$) and measured accurate redshifts for 
  \numspecgood\ (red; spec-$z$).  Our redshift survey has nearly 100\% 
  success to $r = 21$ and has a median redshift of $0.30$.  We 
  obtained redshifts for a large number of both emission-line and 
  absorption-line galaxies.}
\label{fig:z}
\end{figure}

\begin{figure*}
\centering
\includegraphics[width=7in]{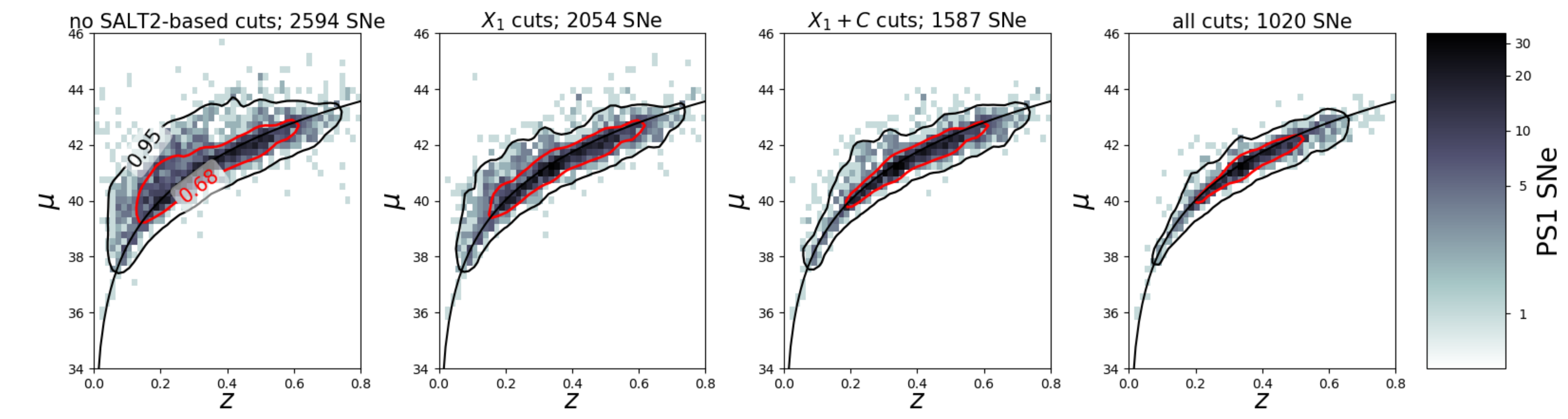}
\caption{Effect of \citet{Betoule14} cuts on the PS1 photometric
  Hubble diagram.  Distance moduli are measured using the Tripp estimator (Eq. \ref{eqn:salt2})
  with nuisance parameters from R14.  Of the 2,594 SNe 
  that are fit by SALT2 and are not possible AGN, shape and color cuts remove 1,007, while 
  $\chi^2$-based fit probability cuts and SNR-type cuts (shape uncertainty 
  and time of maximum uncertainty) remove an additional 567 SNe, leaving 
  \numspecfinal.  Each set of cuts removes a mix of SNe\,Ia
  with poor light 
  curve quality and CC\,SNe.}
\label{fig:cutshubble}
\end{figure*}

During the PS1 survey, many SN host redshifts were measured
using the Hectospec multifiber instrument on the MMT 
\citep{Fabricant05,Mink07}.  Near the end of PS1 operations, we began an 
additional survey with Hectospec to obtain redshifts for as many 
host galaxies as possible.  Redshifts were also obtained 
with the Apache Point Observatory 3.5m 
telescope\footnote{\url{http://www.apo.nmsu.edu/arc35m/}} (APO),
the WIYN telescope,\footnote{The WIYN Observatory is a joint 
facility of the University of Wisconsin-Madison, Indiana 
University, the National Optical Astronomy Observatory, and the 
University of Missouri.} and for the southernmost PS1 field, the
Anglo-Australian Telescope (AAT).  We chose candidate host galaxies
for follow-up in a largely unbiased way; we did not prioritize SNe
based on their magnitudes, colors, or whether or not an SN spectrum had
previously been obtained.  Approximately 600 of our redshifts
come from SDSS \citep{Smee13} or other public 
redshift surveys\footnote{We include redshifts from 2dFGRS \citep{Colless03},
  6dFGS \citep{Jones09}, DEEP2 \citep{Newman13}, VIPERS
  \citep{Scodeggio16}, VVDS \citep{LeFevre05},
  WiggleZ \citep{Blake08} and zCOSMOS \citep{Lilly07}.}.

We used the galaxy size- and orientation-weighted 
$R$ parameter to identify the most likely host galaxy 
for each SN \citep{Sullivan06}.  The isophotal limit of a galaxy 
corresponds to $R\sim3$.  We use the redshift of the host galaxy with the lowest 
$R$ if it has $R\leq5$ following \citet{Sullivan06}.  See 
\citet{Gupta16} for a similar but more rigorous method of identifying 
SN host galaxies.

To estimate the fraction of SNe for which we incorrectly determined which 
galaxy was the host, we compared redshifts derived from the spectroscopic redshifts of SNe to the spectroscopic 
redshifts of their most likely host galaxies.  We found that only 2 of 169 hosts with 
reliable redshifts had evidence of a host galaxy mismatch, $|z_{SN} - z_{host}|
 > 0.02$.  Both of these hosts had multiple large, nearby galaxies with $R<5$.  
This mismatch fraction suggests that 1.2\% $\pm$ 0.5\% of our redshifts are incorrect due 
to mismatched hosts.

Compared to spectroscopically confirmed SNe, it is
unlikely that photometric SNe have a higher
fraction of mismatched hosts.  The spectroscopic 
targeting preferentially followed SNe with a 
larger separation from the center of their host galaxies or SNe 
with fainter hosts, as these SNe
have spectra with less galaxy light contamination.
Just 11\% of photometrically classified SNe are outside the isophotal radii of
their host galaxies compared to 24\% of the 169 SN-host pairs.
However, we also note that the 169 SN-host
pairs have preferentially brighter hosts than the full sample
and have a median redshift of 0.21 compared to the median
redshift of 0.3 for the full sample.  It may be
somewhat easier to mismatch a host galaxy at high-$z$ as
galaxies are more difficult to detect, but we expect this
to be a subdominant effect as \citet{Gupta16} finds the
fraction of mismatched hosts to be approximately
constant at $z < 0.6$ in a DES-like survey (which has similar depth to PS1 templates).

The other source of incorrect redshifts is the measurement 
of velocities from host galaxy spectra.  We measured redshifts by cross-correlating 
our spectra with galaxy templates (the RVSAO package; \citealp{Kurtz98}) and visually 
inspecting the results.  Over the course of the survey, we observed over 
1,500 transient hosts multiple times. For $\sim$250 of these hosts, at 
least one observation yielded a redshift with a high \citet{Tonry79} 
cross-correlation parameter (TDR; $\gtrsim 9-10$). 

By restricting our sample to hosts with TDR $> 4$ and 
redshifts of $0.01 < z < 0.75$, we measure a 
false redshift fraction of \speccontam.  At $z > 0.75$, few SNe
could be discovered by PS1 or have their
host redshifts measured with our program (Figure \ref{fig:z}).
Including mismatched hosts, the total percent of incorrect redshifts we expect is 
\totalcontam.  In \S\ref{sec:sim} we simulate 
this fraction of false redshifts so that this effect
will be incorporated in our BEAMS systematic error budget.  

In total, we observed \numspec\ host galaxies and have \numspecgood\ reliable redshifts.  
The telescopes and instruments comprising 
our redshift survey are summarized in Table \ref{table:z}.
Figure \ref{fig:z} shows the $r$ magnitudes, redshifts, and best-fit 
SED model for the PS1 photometric sample.
87\% of PS1 SNe with detectable host galaxies 
were observed with our redshift follow-up program and reliable redshifts
were measured for 73\% of those galaxies.  We measured redshifts 
for a large number of both emission-line and absorption-line
galaxies.  These data have a median redshift of 0.30.

\subsection{SALT2 Selection Requirements}
\label{sec:dq}
\begin{deluxetable*}{lcccccc}
\tabletypesize{\scriptsize}
\tablewidth{0pt}
\tablecaption{Sequential PS1 Data Cuts}
\tablehead{
\colhead{} &
\colhead{Removed} &
\colhead{Remaining} & \colhead{} &
\colhead{This Cut Only} &
\colhead{Without This Cut} &
\colhead{Comments}}

\startdata
Total candidates&\nodata&5235&&\nodata&\nodata&\nodata\\
Host Sep. $R < 5$&774&4461&&\nodata&\nodata&Likely host galaxy can be identified\\
Good host redshifts&1314&3147&&\nodata&\nodata&\nodata\\
Fit by SALT2&457&2690&&\nodata&\nodata&SALT2 parameter fitting succeeds\\
Possible AGN&96&2594&&2594&1040&Separated from center or no long-term variability\\
$-3.0<X_1<3.0$&540&2054&&2119&1092&SALT2 light curve shape\\
$-0.3<C<0.3$&467&1587&&1903&1215&SALT2 light curve color\\
$\sigma_{\textrm{peakMJD}} < 2$&30&1557&&2630&1021&Uncertainty in time of maximum light (rest frame days)\\
$\sigma_{X_1} < 1$&379&1178&&1930&1386&$X_1$ uncertainty\\
Fit prob. $\ge$ 0.001&158&1020&&2096&1178&$\chi^2$- and N$_{dof}$-based probabilities from SALT2 fitter\\
E(B-V)$_{MW} > 0.15$&0&1020&&2690&1020&Milky Way reddening\\
\enddata
\label{table:cuts}
\end{deluxetable*}

Throughout this work, we use the SALT2.4 model (\citealp{Guy10}, implemented in B14) to 
measure SN light curve parameters.  We use these light curve parameters
to standardize SNe\,Ia and select the SNe\,Ia that can best measure 
cosmological parameters.  The Tripp 
estimator uses SALT2 light curve parameters to infer
the SN distance modulus, $\mu$ \citep{Tripp98}:

\begin{equation}
\mu = m_{B} + \alpha \times X_{1} - \beta \times C - M.
\label{eqn:salt2}
\end{equation}

\noindent $m_B$ is the log of the light curve amplitude,
$X_1$ is the light curve stretch parameter,
and $C$ is the light curve color parameter.  These parameters are all
measured by the SALT2 fitting program, but deriving the distance
modulus from them depends on the nuisance parameters $\alpha$,
$\beta$, and $M$.  $M$ is degenerate with the Hubble constant, H$_0$,
and will be marginalized over during cosmological parameter estimation.

To avoid unexpected biases in our sample selection, we use light 
curve selection requirements (cuts) from 
previous analyses using spectroscopically confirmed SNe.
We make the same series of cuts to PS1 SN light curves as 
\citet{Betoule14} and add one additional cut on the SALT2
fit probability following R14.  These cuts include uncertainty-based cuts that ensure the shape 
and time of maximum light of each SN is well-measured, and shape and color cuts 
that restrict our sample to SNe\,Ia for which the SALT2 model is well-trained.  Our cuts are
summarized in Table \ref{table:cuts} and Figure
\ref{fig:cutshubble}.  Out of \numspecgood\ SNe with reliable host redshifts,
SALT2 fits run successfully on 2,690 SNe
(SALT2 parameter fitting often fails due to lack of light curve
data before or after maximum).  1,020 SNe pass all of our cuts.

Omitting the SALT2 $\sigma_{X_1}$ cut has the largest single
impact on our final sample.  Without it, there would be nearly 1,400 SNe in the sample
but also twice as many SNe with Hubble residuals $>$ 0.5 mag (poorly measured SNe\,Ia or
CC\,SNe).
The cut with the second largest reduction is the cut on $C$, without which there would
be $\sim$1,200 SNe (though many would be CC\,SNe).  Although it may be possible to increase the 
SN sample size with relaxed cuts, the extent to which
SNe\,Ia with low signal-to-noise ratio (SNR) and unusual colors are standardizable is not
well-characterized.

In addition to the \citet{Betoule14} cuts, we implement an additional set 
of cuts to remove possible AGN that were not flagged during the 
PS1 transient search.  We 
tuned our long-term variability criteria to find known AGN in PS1 data.  We found that 
sources where $>$25\% of background epochs have $2\sigma$ 
deviations from 0 are likely AGN (we define background epochs
as $<$20 days before or $>$60 days after the discovery epoch).  
\agn\ SNe with both evidence of long-term variability 
and SN positions within 0.5\arcsec\ of their host centers were removed.  
After light curve cuts, removing likely AGN reduces our sample by just 18 SNe.
To have a sample with uniform selection, we make these cuts (and all cuts) regardless
of whether or not a given SN\,Ia is spectroscopically confirmed.

\subsection{Low-$z$ SNe}
\label{sec:lowzsn}

Cosmological parameter constraints are greatly improved when a large, low-$z$ SN\,Ia
sample is included to anchor the Hubble diagram.  We use the same 197 low-$z$ SNe\,Ia
used in R14 though we anticipate adding additional
low-$z$ SNe in our full cosmological analysis.  These SNe are
spectroscopically confirmed and are assumed to have no CC\,SN contamination.

The R14 PS1 cosmology analysis has a 
low-$z$ sample with higher intrinsic dispersion than the PS1 sample.  
The intrinsic dispersion, $\sigma_{int}$, is defined as the value added in 
quadrature to the SN\,Ia distance modulus uncertainty such that the Hubble 
diagram reduced $\chi^2$ is equal to 1 \citep{Guy07}.  Differences in SN\,Ia
intrinsic dispersion from survey to survey are typical, with the likely source
of the variation including underestimated photometric difference image 
uncertainties and excess scatter from bright host galaxy subtractions
(as seen in R14 and \citealp{Kessler15}).
Redshift evolution of the SN\,Ia population could also play a role.  
We added 0.05 mag in quadrature to the $m_B$ uncertainties of the low-$z$ 
SNe to resolve the discrepancy.  Once added, this additional uncertainty 
term gives both the PS1 and low-$z$ SNe from R14 the same intrinsic dispersion of 
$\sim$0.115 mag.

\section{Simulating the Pan-STARRS Sample}
\label{sec:sim}

To robustly determine how CC\,SN contamination affects PS1 
measurements of $w$, we require a simulation that encapsulates as 
many elements of the PS1 SN survey as possible.
We used the SuperNova ANAlysis software
(SNANA\footnote{\url{http://snana.uchicago.edu/}}; \citealp{Kessler09b}) 
to generate Monte Carlo realizations of the PS1 survey.  SNANA simulates
a sample of SNe\,Ia and CC\,SNe using real observing conditions, host
galaxy noise, selection effects, SN rates, and incorrect redshifts
from host galaxy mismatches or measurement error.  
Simulations assume a flat $\Lambda$CDM cosmology with H$_0 = 70$ km s$^{-1}$ Mpc$^{-1}$, $\Omega_M = 0.3$,
$\Omega_{\Lambda} = 0.7$, and $w = -1$.

We choose not to simulate one significant effect:
the correlation between SN luminosity and host mass
(the host mass bias; \citealp{Kelly10,Lampeitl10}).
We do not simulate the host mass bias
because R14 did not include it (finding it had low significance
in their sample), and we wish to compare our PS1 photometric results directly to those of R14.
This effect has been identified at $>$5$\sigma$ by
\citet{Betoule14}, and we will include it in our future
cosmological analysis with these data.

\begin{figure}
\centering
\includegraphics[width=3.5in]{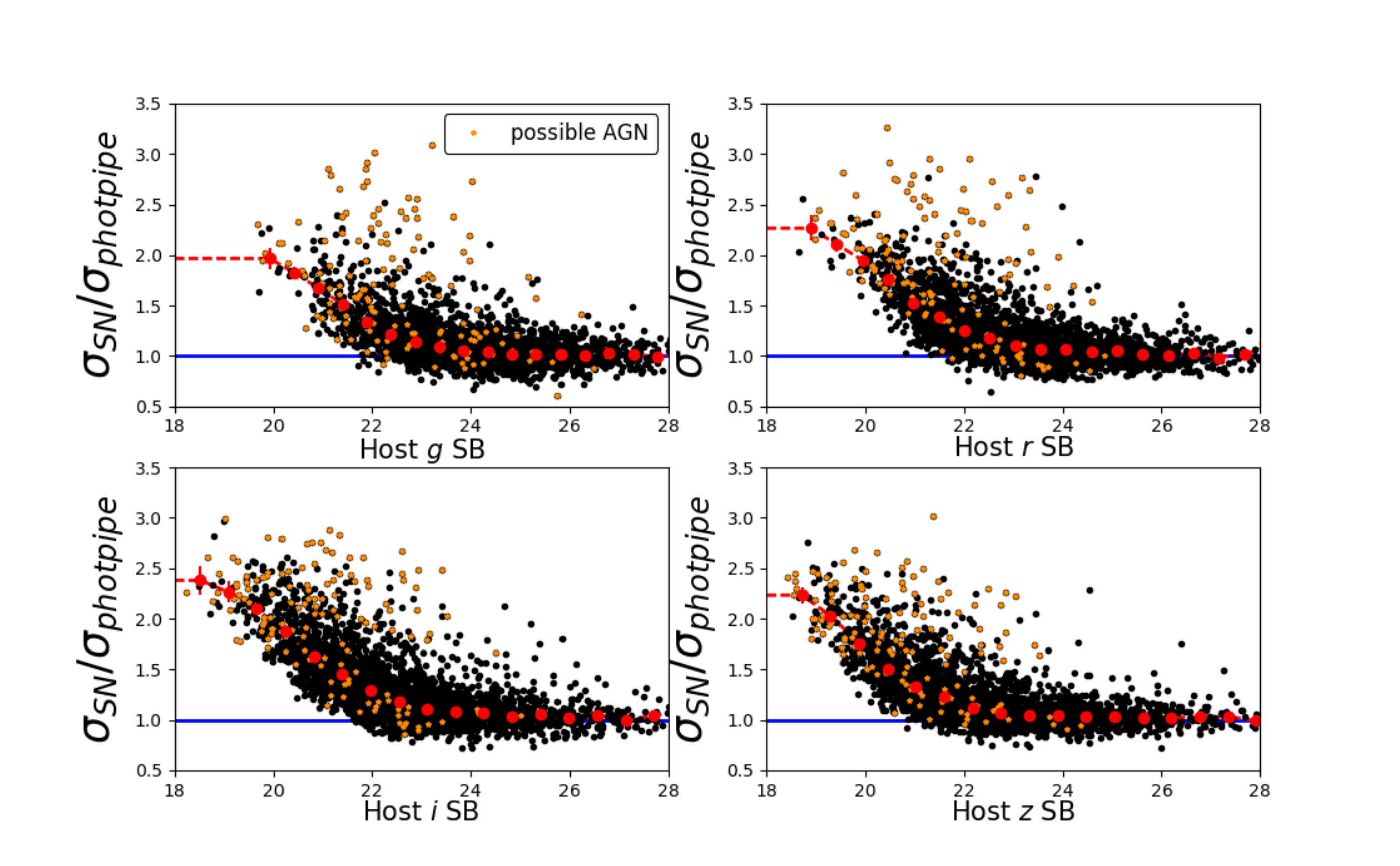}
\caption{Ratio between ``true'' and DAOPHOT-derived photometric noise 
  as a function of host galaxy surface brightness in the 
  $griz_{PS1}$ filters.  We computed the host galaxy surface 
  brightness by averaging over
  one PSF FWHM at the SN location.  We computed the true 
  photometric noise by inflating the
  errors from DAOPHOT (which do not include host galaxy noise)
  such that light curve epochs without SN light had $\chi^2 = 1$.
  Possible AGN (gold stars) comprise many 
  of the outliers in this relation.  We incorporated this 
  relationship into our SNANA simulations to yield an 
  accurate prediction of photometric uncertainties.}
\label{fig:sberr}
\end{figure}

Each major component of our simulation is discussed in detail below:

\begin{enumerate}
\item \textit{Observing conditions}.  SNANA generates SN observations based on a
  simulation library file with observation dates, filters,
  sky noise, zeropoints, and PSF sizes that we measure from PS1 nightly images.
\item \textit{Host galaxies}.  The observed flux scatter of SNe
  found in bright galaxies exceeds what is expected from
  Poisson noise alone (R14; \citealp{Kessler15}).
  To correct for this, SNANA adds host galaxy noise to SN flux uncertainties
  by placing each SN in a simulated host galaxy.  The SN is placed at a
  random location that has been weighted by the galaxy surface brightness profile.  The
  distribution of PS1 host galaxies was determined from PS1
  data; we measured the magnitudes and shape parameters
  of PS1 SN host galaxies using SExtractor, with zeropoints measured from the
  PS1 pipeline.  We then use the noise model from \citet[their Equation 4]{Kessler15}:
  \begin{equation}
    \tilde{\sigma}_{flux} = \sigma_{flux}\times R_{\sigma},
  \end{equation}
  \noindent where $R_{\sigma}$ is a function of
  host galaxy surface brightness (the vertical axis of
  Figure \ref{fig:sberr}).  We determine $R_{\sigma}$ for PS1 by comparing
  host surface brightness to the flux error scaling that gives light curve epochs
  without SN flux a reduced $\chi^2 = 1$.

\begin{figure*}
\centering
\includegraphics[width=7in]{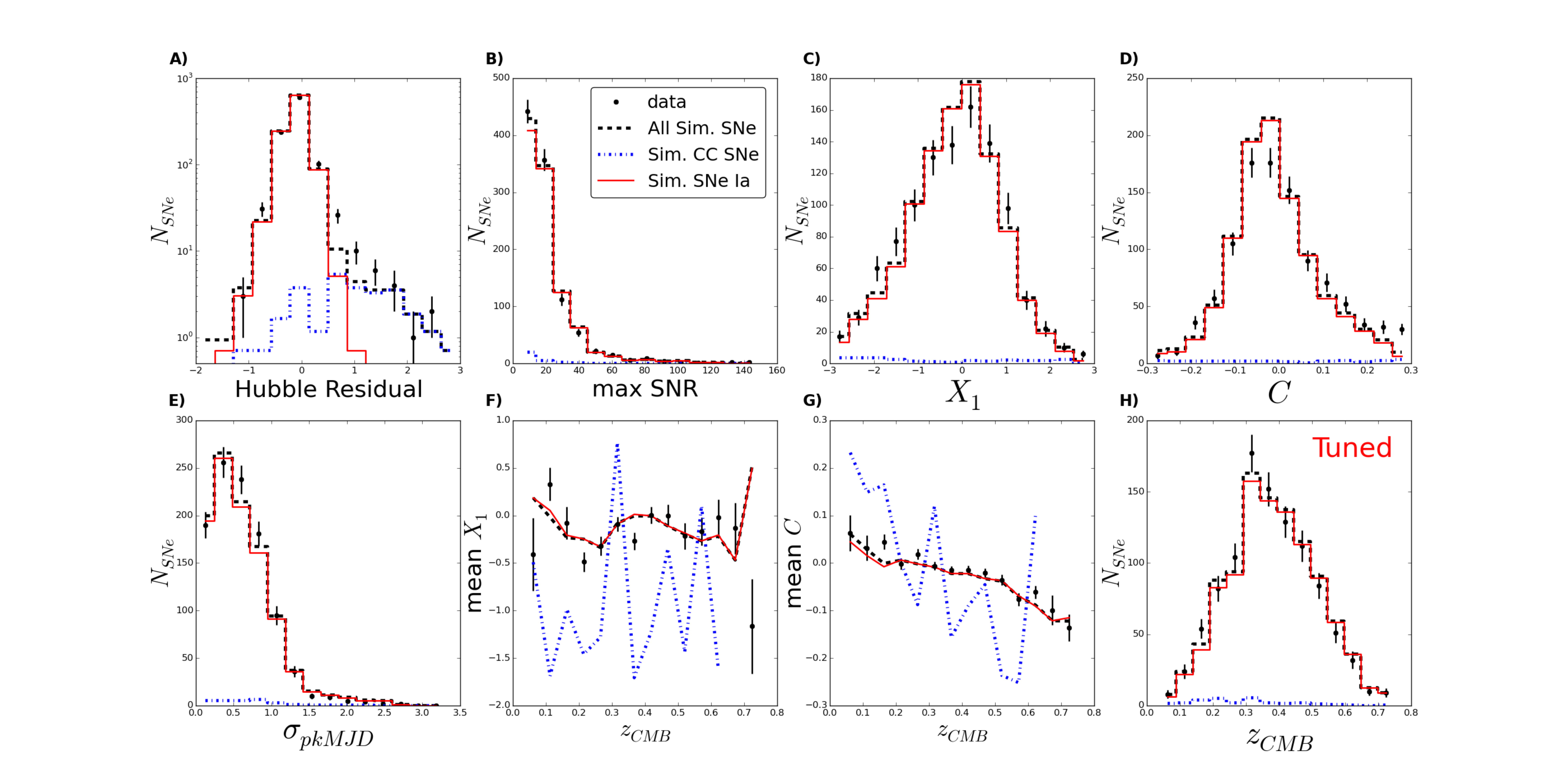}
\caption{SNANA simulations of a PS1 photometric sample compared to PS1
data.  The simulated Hubble residuals (A) of the CC\,SN distribution 
are flatter and fainter than the data.  The simulated SNR (B), shapes (C), 
colors (D), uncertainties (E) and $X_1$/$C$ redshift dependencies 
(F and G) match our data closely, albeit with $\sim$3$\sigma$
discrepancies in the time of maximum uncertainty and SN color.
We tuned the simulated redshift distribution (H) to match our data.}
\label{fig:defaultsim}
\end{figure*}
  
\item \textit{Selection effects}.  Two primary selection effects come into
  play in a photometric SN\,Ia survey.  The first is detection
  efficiency, the fraction of single-epoch detections
  as a function of the photometric SNR.  The detection 
  efficiency is computed by dividing the number of epochs detected 
  by PS1 at a given SNR by the total number of epochs at that SNR.  SNANA uses the
  efficiency vs. SNR, measured by PS1, to determine which simulated epochs are
  detected.  SNANA then applies the PS1 survey requirement of
  three detections to ``discover'' an SN.  The PS1 detection efficiency 
  is $\sim$50\% for epochs with an SNR of 5 in the final light curves.

  The second effect is host galaxy redshift selection.  To model this 
  effect, we incorporated a redshift-dependent ``host galaxy efficiency'' distribution
  in our simulations, which we adjusted such that the redshift distribution 
  of the simulations matched our data.

\item \textit{Uncertainty adjustment}.  SNANA allows its simulated uncertainties
  to be scaled as a function of SNR such that the mean uncertainties 
  in simulations match the mean uncertainties of our data.  In PS1, this 
  requires a modest $\sim 5-10$\% noise increase at low SNR (after excess host 
  galaxy noise is added). This adjustment is necessary due to the non-Gaussian wings of 
  the PS1 PSF and the PSF fitting radius used by the PS1 pipeline.
\item \textit{Mismatched host galaxies and incorrect redshifts}.  As discussed
  in \S\ref{sec:hostz}, we expect \totalcontam\ of our
  redshifts to be incorrect due to mismatched host galaxies and
  redshift measurement uncertainties.  We used SNANA to simulate incorrect host 
  redshifts by assigning false, ``measured'' redshifts to 2.6\% of our SNe.
  These redshifts are drawn from a flat, random distribution
  between $z=0.01$ and $z=0.75$.  This is the range of redshifts
  at which PS1 can discover SNe, with the exception of rare
  superluminous SNe.  Superluminous SNe typically have hosts too faint for our
  follow-up survey to measure their redshifts \citep{Lunnan15}.

  We find that $\sim$50\% of SNe with incorrect 
  redshifts fail our sample cuts, giving a final 
  contamination fraction of $\sim$$1-1.5$\%.  In large part, this reduction is 
  due to cuts on the SALT2 color parameter.  If an SN has an incorrect redshift, 
  SALT2 is twice as likely to infer that its observed-frame colors are inconsistent 
  with normal SNe\,Ia when transformed to the wrong rest frame.

\item \textit{SN\,Ia model}.  The SN\,Ia model used in these simulations 
  is the \citet{Guy10} model with SN\,Ia nuisance parameters 
  from R14 (SALT2 $\alpha = 0.147$, $\beta = 3.13$).
  The parent $X_1$ and $C$ distributions were
  determined by \citet{Scolnic16} for the PS1 spectroscopic sample.  
  We adjusted the parent means of the $X_1$ and $C$ distributions 
  by 1$\sigma$ to better match our data, making $X_1$ lower by 0.17 and 
  $C$ higher by 0.023.  This difference is likely physical; on average,
  $X_1$ is lower and $C$ is higher in massive 
  host galaxies (e.g. \citealp{Childress13}).  Our host follow-up program
  preferentially obtained redshifts of massive galaxies.
\item \textit{CC\,SN templates and diversity}.  CC\,SNe are simulated based 
  on a library of 43 templates in SNANA.  The templates we use
  were originally created for the SN Photometric Classification Challenge 
  \citep{Kessler10} and also used by \citet{Bernstein12}.
  Templates are based on bright, spectroscopically confirmed SDSS, 
  SuperNova Legacy Survey 
  (SNLS; \citealp{Conley11,Sullivan11}), and Carnegie Supernova Project 
  \citep{Hamuy06,Stritzinger11} CC\,SNe with well-sampled light curves.  
  Templates were created from the light curves by warping a 
  model spectrum for each SN subtype to match the light curve fluxes in every 
  broadband filter (see \S\ref{sec:newtmpl}).

  SNANA has 24 II-P templates, 2 IIn 
  templates, 1 II-L template, 7 Ib templates, and 9 Ic templates.
  In this work, we make the assumption
  that reddening in the templates is approximately equal
  to reddening in our data.  This assumption allows us to
  use the \citet{Li11} LFs, which have not been corrected
  for reddening, and SNANA templates, which also include
  intrinsic reddening.  Correcting these templates, the \citet{Li11}
  rates and the \citet{Li11} LFs for reddening are an important avenue for future work.

  We added  a subtype-specific magnitude offset to each CC\,SN template such 
  that the mean simulated absolute magnitude of the subtype matched the mean of
  its \citet{Li11} LF.  By applying a uniform offset 
  to every template in a subtype, the brightness of different templates 
  relative to their subtype is incorporated in our simulations\footnote{We tweaked
    this procedure for SNe\,Ib, which had one anomalously bright template.  All SN\,Ib
    templates were adjusted by individual magnitude offsets such that each template
    matched the mean magnitude of SNe\,Ib given by \citet{Li11}.}.  We 
  also matched the dispersions of the \citet{Li11} LFs by adding an 
  additional, random magnitude offset to each 
  simulated CC\,SN.  This offset was drawn from a Gaussian with a 
  width we adjusted such that the dispersion of the simulated absolute 
  magnitudes for each subtype matched that of \citet{Li11}.

\item \textit{SN Rates}.  SNANA creates a combined SN Ia+CC
  simulation, with each SN type normalized by its 
  rate.  The redshift-dependent SN rates used in this work are
  the same as the baseline model of \citet{Rodney14}.  SNe\,Ia follow 
  measured rates, while CC\,SNe follow the cosmic star formation history.  
  Relative rates of SN types and subtypes are anchored at $z = 0$ by 
  \citet{Li11} and evolve 
  $\propto (1+z)^{\gamma}$, where $\gamma$ is a 
  free parameter tuned to match theory and observations (only a single 
  value for $\gamma$ is needed over the redshift range of PS1).  We used
  $\gamma_{Ia} = 2.15$ and $\gamma_{CC} = 4.5$ \citep{Rodney14}.
\end{enumerate}

\begin{figure}
\centering
\includegraphics[width=3.5in]{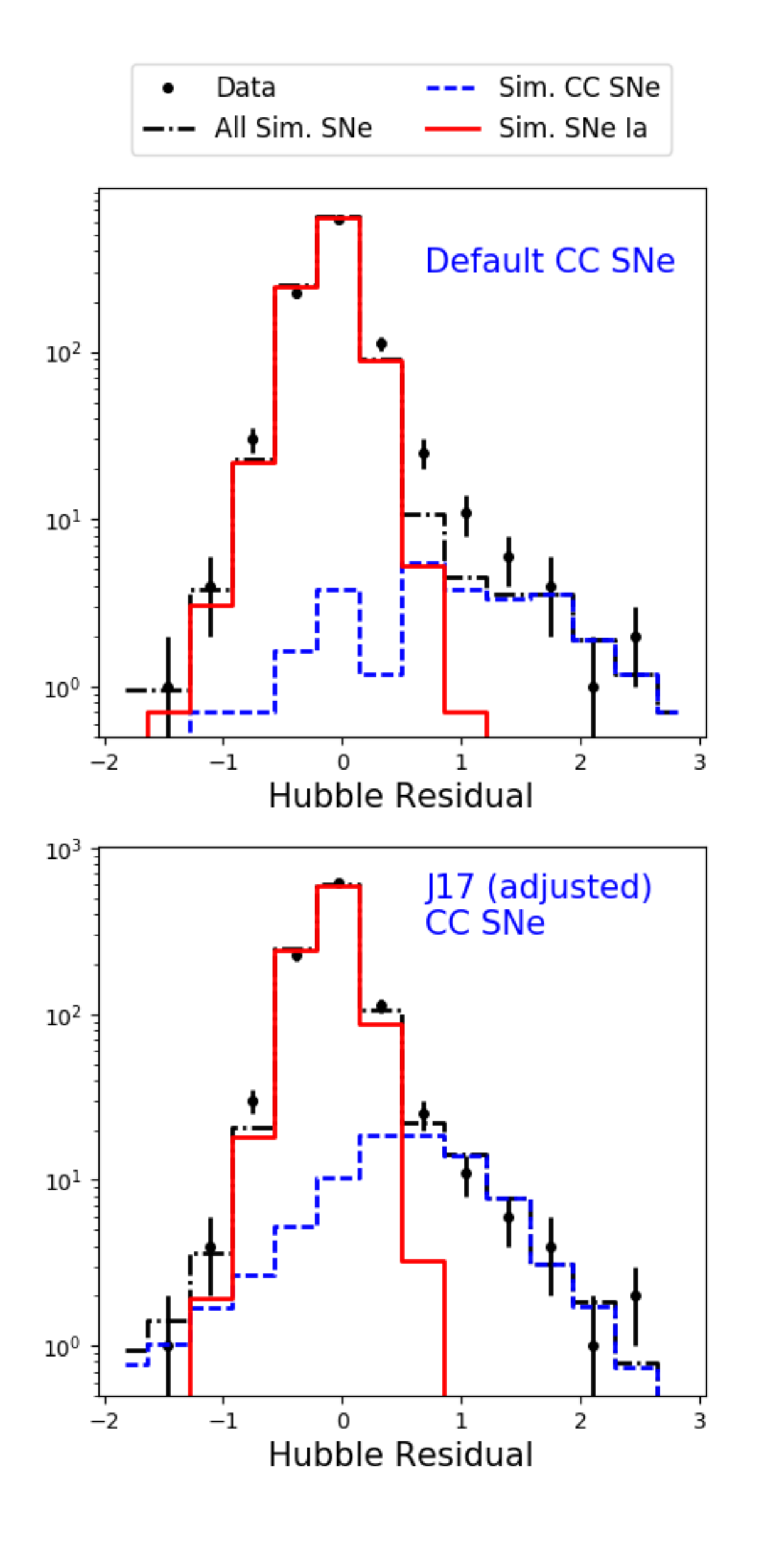}
\caption{Comparison of Hubble residuals before and after 
  empirical adjustments to CC\,SN LFs.  We enlarge Figure 
  \ref{fig:defaultsim}A (top) and compare to our
  adjusted J17 simulations (bottom).  Before empirical adjustments, the 
  simulations contained just 2.4\% CC\,SNe and were a poor match 
  to the data.  After adjustments, the simulations
  have 8.9\% CC\,SNe.  Discrepancies between data and simulations
  in the red end of the SALT2 $C$ 
  distribution can be explained by additional CC\,SNe.}
\label{fig:ccsim}
\end{figure}

Figure \ref{fig:defaultsim} compares our simulations to the 
data after fitting all SNe with the SALT2 
model.  Note that CC\,SN information in this simulation is obtained without
any PS1 analysis or input.  SALT2 fitting 
is an effective way to examine both SNe\,Ia and the light curve parameters of
Ia-like CC\,SNe.  Discrepancies in Figure \ref{fig:defaultsim} indicate
potential biases when measuring cosmological parameters with a
CC\,SN-contaminated sample.

Our simulations agree closely with the data for most light
curve parameters.  The maximum SNR of the simulated light curves matches
the data (\ref{fig:defaultsim}B), as does the distribution of SALT2 
$X_1$ (\ref{fig:defaultsim}C).  However, there 
are too few simulated SNe with red SALT2 colors (\ref{fig:defaultsim}D).
The simulated redshift evolution of $X_1$ and $C$ 
matches the data well (\ref{fig:defaultsim}F and \ref{fig:defaultsim}G).

Though most simulated light curve parameters match our data 
well, the Hubble residuals (\ref{fig:defaultsim}A) show a discrepancy.  We see $\sim$3 times 
more SNe than expected between $0.5 \lesssim \mu - \mu_{\Lambda CDM} < 1.5$ 
mag (these SNe are fainter than SNe\,Ia at their redshifts).  
For this reason, we used light curve-based classifications of our data to
adjust the CC\,SN LFs.
The details of this procedure are discussed in Appendix \ref{app:ccsn}.
We find that the peak of the CC\,SN LF
must be brightened by 1.2 mag for SNe\,Ib/c and 1.1 mag for SNe\,II
in order for our simulations to match our data (Figure \ref{fig:ccsim}).  The dispersion of CC\,SN
templates must be reduced by 55\% for SN\,Ib/c.  
We also add four 1991bg-like SN\,Ia templates and four SN\,IIb template
to SNANA to include a broader range of SN types.

The CC\,SNe LFs in our adjusted simulation are $\sim$5$\sigma$ 
brighter than in \citet{Li11}.  However,
these results do not necessarily imply that the true LFs of 
CC\,SNe show a $\sim$5$\sigma$ inconsistency with \citet{Li11}.  Rather, they 
indicate that our SALT2-based shape and color cuts isolate a region
of CC\,SN parameter space that is not the average.
Although we find it plausible that the CC\,SNe with shapes and colors
most similar to SNe\,Ia have brighter and lower-dispersion LFs than 
CC\,SNe as a whole, further work is 
required to understand the diversity of CC\,SN subpopulations.
Larger sets of high-cadence, high-quality spectral time series 
from which to construct templates are also necessary.  An additional 
factor is that the low statistics in the LOSS volume-limited 
sample require that the shape of the CC\,SN LFs be extrapolated in some way.
We treat CC\,SN LFs as Gaussians, which is most likely a flawed assumption (see 
Figure 16 of \citealp{Li11}).  


\section{Estimating SN\,Ia Distances with BEAMS}
\label{sec:bayes}

\begin{figure}
\centering
\includegraphics[width=3.5in]{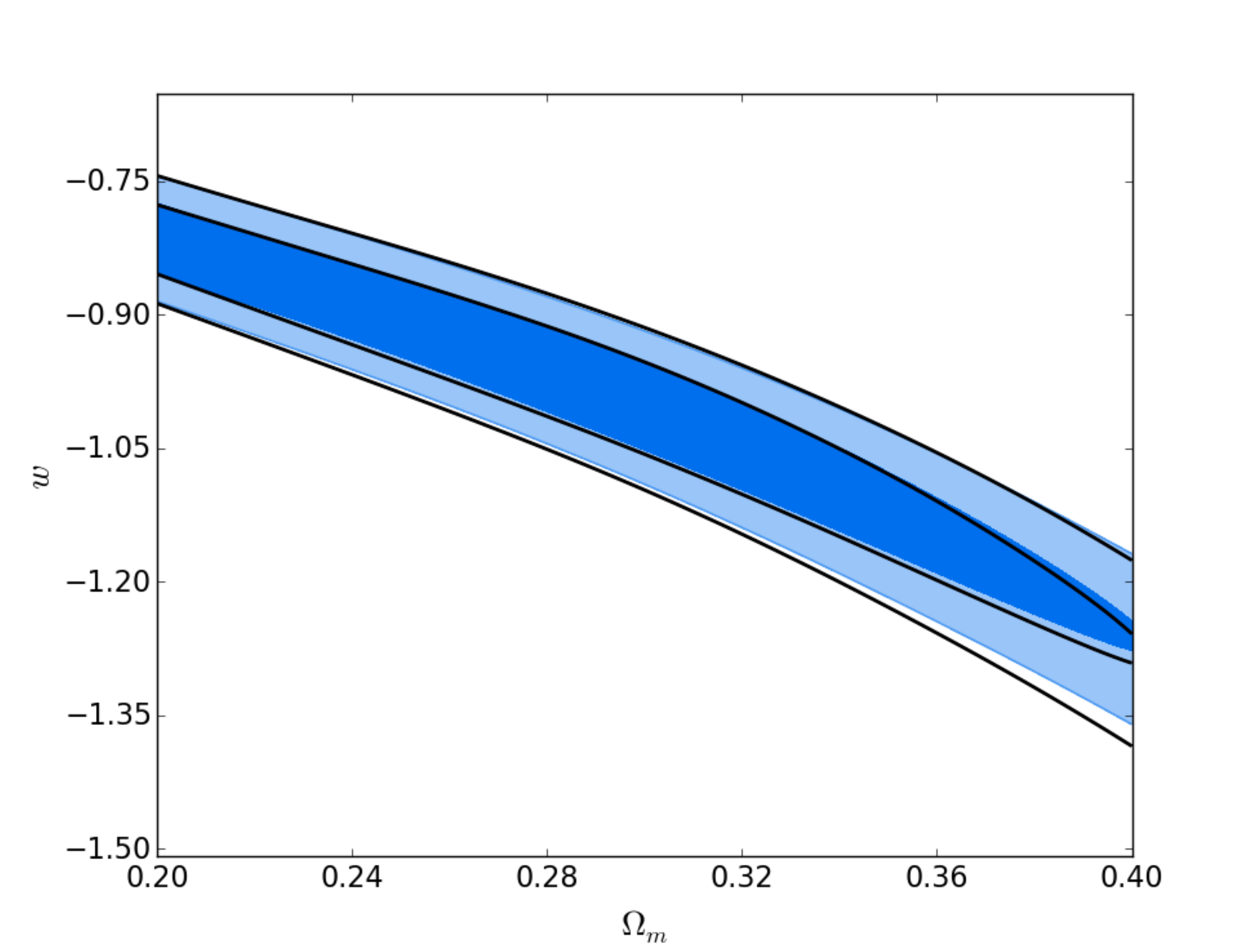}
\caption{Comparing the full SN\,Ia likelihood (filled contours) to the
binned Ia likelihood (black).}
\label{fig:beamsvreal}
\end{figure}

We use the BEAMS
method to obtain SN\,Ia distance measurements that are corrected for
the CC\,SNe contaminating our data (KBH07).  The implementation
of BEAMS suggested in KBH07 solved for distances and cosmological parameters
in a single step; here, we first use BEAMS to solve
for binned SN\,Ia distances and then use CosmoMC \citep{Lewis02}
to determine cosmological parameters.  This procedure will allow us
to more easily combine SN data with complementary CMB and BAO
data in our forthcoming cosmological analysis.  We
summarize the method below.

BEAMS simultaneously determines Ia and CC\,SN distances by sampling a posterior 
probability distribution that includes both SN\,Ia and CC\,SN
populations in the likelihood.  The BEAMS posterior, the probability
of the free parameters $\theta$ given the data, $D$,
is proportional to the product of the individual likelihoods 
for each SN multiplied by the priors on the free parameters:

\begin{equation}
P(\theta|D) \propto P(\theta) \times \prod_{i=1}^N\mathcal{L}_i.
\end{equation}

The simplest suggested likelihood 
from KBH07 uses Gaussian distributions to represent
CC\,SN and SN\,Ia populations:

\begin{equation}
\begin{split}
\mathcal{L}_i = \mathrm{P_i(Ia)} \times \frac{1}{\sqrt{2\pi(\sigma_{i,Ia}^2+\sigma_{Ia}^2)}}\exp(-\frac{(\mu_{i,Ia}-\mu_{Ia}(z_i))^2}{2(\sigma_{i,Ia}^2+\sigma_{Ia}^2)})\\
+  \mathrm{P_i(CC)} \times \frac{1}{\sqrt{2\pi(\sigma_{i,CC}^2+\sigma_{CC}(z_i)^2)}}\\
\times \exp(-\frac{(\mu_{i,CC}-\mu_{CC}(z_i))^2}{2(\sigma_{i,CC}^2+\sigma_{CC}(z_i)^2)}).
\end{split}
\label{eqn:beamslike}
\end{equation}

\begin{figure*}
\centering
\includegraphics[width=7in]{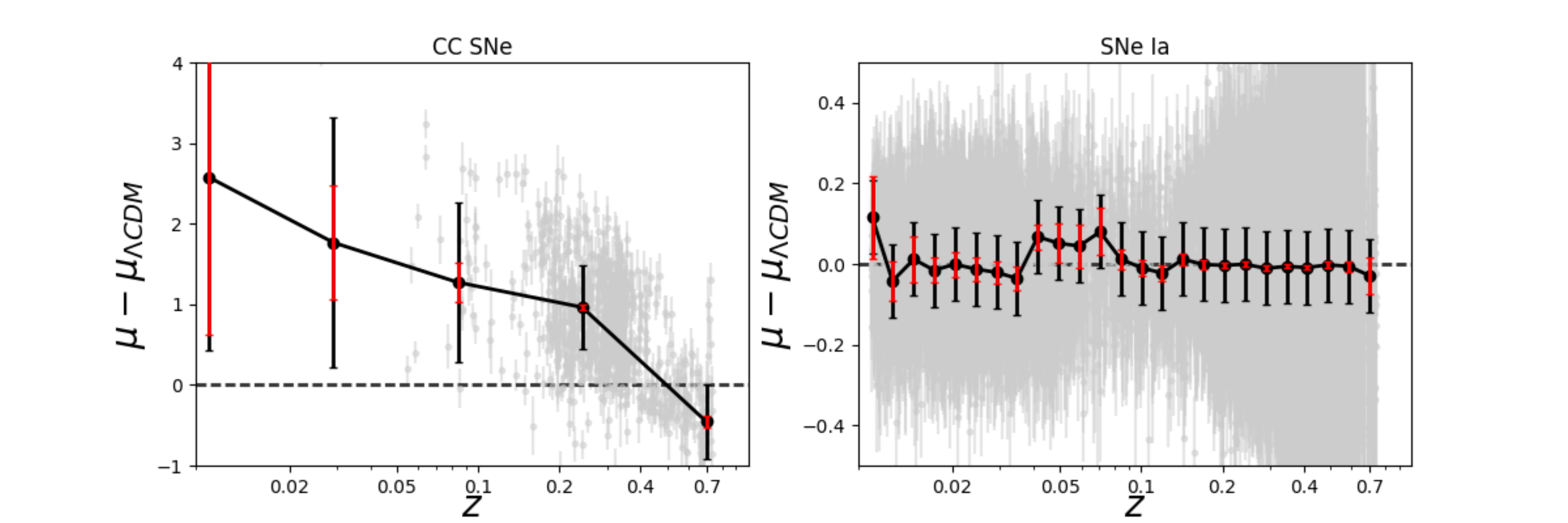}
\caption{Illustration of BEAMS.  Simulated CC\,SNe (left) and SNe\,Ia 
(right) with the redshift-dependent BEAMS parameters $\mu_{CC}$, 
$\mu_{Ia}$ (black points) and $\sigma_{CC}$, $\sigma_{Ia}$ 
(black bars).  Uncertainties on $\mu_{CC}$ and $\mu_{Ia}$ are in 
  red.  We use correct prior probabilities of P(Ia) $=$ 1 for
  SNe\,Ia with correct redshifts and P(Ia) $=$ 0 for all others.}
\label{fig:ccz}
\end{figure*}

\noindent $\mathrm{P_i(Ia)}$ is the prior probability that the $i$th SN 
is of Type Ia.  $\mathrm{P_i(CC)}$, the probability that the SN 
is a CC\,SN, is equal to $1 - \mathrm{P_i(Ia)}$.
$\mu_{i,Ia}$, $\mu_{i,CC}$ and $\sigma_{i,Ia}$, $\sigma_{i,CC}$ 
are the distance modulus and distance modulus 
uncertainties for the $i$th SN, derived using the Tripp 
estimator (Eq. \ref{eqn:salt2}).  We differentiate between 
measured Ia and CC distance moduli from the data because we will allow
the Tripp estimator to use different nuisance parameters
for the SN\,Ia and CC\,SN terms in the likelihood (\S\ref{sec:salt2beams}).
$\mu_{Ia}$, $\sigma_{Ia}$ and 
$\mu_{CC}$, $\sigma_{CC}$ are the means and standard deviations of 
the SN\,Ia and CC\,SN Gaussians, respectively.

The variables $\mu_{Ia}$ and $\mu_{CC}$ are a function 
of the redshift, $z$, of the $i$th SN and 
of cosmological parameters.  The variable $\sigma_{CC}$ is 
redshift dependent as well, primarily due to the changing
mix of CC\,SN subtypes that PS1 is able to discover as a
function of redshift.  We fit for $\mu_{Ia}(z)$, $\mu_{CC}(z)$ and 
$\sigma_{CC}(z)$ by allowing BEAMS to treat them as free parameters at certain fixed 
redshifts $z_b$.  We refer to the set of fixed redshifts 
as ``control points'' following \citet{Betoule14}\footnote{Note that \citet{Betoule14} use
  this method to increase computational efficiency when combining SN\,Ia data with Planck
  priors.  However, their method of reducing SN data to a set of distances
  at redshift control points is well-suited for a BEAMS-like algorithm.}.  Between 
two control points, the distance modulus (and dispersion) 
is interpolated by a linear 
function of log($z$) defined by:

\begin{equation}
\begin{split}
\mu(z) = (1 - \xi)\mu_b + \xi\mu_{b+1}\\
\xi = \mathrm{log}(z/z_b)/\mathrm{log}(z_{b+1}/z_b),
\end{split}
\end{equation}

\noindent where $\mu_b$ is the distance modulus at redshift $z_b$.

\citet{Betoule14} fit to a set of 30 log-spaced 
redshift control points and found that the difference between 
$\Lambda$CDM and the interpolation is always smaller than 1 mmag.  
We used 25 control points for the smaller PS1 redshift range of 
$0.01 < z < 0.7$ (we restrict our sample to $z < 0.7$, as 
very few PS1 SNe can be found at higher redshifts). 
In Figure \ref{fig:beamsvreal}, we compare 
the cosmological constraints from 1,000 individual SNe\,Ia to the 
approximate results derived from the SN\,Ia
distances at 25 control points (P(Ia) $=$ 1 for all SNe\,Ia).
We find that the cosmological constraints 
are nearly identical.

We use 5 log-spaced 
redshift control points for CC\,SNe.  If true SN type probabilities
are known, 5 CC\,SN control points allows BEAMS enough 
flexibility to avoid biasing the Ia likelihood with a poor 
determination of the CC\,SN distribution.  
We allow the intrinsic width of the CC\,SN Gaussian distribution ($\sigma_{CC}$) to vary with 
redshift, but keep the intrinsic width of the SN\,Ia Gaussian fixed.
By using the \texttt{SALT2mu} procedure \citep{Marriner11}, we verified that
the (simulated) uncertainty-weighted dispersion of SNe\,Ia does
not change with redshift for PS1 (this is also a typical assumption
in cosmological analyses; \citealp{Guy10}).
This physically realistic assumption gives BEAMS more leverage to 
discriminate between SNe\,Ia and CC\,SNe, which have much higher 
dispersion than SNe\,Ia.

In total, our baseline implementation of BEAMS has 38 free parameters: 25
SN\,Ia distance moduli at Ia control points, 5 CC\,SN distance moduli at
CC control points, 5 CC\,SN dispersion parameters, 1 SN\,Ia dispersion
parameter,\footnote{Throughout, we have written this dispersion parameter
  as $\sigma_{Ia}$ to distinguish it from $\sigma_{int}$, the global uncertainty
  term used in many previous analyses.  $\sigma_{int}$, defined in \S\ref{sec:lowzsn},
has a different definition from the BEAMS free parameter $\sigma_{Ia}$.},
and the SALT2 nuisance parameters $\alpha$ and $\beta$ which 
are used to compute $\mu_i$ and $\sigma_i$ (discussed below).  BEAMS free 
parameters can be efficiently estimated by sampling the
logarithm of the posterior with a Markov Chain Monte Carlo (MCMC) 
algorithm.  This work uses {\tt emcee}\footnote{\url{http://dan.iel.fm/emcee/current/}}, 
a {\tt Python} MCMC implementation \citep{Foreman13}.  We use {\tt emcee}'s 
Parallel-Tempered Ensemble Sampler to explore the multimodal peaks of the likelihood 
robustly.  Figure \ref{fig:ccz} illustrates the Hubble residual diagram 
from BEAMS using simulated SNe and correct prior probabilities 
(all SNe\,Ia with correct redshifts have P(Ia) $=$ 1 and all other 
SNe have P(Ia) $=$ 0).  Note that if few or no CC\,SNe are in a
given redshift bin, the magnitude and uncertainty of CC\,SN
distances are primarily determined by the priors.

We apply loose Gaussian priors on most BEAMS free parameters, but 
find that with samples of 1,000 SNe or more, the difference between Gaussian
and flat priors is negligible.  For SN\,Ia distances, we apply flat priors.
Though we assume some prior knowledge of the CC\,SN distribution, our 
priors on CC\,SN distance ($\mu_{CC}$ in Eq. \ref{eqn:beamslike})
are very loose; we use broad Gaussians of width 3 mag that are
centered at 2 mag fainter than the SNe\,Ia at each control point.  SALT2 
nuisance parameters have Gaussian priors of width 5 times the 
uncertainties from R14.  Our code is available 
online\footnote{See \citet{jonesbeams17}, with recent updates at https://github.com/djones1040/BEAMS.  Example input
files are also provided.}.

\subsection{SALT2 Light Curve Parameters}
\label{sec:salt2beams}

We use a SALT2 fitting program to measure
SN light curve parameters for our sample.  However, SALT2 parameters
do not directly measure the distance modulus (Eq. \ref{eqn:salt2}).
For BEAMS to measure distances using SALT2 light curve fits, 
the nuisance parameters $\alpha$ and $\beta$ must either be 
fixed to the value from a spectroscopic sample or 
incorporated into BEAMS as free parameters.
We allow $\alpha$ and $\beta$ to be free
parameters here as it is a more 
general test of the method.  Different survey methods, detection 
efficiencies, and selection criteria can significantly bias 
recovered SN parameters \citep{Scolnic16}, which could make it necessary 
for future analyses to be able to fit for these parameters.
In the CC\,SN component of the BEAMS likelihood, we fixed $\alpha$ and
$\beta$ to the nominal value for SN\,Ia spectroscopic samples
(allowing them to float has no effect on
our results).

Because we include $\alpha$ and $\beta$ as free parameters,
the likelihood presented in Eq. \ref{eqn:beamslike} has
a term in the Gaussian normalization factor, $\sigma_i$, that depends on 
$\alpha$ and $\beta$.  The result is a significant bias in the 
derived SN parameters \citep{March11}.  This bias grows for larger SN samples 
(see Appendix B of \citealp{Conley11} and \citealp{Kelly07} for 
details).  The solution adopted in \citet{Conley11} is to neglect the
normalization term when determining $\alpha$, $\beta$, and
$\sigma_{int}$ by using a simple likelihood $\mathcal{L} \propto
\exp(-\chi^2/2)$.  For 1,000 SNe, \citet{Conley11} find that 
the bias from this likelihood is well below the statistical
error.  Though we cannot use
this solution without biasing determinations of the CC\,SN and
SN\,Ia distributions, we use an 
alternative formalism and treat the uncertainties on the 
distance modulus as fixed in the denominator of the 
normalization term (independent of $\alpha$ and $\beta$).  Fixing 
distance modulus uncertainties in the denominator
does not bias $\alpha$, $\beta$, or $w$ and is a very 
modest approximation;
in the PS1 sample, varying $\alpha$ 
and $\beta$ within their 1$\sigma$ errors from R14 gives a 
mean change in uncertainty of only 2 mmag.  No individual 
SN has its uncertainty change by $>$20 mmag.  See
\citealp{Kessler17}, \S8.1 for an alternative solution.

\subsection{Prior Probabilities}
\label{sec:priors}

\begin{figure*}
\centering
\includegraphics[width=7in]{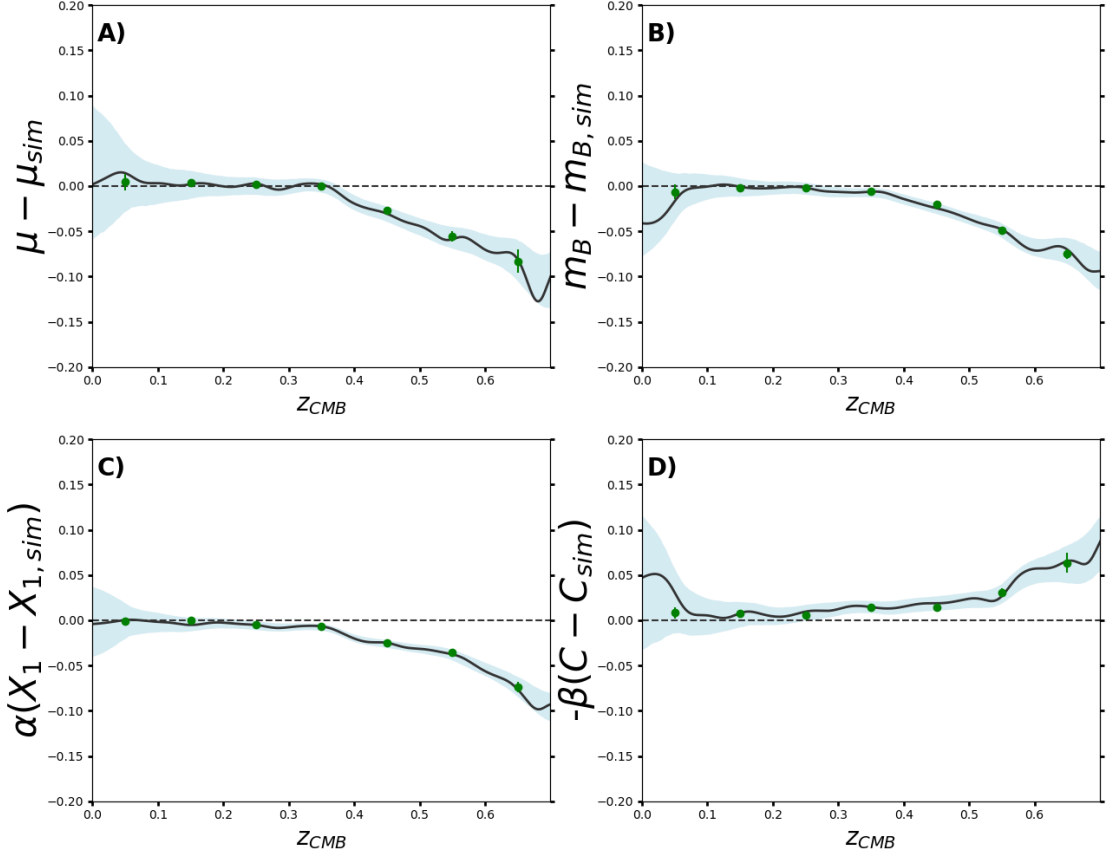}
\caption{Simulated redshift-dependent bias in distance (A), peak B magnitude (B), $\alpha X_1$ (C),
  and $-\beta  C$ (D) for the PS1 photometric sample using 
  non-parametric spatial averaging (black lines with
  95\% confidence intervals in blue) with median bins (points) shown for comparison.
  The PS1 sample has negligible distance (Malmquist) bias until $z \sim 0.3$ 
  and a maximum bias of $\sim$0.1 mag at $z \gtrsim 0.6$.}
\label{fig:malmquist}
\end{figure*}

The BEAMS formalism requires an estimate of the prior 
probability that a given SN is of Type Ia.  This prior can 
be measured by an SN classifier or it can be as simple 
as setting P(Ia) $=$ 1/2 for all SNe.  For our baseline 
analysis, we adopt the PSNID light curve fitter, as implemented
in SNANA \citep{Sako11,Sako14}.
In PSNID, observed SN light curves are fit with perfect, noise-free simulations of the SALT2 
SN\,Ia model and SNANA's CC\,SN templates to determine the
probability that each SN is of Type Ia\footnote{Because
  the simulated CC\,SN models in SNANA are the same as the CC\,SN models
  in the PSNID template library, we used an option in
  PSNID (SNANA v10\_47m and later) that ensures a CC\,SN simulated using a given template
  cannot be classified using a noise-free version of that same template.  This option
  increases the CC\,SN contamination by $\sim$1\%.}.
PSNID estimates P(Ia) from the 
$\chi^2$ of the fit and includes type, redshift, and luminosity priors.  
The set of SNe with P(Ia) $>$ 0.5 has 2.9\% contamination by CC\,SNe
while including 92\% of real SNe\,Ia.


We allow a remapping of the PSNID prior probabilities by adding
two parameters to BEAMS: one that re-normalizes the probabilities 
and a second that shifts them linearly.  The first parameter 
is a scaling factor that corrects for globally skewed prior 
probabilities following \citet{Hlozek12}.
This normalization term allows BEAMS to correct for effects such as 
incorrect redshift-dependent SN rates, inaccurate classifier training, or 
other P(Ia) biases.  The second parameter is a global, linear 
shift in probability to handle incorrect typing near 
P(Ia) $= 0$ or P(Ia) $= 1$ (but requiring $0 <$ P(Ia) $< 1$).  
This is necessary in cases where uncertainty in P(Ia) 
$\simeq$ 1 or P(Ia) $\simeq$ 0 is significant (KBH07).
The relationship between the normalization factor, A, the 
shift parameter, S, and the probability P(Ia) is given by

\begin{equation}
\begin{split}
\tilde{P}\mathrm{(Ia) = \frac{A \times (P(Ia) + S)}{1 - (P(Ia) + S) + A \times (P(Ia) + S)}}\\
0 < \tilde{P}\mathrm{(Ia)} < 1.
\end{split}
\label{eqn:norm}
\end{equation}

\noindent Another 
solution suggested by KBH07 that could be explored in future work is 
adding a probability uncertainty term to the likelihood.

\subsection{Malmquist Bias}

$\mu_{i,Ia}$, the SALT2-derived distance modulus for the $i$th SN,
is subject to Malmquist bias for magnitude-limited surveys such as PS1.
We account for the SN\,Ia Malmquist bias using
PS1 and low-$z$ simulations to determine the redshift-dependent bias of
derived SN\,Ia distances.
We used Monte Carlo simulations of $\gtrsim$10,000 SNe and non-parametric 
spatial averaging to determine and correct for the
trend in distance modulus.  Our spatial averaging algorithm uses
local polynomial smoothing to interpolate the mean distance modulus trend 
across the redshift range.

Our simulations of the spectroscopically confirmed 
low-$z$ SN sample follow R14, who 
use the same $\alpha$ and $\beta$ as our PS1 simulations.
The details of these
low-$z$ simulations and the determination of the spectroscopic
selection function are discussed in detail in \citet[see their
Figure 6 for a comparison between simulations and data]{Scolnic14b}.

Figure \ref{fig:malmquist} shows the simulated, redshift-dependent 
measurement bias in distance modulus, $m_B$, $\alpha X_1$, and 
$-\beta C$.  The average high-$z$ distance modulus 
bias in PS1 is nearly identical to the bias
measured for PS1 spectroscopically confirmed SNe by R14. 
One difference is that the Malmquist bias is
almost negligible in our sample until $z \sim 0.35$.  Some differences
in bias are expected because the R14 bias is dominated by their spectroscopic SN follow-up
selection function.

At $z > 0.5$, we find that the 
bias in $X_1$, $C$, and $\mu$ becomes large as 
flux uncertainties near the epoch of peak brightness are up to a factor of 2
larger than in the lower-$z$ data.  
Greater than 50\% of the $m_B$ and $C$ bias at these redshifts 
is due to our cut on $X_1$ uncertainty, which is effectively an 
SNR cut that increases the selection bias.  Distance 
biases due to cuts on $X_1$ and $C$ are also expected as the data become noisier 
and statistical fluctuations cause more SNe that fall outside the 
luminosity-correlated range to appear on our Hubble diagram 
\citep[see their Figure 4]{Scolnic16}.  Our simulations also show that 
requiring lower $X_1$ uncertainty tends to select narrower measured 
light curve shapes.  Accordingly, Figure \ref{fig:defaultsim}F shows that the 
measured $X_1$ distribution remains largely flat with redshift; 
although SNe with larger $X_1$ values are intrinsically more luminous 
and thus more likely to be discovered, the measurement bias shown 
in Figure \ref{fig:malmquist} has an opposite, and approximately equal, 
effect.

A discussion of systematic error in Malmquist bias determination will be 
presented in our forthcoming cosmological analysis.  This will include
incorporating $\alpha$ and $\beta$ uncertainties, which
can cause differences in the distance bias of $\sim$5 mmag at $z > 0.5$.  Although \citet{Scolnic14b} 
found that the Malmquist bias is not one of the dominant sources of error, the photometric
sample may be subject to different biases than a typical spectroscopic sample
due to its lower average SNR.

We correct all SNe, but only for the SN\,Ia Malmquist bias
(we do not attempt bias corrections based on P(Ia)).  It is
not necessary to correct for the CC\,SN Malmquist bias, as CC\,SNe are not
used to derive cosmological parameters.  However, we implicitly model the CC\,SN
Malmquist bias using BEAMS because BEAMS allows the CC\,SN mean and dispersion
to vary with redshift.

\subsection{Cosmological Parameter Fitting}

Finally, once distance moduli at the 25 redshift control points have
been measured with BEAMS, BEAMS
distances and distance covariance matrices
can be used as inputs into the Cosmological 
Monte Carlo software for cosmological 
parameter fitting (CosmoMC; \citealp{Lewis02}).
For computational efficiency, we did not use the 
full Planck chains in this analysis and instead ran CosmoMC on our 
BEAMS results with a Planck-like prior of $\Omega_{M} = 0.30\pm0.02$.

\section{Cosmological Results from BEAMS}
\label{sec:results}
\begin{figure*}
\centering
\includegraphics[width=7in]{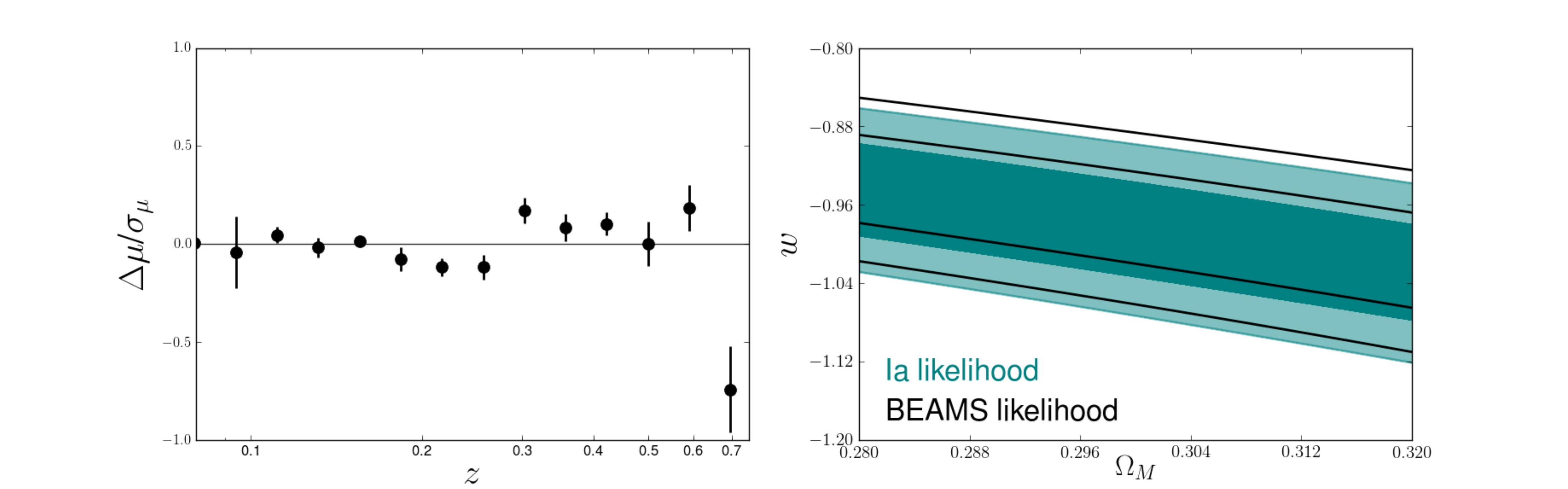}
\caption{Left: average distance modulus bias due to CC\,SN contamination 
(Eq. \ref{eqn:bias}) as a fraction of the statistical 
uncertainty.  Error 
bars are the uncertainty on the median bias from 25 samples.  
The average absolute biases at $z > 0.2$ are $\sim$3 mmag, with
the point at $z \simeq 0.6$ having the largest bias of 6 mmag (with the exception of
the final high-uncertainty point at $z \simeq 0.7$).  There is a slight $z$-dependent
slope, which could bias cosmological parameters, with at 2.3$\sigma$ significance.
Right: 1$\sigma$ cosmological parameter 
likelihood contours from BEAMS compared to the true likelihood 
using a representative sample of 1,000 PS1 SNe.}
\label{fig:baselinebeams}
\end{figure*}

\subsection{Tests with Simulated Data}

We generated 25 simulations of 1,000 PS1 SNe each (25,000 total SNe)
in order to test BEAMS on samples the size of the PS1 photometric sample.
We add simulated low-$z$ samples of 250 SNe\,Ia each, the approximate
number that will be included in our forthcoming cosmological analysis.
The results presented here use the J17 CC\,SN simulations (Appendix
\ref{app:ccsn}), as they have CC\,SN LFs that match our data.

\begin{deluxetable*}{lccccccccccc}
\tabletypesize{\scriptsize}
\tablewidth{0pt}
\tablecaption{Results from BEAMS}
\tablehead{&\multicolumn{5}{c}{PS1 Simulations}&&\multicolumn{5}{c}{PS1 Data}\\*[2pt]
  \cline{2-6}\cline{8-12}\\
&bias&$\sigma_{\textrm{bias}}$\tablenotemark{a}&$\sigma_{\textrm{stat}}$\tablenotemark{b}&bias$/\sigma_{\textrm{stat}}$&$\Delta\sigma_{\textrm{stat}}$&&bias&$\sigma_{\textrm{bias}}$\tablenotemark{a,c}&$\sigma_{\textrm{stat}}$\tablenotemark{b}&bias$/\sigma_{\textrm{stat}}$&$\Delta\sigma_{\textrm{stat}}$}

\startdata

$\mu$\tablenotemark{d}&0.000&0.001&0.031&0.0&0.001 (3\%)&&-0.040&0.019($\pm$0.09)&0.074&-0.4&0.010 (14\%)\\
$\alpha$&0.004&0.000&0.006&0.6&0.000 (3\%)&&0.001&0.001($\pm$0.005)&0.012&0.1&0.001 (5\%)\\
$\beta$&0.088&0.008&0.073&1.2&0.004 (6\%)&&0.199&0.018($\pm$0.10)&0.154&1.4&0.009 (6\%)\\
$w$&-0.005&0.004&0.048&-0.1&0.002 (3\%)&&-0.040&0.012($\pm$0.084)&0.095&-0.4&0.008 (8\%)\\

\enddata
\tablecomments{Bias and increase in uncertainty due to CC\,SN contamination.  All quantities
  shown are taken from the median of 25 samples.  Bias is defined in Eq. \ref{eqn:bias} for
  simulations and Eq. \ref{eqn:databias} for data (bias in data is relative to R14 parameter measurements).}
\tablenotetext{a}{Uncertainty on the median bias.}
\tablenotetext{b}{Statistical uncertainty on each 
  parameter from a single sample.}
\tablenotetext{c}{In parentheses, we show the estimated uncertainty
  on the R14 values.  Because our PS1 data are correlated with R14
  (they share the low-$z$ sample), we take Monte Carlo samples of 
  100 simulated PS1 SNe and combine each with the R14 low-$z$ sample,
  taking the standard deviation of measurements from these combined data as
  the uncertainty.}
\tablenotetext{d}{Averaged over $0.08 < z < 0.7$.}
\label{table:beamsparams}
\end{deluxetable*}





To focus on biases from CC\,SN contamination, we define the
CC\,SN bias $\Delta$ and the increase in statistical uncertainty due to
CC\,SNe, $\Delta \sigma_{stat}$, for a given parameter $P$:

\begin{equation}
\begin{split}
\Delta = P_m - P_{Ia},\\
\Delta \sigma_{stat} = \sigma(P_m) - \sigma(P_{Ia})
\label{eqn:bias}
\end{split}
\end{equation}

\noindent where $P_m$ is the measured parameter from the BEAMS method and
$P_{Ia}$ is the measured parameter from the BEAMS method using SNe\,Ia
alone and setting all prior probabilities equal to one.
For the 25 simulated samples, the
average $w_{Ia}$ value is -1.001$\pm$0.009.  The RMS of $w_{Ia}$ is 0.045,
consistent with the mean statistical uncertainty (0.048).

We compare the Ia-only
distances, SN parameters, and $w$ measurements
against our results from the BEAMS method in
Table \ref{table:beamsparams}.  
Figure \ref{fig:baselinebeams} shows that the
binned distances are biased by less than 20\%
of their uncertainties with the exception of the final control 
point.  Typical biases are $\sim$3 mmag and the
largest average bias from the 25 samples (aside from the final high-uncertainty control point) is
6 mmag at $z \simeq 0.6$.

The SN parameters $\alpha$ and $\beta$ are biased
by 3\%, or 1-1.5 times the average statistical error.
$\sigma_{\textrm{Ia}}$ is 
biased by 4\%, 0.3 times the average statistical error.
Note that $\sigma_{Ia}$ (in Eq. \ref{eqn:beamslike})
is functionally similar to 
the SN intrinsic dispersion, $\sigma_{int}$.
These biases are small enough that they would 
be difficult to measure in real data.  A possible cause of these biases
is that Ia-like CC\,SNe have color laws more consistent with Milky Way dust 
($\beta \sim 4.1$) and different shape-luminosity correlations.  

We find that $w$ has a median bias of \wbias\ due to CC\,SN
contamination, \wbiasper\ of the statistical error on $w$.
While our analysis is consistent with no bias, we assign a
systematic uncertainty on $w$ of 0.005+0.004 $=$ 0.009, though the
true systematic uncertainty could be higher due to uncertainties in
CC\,SN simulations (\S\ref{sec:varresults}).
The statistical uncertainty on $w$ in this case is just 3\% higher than the
statistical uncertainty from SNe\,Ia alone.  This result is consistent 
with KBH07, who find that BEAMS can yield nearly optimal 
uncertainties (we discuss BEAMS uncertainties further in 
\S\ref{sec:beamserr}).

If we compare the bias on $w$ to a na\"{i}ve method 
of measuring $w$ with photometrically classified 
SNe, the advantage of using
BEAMS is obvious.  For our 25 1,000-SN samples, we take
likely SNe\,Ia (P$_{PSNID}(Ia) > 0.5$) and estimate cosmological parameters
assuming that all of these SNe are Type Ia (\citealp{Campbell13}
used a similar method of cutting the sample based on PSNID classifications).  Making 
this cut removes 8\% of the true SNe\,Ia in our sample and 
yields a final sample
contaminated by 2.9\% CC\,SNe.  In spite of
having a sample comprised of $>$97\% SNe\,Ia, the average bias on $w$ is -0.025$\pm$0.004,
a factor of five higher than our BEAMS results.
The bias is $>$50\%
of the statistical uncertainty on $w$
and has 6$\sigma$ significance, while the
BEAMS result is consistent with no bias.
The statistical uncertainty on $w$ from this method is 6\% higher,
compared to 3\% higher from BEAMS.
Even a cut of P$_{PSNID}(Ia) > 0.9$
yields a bias on $w$ of 0.011$\pm$0.003 ($>$3$\sigma$ significance) at the cost of removing 17\% of
real SNe\,Ia.  Furthermore, while BEAMS allows these probabilities to be adjusted
by the method, treating them as fixed in this simplistic method
increases the possibility of biased classifications due to incompleteness in
the CC\,SN template library.
It is clear that BEAMS outperforms this simple cut-based analysis, though
this na\"{i}ve method could still be effective with significantly
improved classification methods.

\subsection{Comparing Real Pan-STARRS Photometric Supernovae to Rest et al. (2014)}
\label{sec:datacosmo}

Rather than analyzing the full PS1 sample, we
analyze 25 random draws of PS1 SNe to compare R14 measurements
$-$ and uncertainties $-$ directly to
measurements from CC\,SN-contaminated samples of the same size.
Because 96 R14 SNe\,Ia pass our sample cuts,
we draw samples of 104 photometric SNe in order that our subsamples
each contain an average of 96 SNe\,Ia (and 8 CC\,SNe; we also use
reprocessed R14 light curves).
We do not explicitly require these random samples to have the same 
redshift distribution as the PS1 spectroscopic sample. 
However, the redshift distribution of the PS1 photometric sample 
is similar to that of R14 (a nearly identical range and median
redshift, though the photometric sample does include more faint
SNe\,Ia with red colors).

For subsamples of PS1 data, we report parameter biases
relative to R14:

\begin{equation}
\begin{split}
  \Delta = P_m - P_{R14}\\
\Delta \sigma_{stat} = \sigma(P_m) - \sigma(P_{R14})
\label{eqn:databias}
\end{split}
\end{equation}

\noindent where $P_{R14}$ and $\sigma(P_{R14})$ refer to
a parameter and its uncertainties from R14.

Although R14 does not have enough SNe to test for small biases 
in $w$, the data still allow for a consistency check that is independent
of the myriad assumptions made in simulations.  In addition, 
the 96 SNe from R14 with low-$z$ SNe can provide 
constraints on the bias of the nuisance parameters $\alpha$, $\beta$, and 
$\sigma_{Ia}$ due to the BEAMS method.  We include low-$z$ SNe
because BEAMS is more robust when it has a spectroscopically confirmed
sample as part of the data and has difficulty measuring accurate 
SN\,Ia dispersions for small samples.

We find that the measured distances, SN nuisance parameters $\alpha$ 
and $\beta$, and $w$ are consistent with R14 (Table 
\ref{table:beamsparams}).  We may be seeing the
same hints of a bias toward higher values of $\beta$ 
that we find in simulations but they have under 2$\sigma$ significance.
The bias in $\alpha$ is not statistically significant (0.1$\sigma$).

The average of $w$ from 25 104-SN samples is 
consistent with the measurements 
from reprocessed R14 light curves (0.4$\sigma$ lower, where
$\sigma$ is the statistical uncertainty
from R14).  The uncertainties on $w$ are 15\% higher 
and distance modulus uncertainties are 14\% higher,
likely due to the lower average SNR of photometric PS1 light curves.
The median SNR at peak is 22 for all PS1 SNe, compared to a
median SNR at peak of 38 for spectroscopically classified SNe.

\section{Results from BEAMS Variants}
\label{sec:varresults}

\begin{figure}
\centering
\includegraphics[width=3.5in]{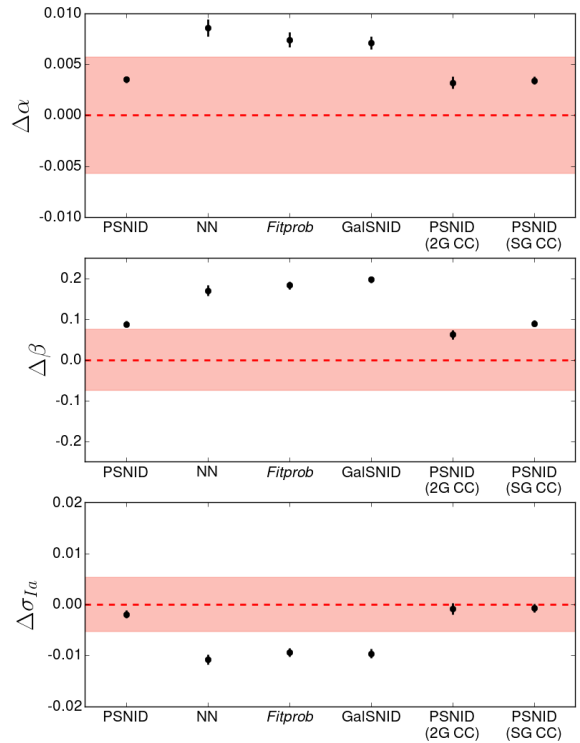}
\caption{Bias in SALT2 $\alpha$, $\beta$, and $\sigma_{Ia}$ measured from
  25 simulations of 1,000 SNe each, with the shaded regions indicating typical 
  uncertainties on each parameter from SN\,Ia-only samples.
  $\sigma_{Ia}$ is too low by $\sim 0.005-0.01$, 
  while $\alpha$ and $\beta$ are too high by $\sim$0.005-0.01 and 
  $\sim$0.1-0.2, respectively ($\sim$1-2$\sigma$).  It is likely that 
  reddened CC\,SNe are responsible for the higher color term (more 
  consistent with Milky Way dust than the SN\,Ia color law).
  ``2G CC'' and ``SG CC'' refer to the two-Gaussian and skewed Gaussian 
  CC\,SN parameterizations, respectively.}
\label{fig:nuisance}
\end{figure}

\begin{figure}
\centering
\includegraphics[width=3.5in]{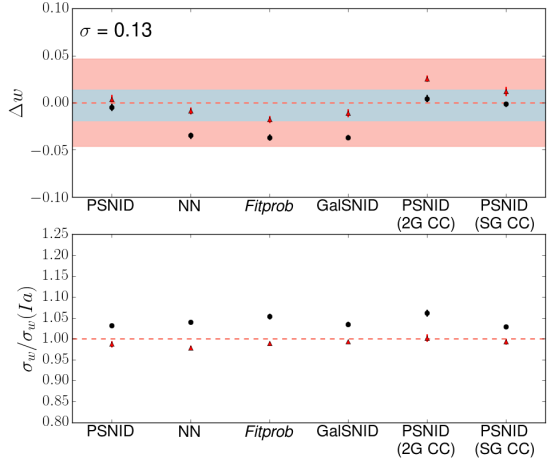}
\caption{$w$ bias (top) and increased uncertainties (bottom)
  due to different
  P(Ia) priors and CC\,SN parameterizations in BEAMS (black points).  We 
  show the median from 25 samples of 1,000 
  simulated SNe.  Red points show the biases
  with $\alpha$ and $\beta$ fixed.
  In the top panel, the statistical error on $w$ from SNe\,Ia
  is shown in the red band and the dispersion of the values
  given in Table \ref{table:wbeams} in blue.
  Red points have lower uncertainties than the Ia-only
  uncertainties because fixing $\alpha$ and $\beta$
  neglects their uncertainties.  ``2G CC'' and ``SG CC'' refer to the 
  two-Gaussian and skewed Gaussian CC\,SN parameterizations, respectively.}
\label{fig:bias}
\end{figure}

\begin{figure}
\centering
\includegraphics[width=3.5in]{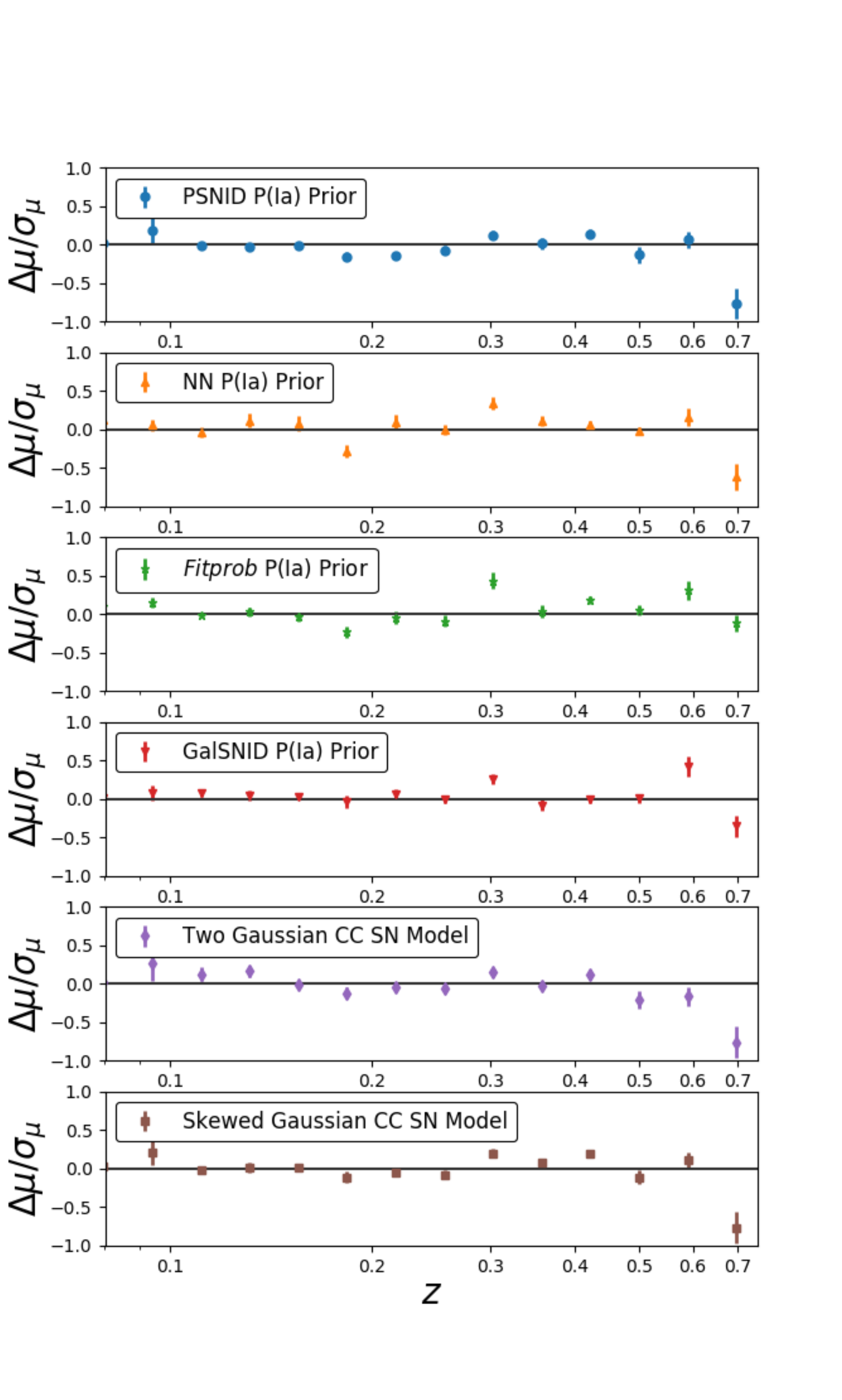}
\caption{Distance bias due to CC\,SN contamination as a fraction of the distance uncertainty
for each BEAMS variant.  Small systematic 
discrepancies begin to appear at $z \geqslant 0.3$.}
\label{fig:beamshubble}
\end{figure}

The BEAMS method measures $w$ with no significant bias
due to CC\,SN contamination and a statistically 
insignificant bias in PS1 data.  However, the reliability of these 
results could depend on the assumptions that we made when 
generating CC\,SN simulations and implementing BEAMS.  
We now expand our study of systematic uncertainties
in simulations by applying alternative SN classification methods,
including ones with less dependence on the accuracy of our CC\,SN simulations,
and adjusting the CC\,SN likelihood model.

\subsection{Analysis Variants}
In total, we test three additional methods of
determining the prior probability P(Ia) (Eq. \ref{eqn:beamslike}) $-$ the Nearest
Neighbor, \fp, and GalSNID classifiers $-$ and
two additional CC\,SN models.  The two additional
CC\,SN models include a two-Gaussian model
and a single, asymmetric Gaussian model.  Nearest Neighbor (NN)
and \fp\ are light curve-based classification methods.  NN uses SALT2 light curve
parameters to classify SNe based on whether they lie
nearer to simulated SNe\,Ia or simulated CC\,SNe in
$X_1$, $C$ and redshift space while
\fp\ uses the $\chi^2$ and degrees of
freedom of the SALT2 light curve fit to measure a probability.
GalSNID \citep{Foley13} uses the fact that, unlike CC\,SNe, many
SNe\,Ia explode in galaxies with
old stellar populations, and thus uses only host galaxy properties
to derive an SN type probability.  We expand the GalSNID method to use
observables from host galaxy spectroscopy in addition to photometric
observations.  PSNID is the best method; NN yields a sample with 6.5\%
contamination at P(Ia) $>$ 0.5 and 3.8\% contamination at P(Ia) $>$ 0.9,
\fp\ yields a sample with 6.1\% contamination at P(Ia) $>$ 0.5 (4.1\% at P(Ia) $>$ 0.9)
and GalSNID gives a sample with 9.3\% contamination (7.2\% at P(Ia) $>$ 0.9; the total
contamination in the sample is 9.7\%).
The details of these variants are given in Appendix \ref{sec:syserr}.

We note that the best approach would be a hybrid
one that takes advantage of all classifiers.  Though we keep these
classifiers as separate here in order to explore the effect of
different classification assumptions, \citet{Kessler17}, for example,
combine a \fp\ $>$ 0.05 cut with the NN classifier.  Combining GalSNID
priors with a light curve-based classifier is another promising option 
for future work.

We test each variant on 25
samples of 1,000 simulated PS1 SNe.  
Though we discuss the ways in which
distances and nuisance parameters are affected by these variants,
we focus primarily on measurements of $w$.  The RMS of these variants gives
an estimate of the systematic uncertainty on $w$, $\sigma_w^{CC}$, an
error which could be reduced in the future by improved SN classification methods.
It could also be reduced by testing our best single classifier on a robust set
of CC+Ia SN simulations that include a larger set of CC\,SN templates and
several methods of adjusting CC\,SN rates and LFs to match the data.

\subsection{Systematic Uncertainty on $w$}

We examine two situations in this section: one where $\alpha$ and $\beta$
are measured by the BEAMS method, and one where $\alpha$ and $\beta$
are fixed to the values measured from spectroscopic samples.
In the case where $\alpha$ and $\beta$ are measured by
the BEAMS method, Figure \ref{fig:nuisance} shows the bias on
$\alpha$ and $\beta$ from each classifier.
$\beta$ biases in particular can cause large distance biases at $z > 0.5$, as the average
SN color at these redshifts is $\sim$-0.1 (for a bias in $\beta$ of 0.2,
$\Delta\beta \times \overline{C} = 20$ mmag).

If $\alpha$ and $\beta$ are fixed, BEAMS requires
very little information to give robust measurements
of $w$.  We test the effect of fixing $\alpha$ and $\beta$ for
all variants and also compare to
the case where BEAMS has minimal prior information: we
set P(Ia) $=$ 1/2 for all photometric SNe while still
fixing P(Ia) $=$ 1 for low-$z$ SNe.
If $\alpha$ and $\beta$ are fixed, the largest
absolute bias on $w$ is -0.018 (the \fp\ classifier)
and the P(Ia) $=$ 1/2 case gives a $w$ bias of only -0.011.
The biases are approximately twice as high
if we instead allow BEAMS to fit for $\alpha$ and $\beta$,
and four times as high for the P(Ia) $=$ 1/2 case,
worse than all other methods (a $w$ bias of -0.043).

Table \ref{table:wbeams} and Figure \ref{fig:bias} 
show the median bias and increase in uncertainty on $w$ due to each P(Ia)
method and CC\,SN model.
Figure \ref{fig:bias} shows the bias before and after fixing $\alpha$ and $\beta$.
We find that alternate CC\,SN models have only a small effect on the
measurement of $w$.  Our lowest $w$ bias of -0.001$\pm$0.003
comes from the skewed Gaussian CC\,SN model; however,
the results from these three CC\,SN treatments are
statistically consistent (with the exception of the two-Gaussian
model with $\alpha$ and $\beta$ fixed, which appears to have
difficulty robustly measuring both CC\,SN Gaussians).

\begin{deluxetable*}{lcccc}
\tabletypesize{\scriptsize}
\tablewidth{0pt}
\setlength{\tabcolsep}{1pt}
\tablecaption{Cosmological Results from BEAMS Variants}
\tablehead{Method&$\Delta w$\tablenotemark{a}&$\sigma_{\textrm{stat}}$\tablenotemark{b}&$\Delta w/\sigma_{\textrm{stat}}$&$\Delta \sigma_{\textrm{stat}}$}

One Gaussian\tablenotemark{c}&-0.005$\pm$0.004&0.050&-0.1&0.002 (3\%)\\*[2 pt]
Two Gaussians\tablenotemark{c}&0.004$\pm$0.004&0.051&0.1&0.003 (6\%)\\*[2 pt]
Skewed Gaussian\tablenotemark{c}&\textbf{-0.001$\pm$0.003}&0.050&-0.0&0.001 (2\%)\\*[2 pt]
\\
&\multicolumn{4}{c}{P(Ia) Method\tablenotemark{d}} \\*[2 pt]
&$\Delta w$&$\sigma_{\textrm{stat}}$&$\Delta w/\sigma_{\textrm{stat}}$&$\Delta \sigma_w$\\*[2 pt]
\tableline\\
PSNID&\textbf{-0.005$\pm$0.004}&0.050&-0.1&0.002 (3\%) \\*[2 pt]
NN\tablenotemark{e}&-0.009$\pm$0.004&0.047&-0.2&-0.001 (-2\%) \\*[2 pt]
$Fitprob$\tablenotemark{e}&-0.018$\pm$0.004&0.047&-0.4&-0.001 (-1\%) \\*[2 pt]
GalSNID\tablenotemark{e}&-0.011$\pm$0.004&0.048&-0.2&-0.000 (0\%) \\*[2 pt]
\enddata
\tablenotetext{a}{The median bias on $w$ and its uncertainty.}
\tablenotetext{b}{The statistical uncertainty on $w$ from a single sample of 1,000 PS1 SNe.}
\tablenotetext{c}{Using PSNID for the P(Ia) prior probabilities.}
\tablenotetext{d}{Using a single-Gaussian CC\,SN model.}
\tablenotetext{e}{For these classifiers, we keep $\alpha$ and $\beta$ fixed to their known values.  $\Delta\sigma_w$ is negative in some cases, because fixing $\alpha$ and $\beta$ neglects the contribution of nuisance parameter uncertainties to the uncertainty on $w$.}
\tablecomments{Bias of $w$ in simulations from each CC\,SN model and
prior probability method.  We take the median of 25 samples of 
1,000 PS1 SNe.
For each increase in uncertainty ($\Delta \sigma_w$), 
we show its percent increase in parentheses.  Methods 
with the lowest bias are 
highlighted in bold.}
\label{table:wbeams}
\end{deluxetable*}

Using all variants, $\sigma_w^{CC}$ has 
an average value of \simsysp\ (\simsyspper\ of the statistical 
error) if $\alpha$ and $\beta$ are fixed for the NN, GalSNID,
and \fp\ classifiers (these classifiers give twice the bias on $\alpha$
and $\beta$ as PSNID does).  The uncertainty is due to the 
dispersion of the systematic uncertainty from sample to sample.
BEAMS distances (Figure \ref{fig:beamshubble}) and nuisance 
parameters (Figure \ref{fig:nuisance}) are consistent to within 1$\sigma$, 
regardless of the method.

We note that in some cases fixing $\alpha$ and $\beta$ may
subject the sample to additional systematic
uncertainty.  For example, $\alpha$ and $\beta$
could be different in a photometric sample because the host
galaxy spectroscopic follow-up selects bright
hosts.  Host properties correlate with shape and color, which in turn
can affect the measured $\alpha$ and $\beta$ \citep{Scolnic14}.
However, these biases are well-known and can in principle be simulated and corrected
for (see \citealp{Scolnic16}).

Current measurements of $w$ (e.g. B14) have approximately
equal statistical and systematic uncertainties.
Therefore, a measurement of $w$ biased by less than
about half the statistical uncertainty (0.024 in this work),
such as the value of $\sigma_w^{CC} = 0.014$ measured here,
does not prohibit a robust measurement of
$w$.  Any bias larger than that
$-$ such as the alternative classifiers discussed in this section
without $\alpha$ and $\beta$ fixed $-$
will dominate the systematic error budget
and make it unlikely that photometric SN
samples can be competitive with spectroscopically
classified samples.  For future surveys, such as DES and
LSST, this bias may be approximately equal to the statistical
error and must be reduced through improved classification
methods or a better understanding of CC\,SNe to yield
accurate results.

\section{Discussion}
\label{sec:discussion}

The PS1 photometric SN sample is the largest SN\,Ia
sample, but using it to optimally measure cosmological parameters
$-$ particularly if the nuisance parameters $\alpha$ and $\beta$ are
unknown or observationally biased $-$ requires accurate SN type
probabilities.  These in turn rely 
on our understanding of the PS1 sample and the CC\,SNe in it. 
Evaluating how our incomplete knowledge of CC\,SNe could bias 
the results is difficult. In this section, we discuss how
CC\,SN simulations could be improved in the future.  
We also present alternatives to our implementation of BEAMS 
and measure the degree to which different methods
and priors affect the statistical uncertainty on $w$.

\subsection{Generating Reliable CC\,SN Simulations}
\label{sec:ccsimdisc}

Evaluating the reliability of our method would be subject 
to fewer uncertainties if CC\,SN simulations were 
more robust.  These simulations are currently subject to
two primary limiting factors: the assumption that
the CC\,SN LF is Gaussian with measured mean and RMS from \citet{Li11}
and the limited CC\,SN template diversity.

\begin{figure}
\centering
\includegraphics[width=3.5in]{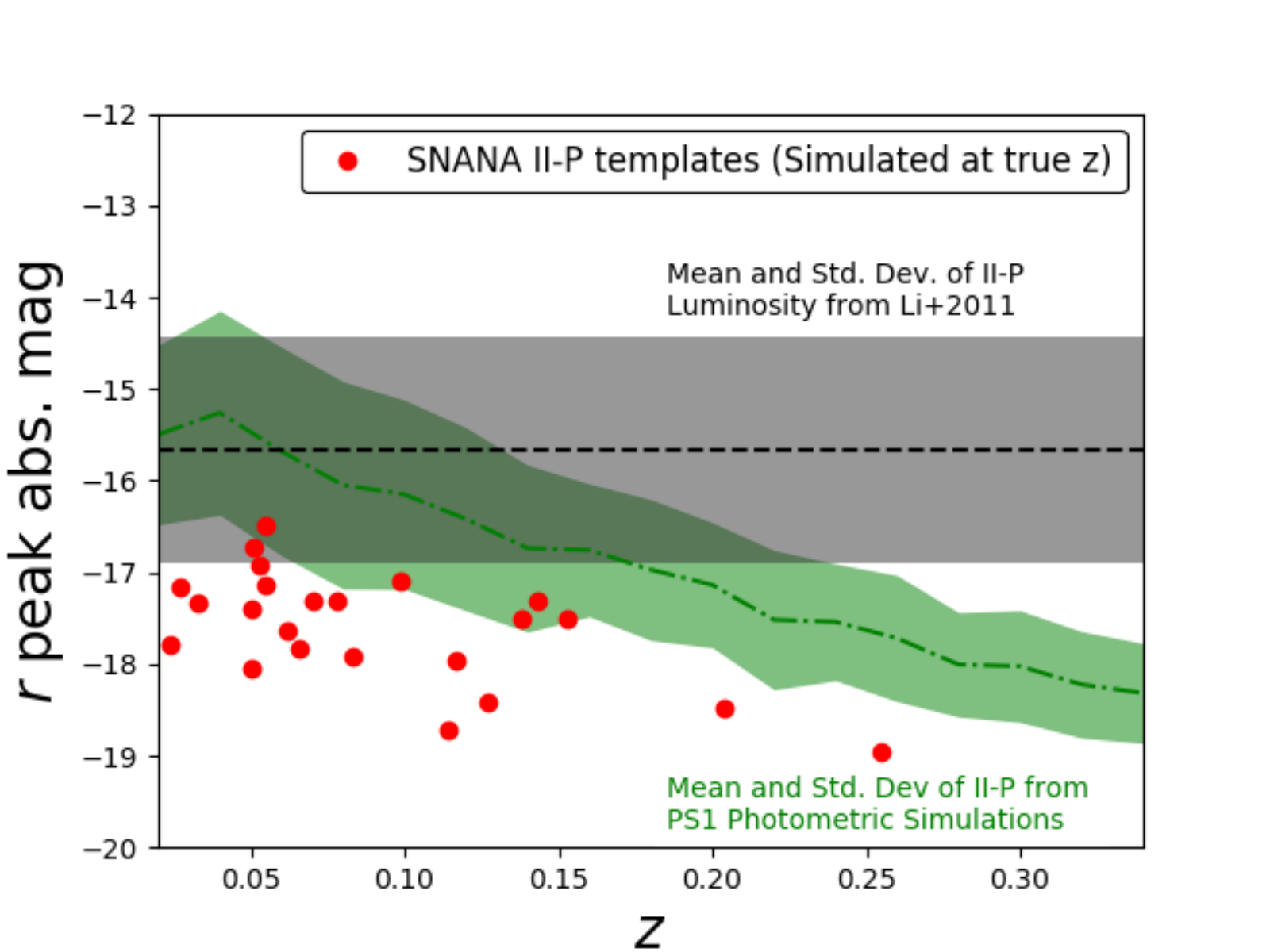}
\caption{In this work, SNANA II-P templates from SDSS (red)
  are made fainter to match \citet{Li11} LFs (gray) and then
  used to generate simulations of the PS1 survey (green).
  SNANA II-P templates are typically 2-3$\sigma$ brighter than the mean
  magnitude of the population from \citet{Li11}.}
\label{fig:ccsne}
\end{figure}

Figure \ref{fig:ccsne} shows that the assumption of
the shape of CC\,SN LFs could have a strong
impact on the fraction of bright CC\,SNe.
While the Malmquist bias for SNe\,Ia is $\sim$0.1 mag at maximum, 
Type II SNe observed at the median PS1 survey redshift are up to 
3 magnitudes $-$ and 2-3 standard deviations $-$ brighter 
than the peak of their LF.  Determining the frequency of such
bright CC\,SNe requires measuring the shapes
of their LFs with better precision than what is 
currently available from volume-limited surveys such as 
\citet{Li11}.  Due to low statistics, 
our current simulations treat the LFs of each SN subtype as 
Gaussian, a flawed assumption.

Generating more robust simulations also requires additional, 
diverse CC\,SN templates.  Our simulations sample the 
luminosity, shape, and color distribution of most CC\,SN subtypes with just 
a few templates.  In addition, the luminosity distribution of
these templates is heavily biased; nearly all CC\,SNe 
currently used as templates are much brighter than the mean luminosity 
of their subtypes.  Our method makes these bright templates
fainter to match the \citet{Li11} LFs, implicitly assuming
that faint CC\,SNe have similar light curves to bright 
CC\,SNe.  A better approach would be based on CC\,SN 
templates that sample the full range of luminosity 
space for CC\,SNe.

We note that additional high-SNR CC\,SN light curves and spectra
exist, but require careful smoothing, interpolation, and
spectral mangling to be a reliable addition to the SNANA template 
library.  We have added 
SNe\,Ia-91bg and SN\,IIb templates to SNANA (Appendix \ref{sec:newtmpl}), but assembling and 
mangling all available CC\,SN light curves and templates is beyond 
the scope of this work.

In the absence of additional templates and improved LF measurements,
we can use GalSNID and \fp\ classifications to give measurements of
$w$ some degree of independence from these sources of uncertainty.
Though these classifiers are sub-optimal compared to classifiers such as PSNID,
they give a unique set of probabilities that do not rely 
on simulations for training (though \fp\ is implicitly dependent
on the nature of CC\,SN light curves contaminating our sample).
\fp\ and GalSNID \textit{explicitly} depend on 
simulations only through their rates priors.  Adjusting these
priors by a factor of 2 biases $w$ by $\sim$20\% of the
statistical uncertainty or less.

\subsection{Alternatives in Implementing BEAMS}
\label{sec:beamsvardisc}

In determining cosmological parameters with the BEAMS method, we made a set of
choices with a modest number of free parameters that reproduced
the full cosmological parameter likelihoods.  We found that most choices, e.g.
varying priors or adding additional CC\,SN bins, made little
difference provided that we had a large number of MCMC steps and few enough
parameters.

Two additional choices can improve the systematic error due to CC\,SN contamination.
First, though fixing $\alpha$ and $\beta$ does not improve the 
accuracy of the BEAMS method when using PSNID priors,
it does improve the accuracy when using the NN, \fp\ and GalSNID, methods
with less accurate classifications.  With $\alpha$ and $\beta$
fixed, NN, \fp\ and GalSNID are competitive with the
more sophisticated light curve based methods.  If we choose to either keep
$\alpha$ and $\beta$ fixed when measuring $w$ from these classifiers,
we find that the $\sigma_w^{CC}$ decreases by $\sim$30\% on average.  In Pan-STARRS,
spectroscopically confirmed SNe can measure these parameters with low
uncertainty, and fixing them for our future cosmology
analysis in some or all methods could be advantageous.

The second method of improving BEAMS is by cutting additional likely CC\,SNe
from the sample.  Following \citet{Kessler17}, we tested
a cut on the NN prior probability by requiring $0.5 < P_{NN}(Ia) < 1$.
Our simulations show that this cut removes $\sim$33\% of 
contaminants but just 5\% of SNe\,Ia.  The rejected sample has $\sim$40\%
CC\,SN contamination.
We found that an NN probability cut yields no improvements to our
results using the NN classifier.  However, when this cut is
added to our other classification
methods, it reduces $\sigma_w^{CC}$ by $\sim$30\% 
on average.  We have not included this cut in
our systematic error analysis (\S\ref{sec:varresults}) as
it makes our classification methods more correlated
and adds an additional dependence on uncertain simulations 
to the measured systematic error.  However, it is likely that 
this cut will increase the consistency of the full PS1 
cosmological results.  \citet{Kessler17} use a hybrid
classification approach by requiring \fp\ $>$ 0.05.
In our simulations, this cut reduces the CC\,SN
contamination by an additional 30\% compared to using the
NN classifier alone.

A third option for BEAMS is to estimate SN\,Ia distances
with a stricter CC\,SN model.  \citet{Kessler17} adopt an
approach where BEAMS CC\,SN
distributions are determined directly from simulations.  For our PS1
analysis, we have adopted a more general approach to CC\,SNe at the cost of
several additional parameters to marginalize over and a simpler form
of the likelihood (\citealp{Kessler17} also suggest free
CC\,SN parameters as a possible improvement to their method).
Tests show our parameterization is capable of marginalizing over the simulated CC\,SNe such
that the Ia likelihood is recovered, and our method is slightly more
general than a simulation-based method.  A
simulation-based mapping of CC\,SNe may be more robust, but 
validating it thoroughly is beyond the scope of this paper.  In particular,
the influence of inaccurate simulations on its recovered results must be
explored fully.

\subsection{Uncertainties in BEAMS Distances}
\label{sec:beamserr}

By setting P(Ia) $=$ 1/2 for all photometric SNe,
the BEAMS method measures $w$ with a bias of -0.01, 0.2 times the 
statistical uncertainty on $w$.  The statistical uncertainty on $w$
from setting P(Ia) $=$ 1/2, even with
no prior information as to which SNe are of Type Ia, is 
just 5\% higher than using SNe\,Ia alone (comparing to
SNe\,Ia alone in the case where $\alpha$ and $\beta$ are
fixed to known values).  This is primarily due
to two factors: the loose priors we employ and the fact that 
we include a sample of low-$z$ spectroscopically confirmed 
SNe\,Ia for which P(Ia) is fixed to 1.  These low-$z$ SNe\,Ia 
help to set the SN\,Ia dispersion and the SN parameters 
$\alpha$ and $\beta$, which are fixed as a function of redshift.

If we remove the low-$z$ sample, the distance and SN parameter
biases increase.  Distance uncertainties, which are higher 
by just $\sim$5\% when using the P(Ia) $=$ 1/2 prior, increase
by nearly 50\%.  Nevertheless, BEAMS does remarkably
well at determining the Gaussian distributions of SNe\,Ia 
and CC\,SNe with relatively little information.
This is helped by the fact that because SNe\,Ia have a factor
of $\sim$20 lower dispersion than CC\,SNe, a loose prior 
on BEAMS free parameters is sufficient to find the most probable
Gaussian distributions.  

If we use a more flexible CC\,SN model (a two-Gaussian or
skewed Gaussian CC\,SN model), the requirements
on our prior probabilities must become more stringent to
yield precise distances.  In the case of the two-Gaussian model,
prior probabilities can no longer be renormalized
or shifted (Eq. \ref{eqn:norm}) $-$ these are parameters
that can greatly improve 
the results for alternative prior probability methods.  Second, 
our prior probabilities must be significantly
more accurate to yield results with low uncertainties.  With the 
two-Gaussian CC\,SN model, the uncertainty on $w$ increases by 20\% when 
using GalSNID priors and by 100\% when setting P(Ia) = 1/2 for all 
photometric SNe.  Using the skewed Gaussian model, the GalSNID and 
P(Ia) $=$ 1/2 priors increase the uncertainties by 
13\% and 27\%, respectively.


Fortunately, a single-Gaussian model for CC\,SNe appears to
yield unbiased distances even though the simulated distribution
is not perfectly Gaussian.
In essence, BEAMS attempts only to 
determine the Gaussian distributions of two types of SNe and 
fortunately, those distributions are relatively well-separated 
in dispersion even if they are not always well-separated in distance.


\section{Conclusions}
\label{sec:conclusions}

We measured spectroscopic redshifts for \numspecgood\ SN host galaxies in
Pan-STARRS, over 1,000 of which are cosmologically useful, likely
SNe\,Ia.  When combined with the full PS1 spectroscopic sample
(Scolnic et al. in prep.), we will have \numsnetotal\
cosmologically useful SNe\,Ia from PS1.

We find that currently available CC\,SN templates and luminosity
functions are biased or incomplete.  Our results suggest there are
too few bright CC\,SNe in our simulations.

We generate 25 simulations that closely 
resemble the PS1 sample.  Each has 1,000
photometric PS1 SNe and 250 low-$z$
spectroscopically confirmed SNe\,Ia.
These simulations show that our method can measure $w$
with a bias due to CC\,SN contamination as low as -0.001$\pm$0.003.
This equates to a systematic
uncertainty on $w$ of just 0.004, 8\% of the statistical uncertainty,
but this uncertainty could be affected by incomplete knowledge of the CC\,SN distribution.
The SN\,Ia dispersion, $\sigma_{Ia}$, is
biased by -0.005 ($\sim$0.5$\sigma$), the SALT2 shape 
parameter $\alpha$ is biased by $\sim$0.005 ($\sim$1$\sigma$),
and the color parameter $\beta$ is biased by $\sim$0.1 
($\sim$1.5$\sigma$).  The statistical uncertainties on $w$ are nearly
equivalent to those using only SNe\,Ia.

Using several variants of the method and a CMB-like prior on $\Omega_M$, we 
estimate the systematic error introduced by CC\,SN contamination
to be 0.014$\pm$0.007 (29\% of the statistical error).
This systematic error would constitute only
a 3\% increase on the uncertainty on $w$ in a JLA-like
analysis with CMB priors ($\sigma_w = 0.057$ (stat+sys), and
$\sqrt{0.057^2 + 0.014^2} =$ 0.059).
However, this systematic error assumes that $\alpha$ and $\beta$
can be fixed to known values from a spectroscopic sample
for the alternate classification methods.
If $\alpha$ and $\beta$ are fixed,
our least accurate classifiers $-$ including
an uninformative prior probability P(Ia)$=$1/2 for all simulated PS1 SNe
$-$ give a median bias on $w$ between -0.01 and -0.02.
Systematic error could be reduced further by using a cut on 
prior probabilities from one variant to reduce CC\,SNe in 
the sample for the other variants.
We caution that due to uncertainties in CC\,SN simulations
and statistical fluctuations, the
CC\,SN contamination systematic affecting our forthcoming
cosmological results may be somewhat lower or higher
than the one estimated in this work.  However, that analysis
will also include a subset of PS1 SNe with known (spectroscopic)
classifications as part of the data,
a scenario that will likely reduce the systematic 
uncertainty due to CC\,SN contamination.

Included in these variants are a total of four different 
classification methods to measure cosmology,
including a host galaxy spectrum-based version of GalSNID \citep{Foley13}
that we introduce in this work (see Appendix \ref{sec:galsnid}).  GalSNID is based only on 
SN\,Ia host galaxy observables and a rates prior.  GalSNID 
provides a method of measuring $w$ from photometric data that 
does not depend on SN light curves and training on simulated data.  
Machine learning techniques may be able to improve on the efficiency 
of this method in the future.  We caution that even with these multiple variants,
if CC\,SN simulations are inaccurate it could cause the
systematic error to be underestimated in real data.
Additional CC\,SN templates and a better measurement of
the shape of CC\,SN LFs could help to
ameliorate these concerns.

By drawing random samples from real PS1 data, we tested whether 
the BEAMS method can work on real data within the confidence intervals of 
\citet{Rest14}.  We found that our measurements of $w$ were 
fully consistent with those of \citet{Rest14}, as were the SN nuisance parameters
$\alpha$ and $\beta$.

Though our results are robust, $w$ is an extremely 
sensitive measurement and the burden of proof
for BEAMS is high.  Future validation tests could include SDSS and
SNLS photometric data, as well as simulated tests with a variety of
CC\,SN LFs.  Additional light curve classification methods could also
help to improve the reliability of the BEAMS method.

Future SN\,Ia samples from DES and LSST will be unable to rely solely on
spectroscopic classification to measure cosmological parameters.  
With the light curve classification and Bayesian methodologies 
presented here, we validate some of the techniques that will be 
used in future surveys, and anticipate that PS1 photometric 
SNe can provide a robust measurement of $w$ using the largest 
SN\,Ia sample to date.

\acknowledgements
We would like to thank the anonymous referee for many
helpful suggestions.
This manuscript is based upon work supported by the National
Aeronautics and Space Administration under Contract No.\ NNG16PJ34C
issued through the {\it WFIRST} Science Investigation Teams Programme.
R.J.F.\ and D.S.\ were supported in part by NASA grant 14-WPS14-0048.
The UCSC group is supported in part by NSF grant AST-1518052 and from
fellowships from the Alfred P.\ Sloan Foundation and the David and
Lucile Packard Foundation to R.J.F.  This work was supported in part
by the Kavli Institute for Cosmological Physics at the University of
Chicago through grant NSF PHY-1125897 and an endowment from the Kavli
Foundation and its founder Fred Kavli.  D.S, gratefully acknowledges
support from NASA grant 14-WPS14-0048. D.S. is supported by NASA through
Hubble Fellowship grant HST-HF2-51383.001 awarded by the Space Telescope
Science Institute, which is operated by the Association of Universities
for Research in Astronomy, Inc., for NASA, under contract NAS 5-26555.

Many of the observations reported here were obtained at the MMT Observatory, 
a joint facility of the Smithsonian Institution and the 
University of Arizona.  This paper uses data products produced by 
the OIR Telescope Data Center, supported by the Smithsonian 
Astrophysical Observatory.  Additional data are thanks to the Anglo Australian
Telescope, operated by the Australian Astronomical Observatory, through
the National Optical Astronomy Observatory (NOAO PropID: 2014B-N0336; 
PI: D. Jones).  We also use data from observations at Kitt Peak National 
Observatory, National Optical Astronomy Observatory,
which is operated by the Association of Universities for Research in 
Astronomy (AURA) under a cooperative agreement with the National 
Science Foundation.  Also based on observations obtained with the Apache 
Point Observatory 3.5-meter telescope, which is owned and operated 
by the Astrophysical Research Consortium.

The computations in this paper used a combination 
of three computing clusters.  BEAMS analysis was performed 
using the University of Chicago Research Computing Center 
and the Odyssey cluster at Harvard University.  We are grateful 
for the support of the University of Chicago Research Computing Center for 
assistance with the calculations carried out in this work.  The 
Odyssey cluster is supported by the FAS Division of Science, Research 
Computing Group at Harvard University.  Supernova light curve 
reprocessing would not have been possible without the Data-Scope project 
at the Institute for Data Intensive Engineering and Science at 
Johns Hopkins University.

Funding for the Sloan Digital Sky Survey IV has been provided by the
Alfred P. Sloan Foundation, the U.S. Department of Energy Office of
Science, and the Participating Institutions. SDSS- IV acknowledges
support and resources from the Center for High-Performance Computing at
the University of Utah. The SDSS web site is www.sdss.org.

SDSS-IV is managed by the Astrophysical Research Consortium for the 
Participating Institutions of the SDSS Collaboration including the 
Brazilian Participation Group, the Carnegie Institution for Science, 
Carnegie Mellon University, the Chilean Participation Group, the 
French Participation Group, Harvard-Smithsonian Center for Astrophysics, 
Instituto de Astrofísica de Canarias, The Johns Hopkins University, 
Kavli Institute for the Physics and Mathematics of the Universe 
(IPMU) / University of Tokyo, Lawrence Berkeley National Laboratory, 
Leibniz Institut für Astrophysik Potsdam (AIP), Max-Planck-Institut 
für Astronomie (MPIA Heidelberg), Max-Planck-Institut für Astrophysik 
(MPA Garching), Max-Planck-Institut für Extraterrestrische Physik (MPE), 
National Astronomical Observatory of China, New Mexico State University, 
New York University, University of Notre Dame, Observatório Nacional / MCTI, 
The Ohio State University, Pennsylvania State University, Shanghai 
Astronomical Observatory, United Kingdom Participation Group, Universidad 
Nacional Autónoma de México, University of Arizona, University of Colorado 
Boulder, University of Oxford, University of Portsmouth, University of 
Utah, University of Virginia, University of Washington, University of 
Wisconsin, Vanderbilt University, and Yale University.

This research makes use of the VIPERS-MLS database, operated at CeSAM/LAM, 
Marseille, France. This work is based in part on observations obtained 
with WIRCam, a joint project of CFHT, Taiwan, Korea, Canada and France. 
The CFHT is operated by the National Research Council (NRC) of Canada, 
the Institut National des Science de l’Univers of the Centre National de 
la Recherche Scientifique (CNRS) of France, and the University of Hawaii. 
This work is based in part on observations made with the Galaxy Evolution 
Explorer (GALEX). GALEX is a NASA Small Explorer, whose mission was 
developed in cooperation with the Centre National d’Etudes Spatiales 
(CNES) of France and the Korean Ministry of Science and Technology. GALEX 
is operated for NASA by the California Institute of Technology under NASA
contract NAS5-98034. This work is based in part on data products produced at 
TERAPIX available at the Canadian Astronomy Data Centre as part of the 
Canada-France-Hawaii Telescope Legacy Survey, a collaborative project of 
NRC and CNRS. The TERAPIX team has performed the reduction of all the 
WIRCAM images and the preparation of the catalogues matched with the 
T0007 CFHTLS data release.

Funding for the DEEP2 Galaxy Redshift Survey has been
provided by NSF grants AST-95-09298, AST-0071048, AST-0507428, and 
AST-0507483 as well as NASA LTSA grant NNG04GC89G.  This research uses data 
from the VIMOS VLT Deep Survey, obtained from the VVDS database operated 
by Cesam, Laboratoire d'Astrophysique de Marseille, France.  zCosmos data 
are based on observations made with ESO Telescopes at the La Silla or 
Paranal Observatories under programme ID 175.A-0839.

\appendix
\section{The Dearth of Simulated CC\,SNe}
\label{app:ccsn}

\subsection{Core-collapse SNe}
\label{sec:non1a}

There are a few potential explanations for the difference
in Hubble residuals (0.5 $<$ $\mu-\mu_{\Lambda CDM}$ $<$ 1.5)
between simulations and data.  In this 
appendix, we attempt to identify the cause of the discrepancy.

First, a large percentage ($\gtrsim$20\%) of inaccurate
SN\,Ia redshifts could explain the data.  However, in 
addition to disagreeing with
our measurements, this would give too 
many simulated SNe with very bright and very faint 
Hubble residuals.  Requiring a high TDR minimum
and a small separation between the 
SN location and host galaxy center in our data
does not resolve the conflict.

A second option is that the relative rates or magnitude
distributions from \citet{Li11} are 
erroneous or are biased by the targeted nature of the survey 
(LOSS searched for SNe in a set of pre-selected bright galaxies).
These rates also do not take into account that the relative
fractions of different CC\,SN subtypes could change with redshift.
Modest adjustments, such as ``tweaking'' the mean
magnitudes or dispersions of CC\,SNe by $\lesssim$0.5 mag,
cannot explain the discrepancy.  Simulating CC\,SNe using 
LFs from \citet{Richardson14}, which 
are typically $\sim$0.3-1.0 mag brighter
than those of \citet{Li11}, produces far too many
bright CC\,SNe compared to our data.  The effect of weak 
lensing on the data is expected to be an order of magnitude 
less than the size of the offset we see here \citep{Smith14}.
It is also unlikely that strongly lensed SNe
contribute significantly to the discrepancy \citep{Oguri10}.

By reclassifying LOSS SNe, \citet{Shivvers17} 
recently found that SN\,Ib relative rates were more 
than double the fraction found by \citet{Li11}.  This 
change could reduce the Hubble residual discrepancy by half or more.  
However, \citet{Shivvers17} determined these rates by reclassifying
a number of LOSS SNe\,Ic as SNe\,Ib, which in turn means
that the SN\,Ib LF should be made fainter.  Making the SN\,Ib 
LF fainter will increase the discrepancy in Hubble residuals.  
We continue to use \citet{Li11} in this work, as 
we can be sure that the LFs and relative rates are self-consistent.

Finally, we consider that our results could be biased if SNANA
templates have lower average reddening
than PS1 data.  There are likely substantial
differences between the reddening distribution of the templates and 
the data.  However, we find that adding additional reddening to our simulations
tends to make the magnitude distribution of CC\,SNe broader
(we approximately adjust the \citealp{Li11} LFs for dust
following \citealp{Rodney14}).
This \textit{increases} the discrepancy between
simulations and data.  Correcting for the unknown intrinsic
reddening of these templates is an important future objective that
can allow SNANA simulations to be more realistic.  See \S\ref{sec:ccsimdisc}
for further discussion of biases in our simulations and templates.

\subsubsection{Adding New Supernova Templates to SNANA}
\label{sec:newtmpl}

Several CC\,SN or peculiar Ia subtypes are 
missing from the SNANA simulation library but
could be present in the PS1 data.  
Missing SN types include superluminous SNe, SNe IIb, SNe
Ibc-pec, and peculiar, faint SNe\,Ia such as 1991bg-like SNe\,Ia 
(Ia-91bg) and SNe\,Iax \citep{Foley13b}.  Superluminous SNe are unlikely 
to help resolve the discrepancy, as they are brighter than SNe\,Ia 
and occur preferentially in faint hosts for which redshifts are 
difficult to measure \citep{Lunnan15}.
SNe Ibc-pec have similar LFs to SNe\,II-P but are 
much less common, so it is unlikely that many would 
fall on the Hubble diagram so near the SN\,Ia distribution.  SNe\,Iax 
are red, fast-declining SNe that may be relatively common but have faint 
(albeit uncertain) LFs more similar to SNe II-P and Ibc-pec.
These also tend to be poorly fit by SALT2, and would frequently fail our cuts.

SNe IIb and SNe\,Ia-91bg both have LFs
only $\sim$1 mag fainter than SNe\,Ia, though they are relatively uncommon and
would need a high fraction to pass SALT2 light curve cuts to be 
major contributors to our Hubble diagram.  We investigated 
their impact by adding Ia-91bg and IIb templates to SNANA.  

To simulate CC\,SNe over a wide range of redshifts and passbands,
SNANA templates require relatively high-SNR, high-cadence 
spectral and photometric sampling, which exists for a paucity
of CC\,SNe.  Simulating SN light curves at high redshift often
necessitates near-ultraviolet data as well.  To create a template, 
an interpolated, flux-calibrated spectral time series is
``mangled'' to match the observed photometry by using 
wavelength-dependent splines with knots at the effective 
wavelengths of the photometric filters.  Least-squares fitting
determines the the best-fit spline that scales the spectrum to match
the photometry.  \citet{Hsiao07} describes the ``mangling'' procedure
in detail.

To improve the SNANA CC\,SN simulation, we add four SN\,IIb
templates $-$ SNe 1993J, 2008ax, 2008bo, and 2011dh $-$ 
using spectra and light curves consolidated by the Open Supernova Catalog 
\citep{Guillochon17}\footnote{References for the spectra and photometry are listed here.
  SN 1993J: \citet{Richmond96,Metlova95,Barbon95,Jerkstrand15,Modjaz14}.
  SN 2008ax: \citet{Modjaz14,Brown14,Taubenberger11,Tsvetkov09,Pastorello08}.
  SN 2008bo: \citet{Modjaz14,Brown14,Bianco14}.
  SN 2011dh: \citet{Ergon15,Ergon14,Shivvers13,Arcavi11}.  Secondary sources: \citet{Yaron12,Richardson01,Silverman12} and the Sternberg Astronomical Institute Supernova Light Curve Catalogue.}.  Each of these templates 
have well-sampled spectra and optical light curves.  We also add Ia-91bg templates 
using the SN 1991bg spectrum from \citet{Nugent02}\footnote{\url{https://c3.lbl.gov/nugent/nugent_templates.html}}, 
warped to match SNe\,Ia-91bg with well-sampled light curves before and after 
maximum (SNe 1991bg, 1998de, 1999by, 2005bl\footnote{References for the photometry are listed here.
  SN 1998de: \citet{Silverman12,Ganeshalingam10,Modjaz01}.
  SN 1999by: \citet{Silverman12,Ganeshalingam10,Garnavich04}.
  SN 2005bl: \citet{Contreras10}.
  Secondary sources: the Sternberg Astronomical Institute Supernova Light Curve Catalogue.}).
Using multiple SN templates helps us obtain better sampling of the shape-luminosity relation for SNe 91bg (steeper
than the relation for normal SNe\,Ia; \citealp{Taubenberger08}).

\begin{figure*}
\centering
\includegraphics[width=7in]{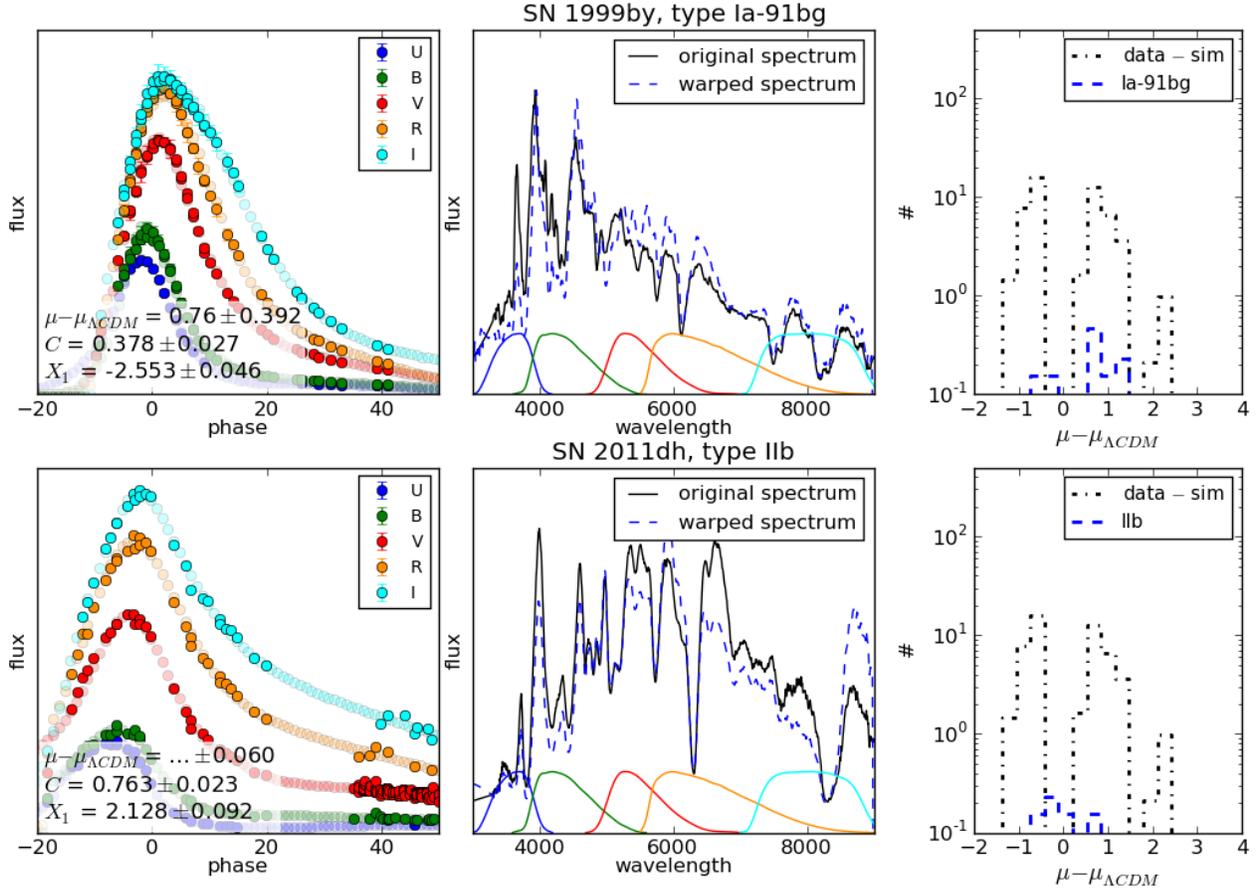}
\caption{New templates for SNe Ia-91bg (top panels) and IIb (bottom panels) were
  added to SNANA by
  mangling a template spectrum to match light curve data.  From
  left to right, we show the interpolated SN light curves (light 
  shading indicates interpolated points), the warped
  template spectra at peak brightness, and the Hubble 
  residuals of \textit{all} templates of the new subtype. 
  We compare the Hubble residuals of the new templates to the 
  difference between the data and our simulations; the new 
  templates cannot explain the discrepancy 
  we observe.  Because SN 2011dh has $z < 0.01$, its distance
  modulus residual is not shown in the left panel.}
\label{fig:IIb_91bg}
\end{figure*}

Figure \ref{fig:IIb_91bg} shows the interpolated light curves, 
mangled spectra, and Hubble residual histograms for SNe\,IIb and Ia-91bg.  For 
Ia-91bg, we assume their rates have the same redshift dependence 
as SNe\,Ia.  SNe\,Ia-91bg have magnitude distributions that could explain the 
data, but their rates are inconsistent with the data.
SNe\,IIb are far too rare, as nearly all simulated SNe\,IIb have measured
colors that are too red to be SNe\,Ia.  Though we find that Ia-91bg and IIb 
SNe are not frequent enough to resolve the difference between PS1 data and 
simulations, we incorporate these subtypes in our simulations hereafter.

\subsubsection{Measuring CC\,SN Luminosity Functions with PSNID}
\label{sec:cclfpsnid}

\begin{figure}
\centering
\includegraphics[width=3.5in]{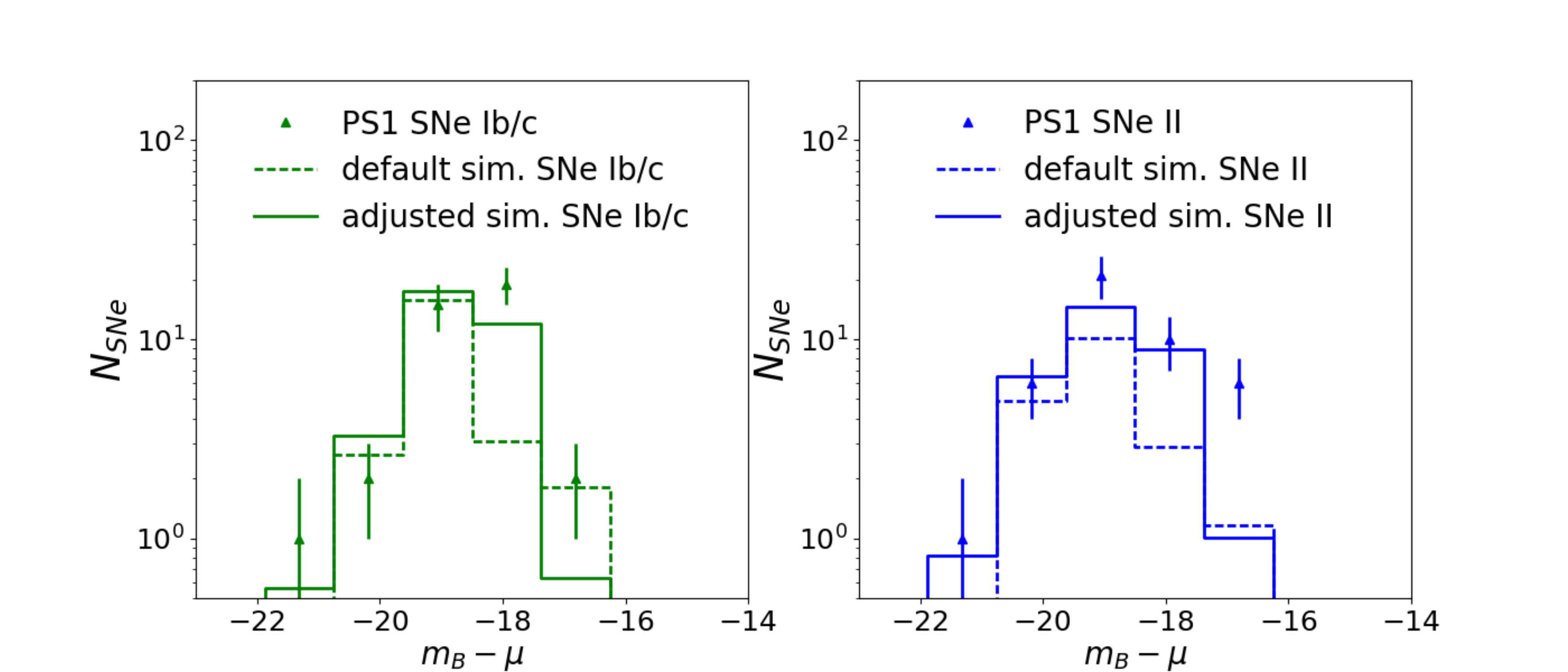}
\caption{Empirical adjustments to SNANA simulations motivated by PSNID
classifications, shown using histograms of SN absolute magnitude
(SALT2 $m_B - \mu_{\Lambda CDM}$).  PSNID-classified 
PS1 SNe and PSNID-classified simulations suggest that SNe\,Ib/c, 
after shape and color cuts, are brighter than expected.  Our adjusted 
simulations (solid lines) match the data after we reduce the simulated 
dispersions and brighten LFs by $\sim$1 mag.}
\label{fig:psnid}
\end{figure}

\begin{figure}
\centering
\includegraphics[width=3.5in]{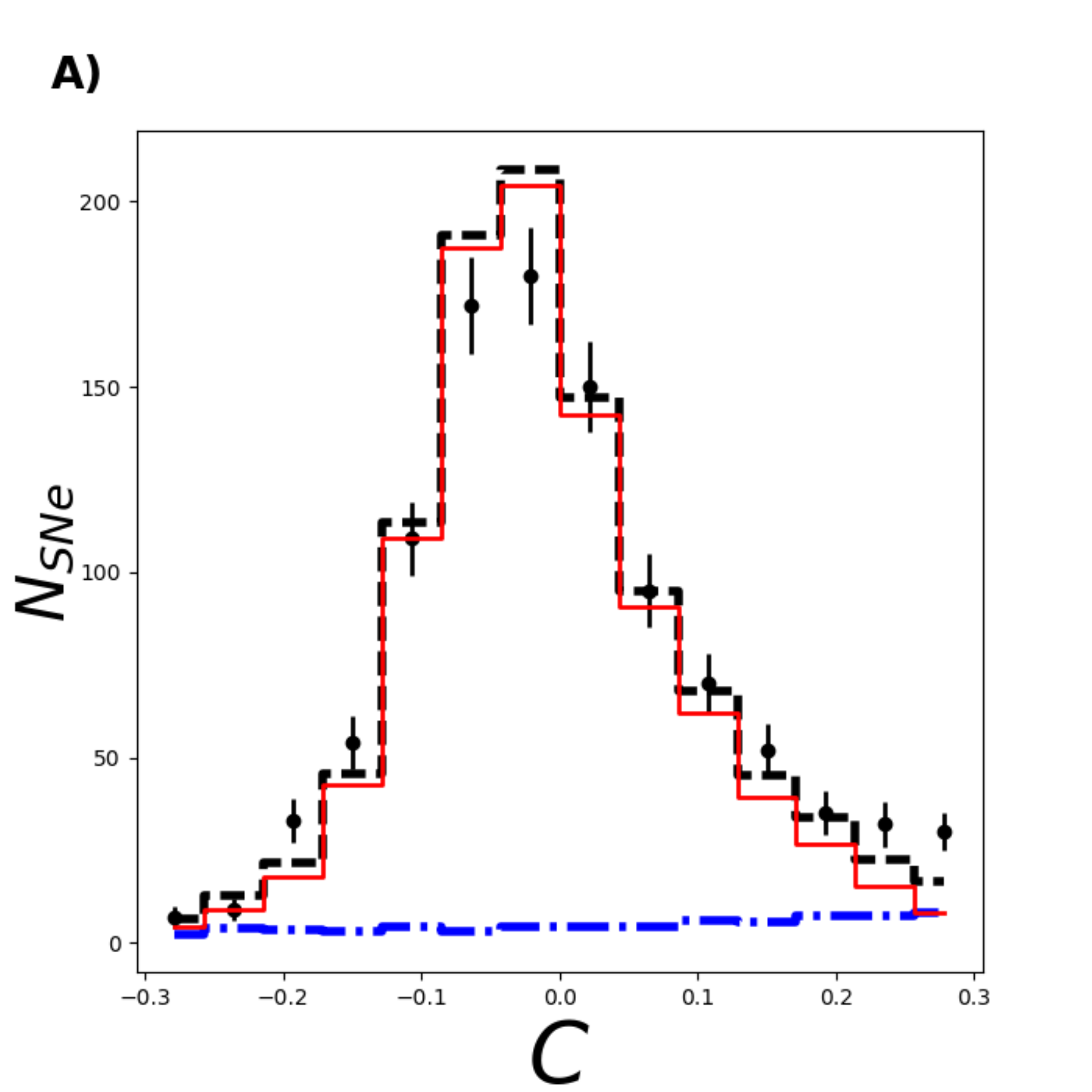}
\caption{Simulated SALT2 $C$ (right), compared to data in
  the J17 (adjusted) simulations.  Compared
  to the original simulations (Figure \ref{fig:defaultsim}D and
  \ref{fig:defaultsim}E), the red end of the
  $C$ distribution is more consistent with our data in the J17 simulations.}
\label{fig:j16sim}
\end{figure}

There is an additional procedure by which PS1 data can inform CC\,SN LFs: we 
use the PSNID light curve classifier \citep{Sako11,Sako14} to 
separate the likely contributions of SNe\,Ia, Ib/c, and II.
The SNANA implementation of PSNID compares the 
SALT2 SN\,Ia model and SNANA's CC\,SN templates 
to the observed data.  PSNID determines the fit $\chi^2$- 
and prior-based probability that a given SN is Type Ia, Type Ib/c, 
or Type II.  Though the set of templates we use for PSNID is the
same set we use to generate CC\,SNe in our simulations,
broad priors allow these templates to be shifted 
in magnitude and extinction to fit our data.

We compare PSNID's classifications of PS1 data and 
simulations by examining the distribution of $m_B - \mu_{\Lambda CDM}$, a proxy for
absolute magnitude at peak (Figure \ref{fig:psnid}).  
We find that likely SNe\,Ib/c are much brighter and have lower dispersion than 
the simulations.  To bring our simulations into agreement with the data, we 
adjusted the simulated SN\,Ib/c and II distributions
such that the mean and standard deviations of the \textit{simulated} 
SNe that PSNID classified as Type Ib/c and II matched the mean 
and standard deviations of the \textit{real} PS1 SNe that PSNID 
classified as Type Ib/c and II.  This 
requires reducing the dispersion of CC\,SN 
templates by 55\% for SNe Ib/c. 
It also requires brightening the simulated LFs by 1.2 mag for
SNe\,Ib/c and 1.1 mag for SNe\,II.
We made shape and color cuts (\S\ref{sec:dq}) in this analysis but neglected 
$\sigma_{X_1}$ and $\sigma_{\textrm{peakMJD}}$ cuts to increase 
our SN statistics.

Figure \ref{fig:psnid} shows the distributions of
PSNID-classified PS1 SNe (P(SN Type) $>$ 95\%) compared to our simulations before 
and after absolute magnitude and dispersion adjustments.
We apply shape and color cuts but neglect additional cuts to 
increase our CC\,SN sample size.

After these adjustments, simulated CC+Ia SNe are consistent 
with our data.  Figure \ref{fig:ccsim} shows 
Hubble residual histograms before and after our PSNID-based adjustments.  
After correction, CC\,SNe are 8.9\% of our final sample and SNe\,Ia-91bg 
comprise 0.2\%.  Additional CC\,SNe can explain the red tail of the 
SALT2 $C$ distribution in Figure \ref{fig:defaultsim}C (Figure \ref{fig:j16sim}).
No CC\,SN rate adjustments were made.
Although the simulated absolute magnitudes have been brightened by 
$\sim$1 mag, CC\,SN in the adjusted simulations are only 
$\sim$0.5 mag brighter than the original simulation on average.  This 
is because as we brighten the CC\,SN distribution, the number 
of detectable faint SNe $-$ which are nearer to the peak of the 
LF, and thus occur more frequently $-$ increases, reducing
the mean absolute magnitude.  Note that the $\sim$2-3$\sigma$
discrepancy on the left (bright) side of the Hubble diagram can
be reduced by simulating a nominal host mass correction, which tends to
very slightly broaden the simulated distribution of SNe\,Ia.

Our adjusted simulation matches the Hubble residuals 
of the PS1 data.  It also resolves the discrepancies in the 
PS1 $C$ distribution (Figure
\ref{fig:j16sim}).  Hereafter, we refer to the adjusted simulation, which
adds new CC\,SN templates and uses PSNID to infer the 
true SN\,Ib/c distribution, as the J17 simulation\footnote{Templates and simulation input files for this simulation
have been added to the SNANA library.}.

\section{Expanding the BEAMS Method}
\label{sec:syserr}

\begin{figure*}
\centering
\includegraphics[width=7in]{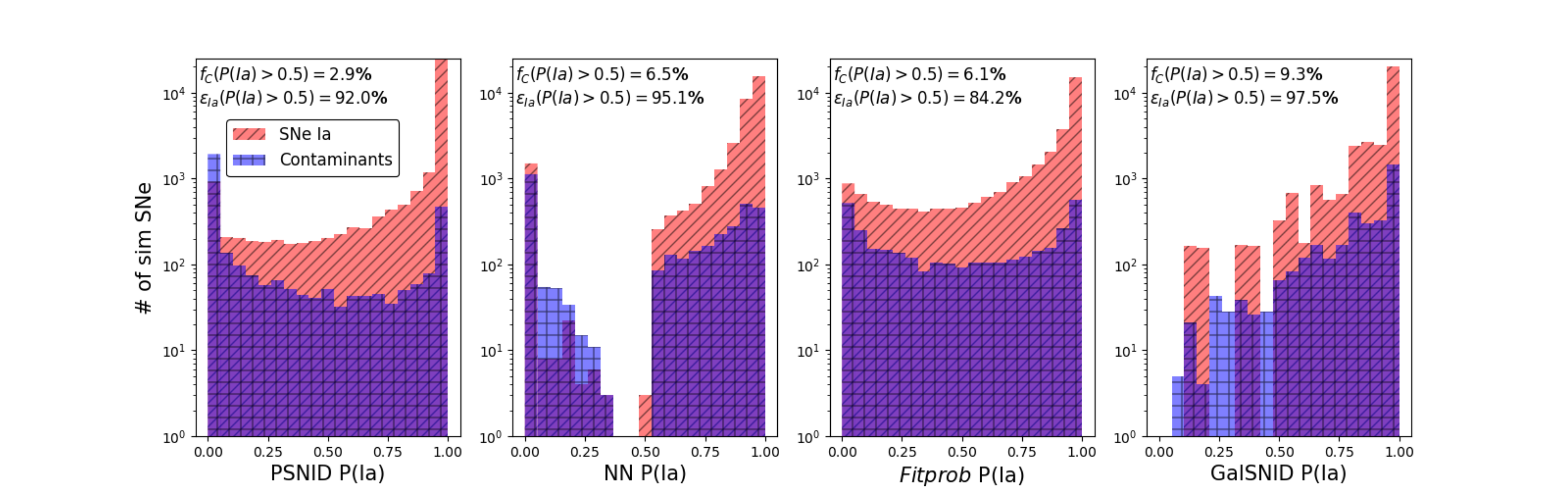}
\caption{Simulated prior probabilities from the four classification methods
  discussed in this work for SNe\,Ia (red) and contaminants (blue; includes CC\,SNe
  and SNe\,Ia with incorrect redshifts).  For each method, we show the percentage of contaminants 
  $f_C$ and the fraction of SNe\,Ia included, 
  $\epsilon_{Ia}$, in a P(Ia) $>$ 0.5 sample.}
\label{fig:prob}
\end{figure*}

We discuss the methodology behind alternative BEAMS variants in this section.
The results from these variants are given in \S\ref{sec:varresults}.


\subsection{Additional P(Ia) Priors}

In addition to the PSNID prior probabilities in our baseline method,
we use three additional methods of estimating P(Ia): \fp,
NN, and GalSNID.  The effectiveness of each method is 
illustrated in Figure \ref{fig:prob}.  The 
NN, \fp, and PSNID classifiers all determine 
probabilities by fitting to the photometric SN light curve.  \fp\ 
relies on only the SALT2 model for fitting, while PSNID and NN depend on 
CC\,SN simulations for templates and training, respectively.
GalSNID uses host galaxy information and depends 
on SNANA simulations only through the SN rates prior.

\subsubsection{Nearest Neighbor}
\label{sec:nn}

The Nearest Neighbor (NN) classifier \citep{Sako14} uses a set of observables to define how 
close a given SN is to the CC\,SN and SN\,Ia populations.  In our 
implementation, we use the SALT2 color ($C$), 
stretch ($X_1$), and redshift ($z$).  The equation

\begin{equation}
d_i^2 = \frac{(z-z_i)^2}{\Delta z_{\textrm{max}}^2} + \frac{(C-C_i)^2}{\Delta
  C_{\textrm{max}}^2} + \frac{(X_1 - X_{1,i})^2}{\Delta X_{1,\textrm{max}}^2}
\end{equation}

\noindent defines a list of NN distances between the $i$th SN and
simulated training data.  For the $i$th SN, neighbors are defined as all simulated
events with $d_i < 1$.
NN training finds the parameters $\Delta C_{\textrm{max}}$, $\Delta
X_{1,\textrm{max}}$ and $\Delta z_{\textrm{max}}$ that optimize
the classification metric (efficiency $\times$ purity) of simulated
training data.  NN is an efficient and accurate classifier in 
PS1 simulations: the set of SNe with P$_{NN}$(Ia) $>$ 0.9 has
3.8\% contamination compared to 9.7\% contamination for the full sample
(including CC\,SNe and SNe\,Ia with incorrect redshifts).
This set includes 74\% of all SNe\,Ia.
See \citet{Kessler17} for details
on the NN classification method.


\subsubsection{\textit{Fitprob}}
\label{sec:fitprob}

The \fp\ method estimates P(Ia) from the $\chi^2$ and 
number of degrees of freedom of the 
SALT2 light curve fit (the SALT2 fit probability).  Because 
the SALT2 fit $\chi^2$ has no knowledge about the 
relative frequency of different SN types, we multiplied 
P$_{\textrm{fp}}$(Ia), the \fp\ probability, by a redshift-dependent 
SN rates prior, P(Ia$|z$).  P(Ia$|z$) is the number of SNe\,Ia divided by
the total number of SNe at a given redshift (after sample cuts; measured 
using the J17 simulations):

\begin{align}
& \mathrm{\tilde{P}_{fp}(Ia)} = \frac{\mathrm{P(Ia}|z)\mathrm{P_{fp}(Ia)}}{\mathrm{P(CC}|z)(1 - \mathrm{P_{fp}(Ia)}) + \mathrm{P(Ia}|z)\mathrm{P_{fp}(Ia)}} \\
& \mathrm{P(CC}|z) = 1 - \mathrm{P(Ia}|z). \nonumber
\end{align}

\noindent Compared to the PSNID (baseline) classifier, \fp\ has 
twice the fraction of contaminants at P(Ia) $>$ 0.5.  The fraction of CC\,SNe
with high P(Ia) is also higher by a factor of $\sim$2.

\subsubsection{GalSNID}
\label{sec:galsnid}

SNe\,Ia have much longer average delay times between progenitor formation and
explosion than CC\,SNe. Because of this, SNe\,Ia are the only SN type found in
early type hosts.  This allows methods such as GalSNID \citep{Foley13} to classify SNe with 
host galaxy information.  The GalSNID method in \citet{Foley13} is 
based on photometric information and is highly dependent on host morphology.
Because measuring galaxy morphologies at typical PS1 redshifts requires 
$\sim$0.1\arcsec\ image resolution, we modified the method by adding spectral 
observables.  Though GalSNID is a very inefficient classifier, it measures
SN\,Ia probabilities in a way that is only minimally subject to light
curve and LF uncertainties.

To train GalSNID, we used 602 host galaxy spectra from the Lick Observatory Supernova 
Search (LOSS; \citealp{Leaman11}) and 354 host galaxy 
spectra of PS1 spectroscopically confirmed SNe.
The equivalent widths of spectral emission lines, and H$\alpha$ in 
particular, correlate with SN type.  Another useful diagnostic 
is the template that cross-correlates best with the observed host spectrum.  
Finally, we include host galaxy $R$ (labeled effective offset in
\citealp{Foley13}), $B-K$ colors and absolute $K$ magnitudes
from \citet{Foley13}.

We trained GalSNID on spectral information using LOSS host galaxy spectra
 and spectroscopically confirmed PS1 SNe for which we have host galaxy 
spectra.  Relative to the PS1
spectroscopic sample, LOSS has a greater number of total SNe, and a
greater diversity and number of CC\,SNe on which to train the
data.

Spectra for $\sim$1/3 of the LOSS sample are available from SDSS/BOSS
(\citealp{Alam15}; 297 spectra), and we found an additional 
$\sim$1/3 (305 spectra) by querying the NASA/IPAC Extragalactic 
Database.  In total, 67\% of the 905 SNe 
discovered by LOSS have host galaxy spectra.  In general, 
the SNR of these data are high (much higher on average than 
our redshift survey data).

PS1 spectroscopically classified 520 SNe, of which $\sim$150 are
CC\,SNe and the rest are SNe\,Ia.  Of the CC\,SNe, $\sim$30 are 
SNe IIn (Drout et al. in prep), 76 are II-P or II-L
\citep{Sanders15} and $\sim$20-30 are SNe Ib or Ic.  We obtained host 
galaxy spectra for 354 of these SNe.

We searched for a number of prominent, observational galaxy
diagnostics that correlate with the age of the host, and found that
the equivalent widths of bright emission lines such as OII, OIII,
H$\alpha$ and H$\beta$ are measurable in many of our spectra.
We required continuum SNR $>$ 5 near a given line measurement
for an observable to be used in training or classification.
As a way to incorporate additional information in a single diagnostic,
we included the best-matched spectral template based on cross-correlation
as an observable.

Although these diagnostics are correlated, in this work we
follow \citet{Foley13} in treating them as independent.  Final 
probabilities for a given SN can therefore be computed 
by multiplying the probability of a Ia given each 
observable \citep{Foley13}:

\begin{equation}
\mathrm{P(Ia|D)} = k^{-1}\mathrm{P(Ia|}z)\prod_{i=1}^{N}\mathrm{P(D_i|Ia)},
\end{equation}

\noindent where $N$ is the number of observables and P(D$_i|$Ia) is 
the probability of an observable given that the SN is Type Ia
(Table \ref{table:galsnid}). P(D$_i|$Ia) is easy to compute; it is 
the fraction of SN\,Ia host galaxies that have observable D$_i$.  
P(Ia$|z$) is a rates prior informed by our SNANA simulations.  $k$
is a normalization factor that requires
$\textrm{P(Ia}|\textrm{D}) + \textrm{P(CC}|\textrm{D}) = 1$.
See \citet{Foley13} for additional details on the methodology.
In the future, machine learning techniques may be able to improve our 
results by relaxing the assumption that observables are uncorrelated.

\begin{figure}
\centering
\includegraphics[width=3.5in]{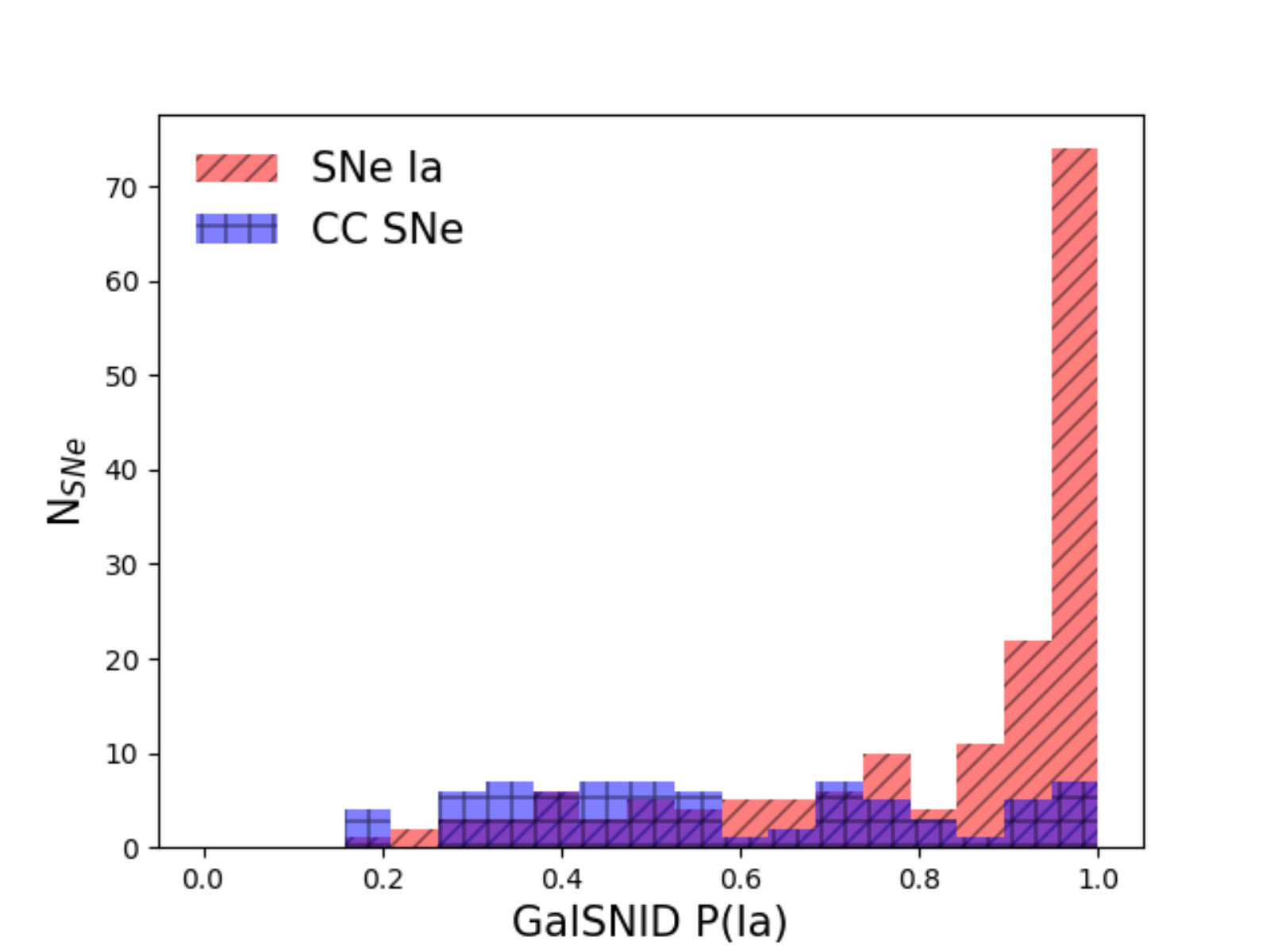}
\caption{GalSNID classifications of spectroscopically
  classified CC\,SNe and SNe\,Ia in Pan-STARRS, neglecting rates priors.}
\label{fig:probreal}
\end{figure}

The probabilities from our LOSS+PS1 training sample are provided in 
Table \ref{table:galsnid}.  We also include the effective offset,
$B-K$ colors, and $K$ absolute magnitudes using probabilities
measured from \citet{Foley13} and SED fits using PS1 host galaxy 
photometry.  Note that because H$\alpha$ and H$\beta$ are almost perfectly 
correlated (the correlation coefficient is 0.94), we do not use H$\beta$
as an observable when H$\alpha$ is present in optical spectra ($z \lesssim 0.35$).
Figure \ref{fig:prob} shows the GalSNID probabilities of SNe\,Ia and
CC\,SNe in PS1 and our simulations (we redshift and add noise to 
LOSS spectra to determine simulated GalSNID probabilities).
Figure \ref{fig:probreal} shows GalSNID probabilities for
real spectroscopically classified PS1 SNe.

To create GalSNID probabilities for the simulated sample, we 
artificially redshifted LOSS host galaxy spectra, added noise to 
make them consistent with the SNR of PS1 host spectra, and used 
GalSNID to measure the probability that each host observed an SN\,Ia.
We took the distributions 
of GalSNID probabilities for the redshifted, noisy spectra corresponding 
to LOSS SNe II, Ib/c, and Ia hosts in each simulated 
redshift bin and assigned the probabilities drawn from 
those distributions to simulated SNe II, Ib/c, and Ia. This gave 
our simulated SNe the same probability distributions as the 
redshifted LOSS data.
Figure \ref{fig:prob} shows that GalSNID is a relatively 
imprecise classifier, but it provides constraints that are 
independent of SN light curves and their associated uncertainties.
We have not taken into account 
the redshift evolution of SN host galaxies in this work.

On PS1 data, GalSNID is by far the least efficient classifier.  Because 
classifications are highly influenced by the rates prior, GalSNID 
considers just 5\% of contaminants to be likely CC\,SNe.  If we set a 
higher threshold of P(Ia) $>$ 0.9, GalSNID removes $\sim$25\% of CC\,SNe and 
keeps $\sim$70\% of SNe\,Ia.  GalSNID is also most effective at 
$z \lesssim 0.35$, where H$\alpha$ is present in our optical 
spectra (the best indicator of SN type in our spectra).
Unfortunately, the largest SN\,Ia distance biases
are at $z > 0.4$, where the CC\,SN distribution becomes blended
with the SN\,Ia distribution.

GalSNID would also be useful as an additional prior on SN type in conjunction 
with other methods.  However, due to the uncertainty in CC\,SN models 
and LFs, in the present analysis we consider it most
powerful as a stand-alone tool that can measure
SN\,Ia probabilities without using light curve data.

\subsection{Varying the CC\,SN Model}
PS1 and other spectroscopic data show that SNe\,Ia are
well-represented by a Gaussian Hubble residual model, but CC\,SNe are not.
We investigated replacing the CC\,SN likelihood in 
Eq. \ref{eqn:beamslike} with two likelihoods that are more 
consistent with our CC\,SN simulations.  We tested a 
two-Gaussian model with ten additional free parameters 
for CC\,SNe (the means and standard deviations of the second 
Gaussian at five redshift control points).  We also tested a single, 
asymmetric Gaussian model with five additional free parameters 
(skewness at each CC\,SN control point).

If we allow BEAMS to shift and/or rescale the prior
probabilities that an SN is of Type Ia (Eq. \ref{eqn:norm}), 
BEAMS can give unphysical results.  
The alternative CC\,SN models are significantly more flexible and
that flexibility must be constrained by accurate, fixed prior 
probabilities such as those from NN (see \S\ref{sec:beamserr}).
We fix the parameters 
that allow BEAMS to adjust the priors ($A = 1$ and
$S = 0$ in Eq. \ref{eqn:norm}) or else the 
uncertainties on SN\,Ia distances will inflate to 
$>$0.1 mag for even our best-measured redshift
control points.

\begin{deluxetable}{llll}
\tablewidth{0pt}
\tablecaption{Probability of Host Properties Given Type\label{t:prop}}
\tablehead{
\colhead{Bin} &
\colhead{$P(D_{i} | {\rm Ia})$} &
\colhead{$P(D_{i} | {\rm Ibc})$} &
\colhead{$P(D_{i} | {\rm II})$}}

\startdata

\tableline\\*[0pt]
\multicolumn{4}{c}{Cross-Correlation Template} \\
\tableline\\*[0pt]
Absorption  & 0.502 \err{0.054}{0.048} & 0.256 \err{0.069}{0.055} & 0.286 \err{0.036}{0.033}\\*[4pt]
Ellipt+A stars  & 0.431 \err{0.050}{0.045} & 0.598 \err{0.097}{0.085} & 0.609 \err{0.051}{0.048}\\*[4pt]
Late-type  & 0.029 \err{0.017}{0.012} & 0.037 \err{0.035}{0.020} & 0.030 \err{0.015}{0.010}\\*[4pt]
Emission  & 0.029 \err{0.017}{0.012} & 0.098 \err{0.047}{0.034} & 0.071 \err{0.021}{0.016}\\*[4pt]
\tableline\\*[0pt]
\multicolumn{4}{c}{H$\alpha$ Equivalent Width} \\
\tableline\\*[0pt]
$<$-5.0 & 0.005 \err{0.011}{0.004} & 0.000 \err{0.033}{0.000} & 0.000 \err{0.010}{0.000}\\*[4pt]
-5.0  -- 0.0 & 0.323 \err{0.045}{0.039} & 0.054 \err{0.051}{0.029} & 0.116 \err{0.031}{0.024}\\*[4pt]
0.0  -- 5.0 & 0.219 \err{0.038}{0.033} & 0.250 \err{0.086}{0.066} & 0.217 \err{0.039}{0.034}\\*[4pt]
5.0  -- 10.0 & 0.095 \err{0.026}{0.022} & 0.125 \err{0.067}{0.046} & 0.143 \err{0.033}{0.027}\\*[4pt]
$>$10.0 & 0.358 \err{0.047}{0.042} & 0.571 \err{0.120}{0.100} & 0.524 \err{0.058}{0.052}\\*[4pt]
\tableline\\*[0pt]
\multicolumn{4}{c}{H$\beta$ Equiva1ent Width} \\
\tableline\\*[0pt]
$<$-5.0 & 0.000 \err{0.007}{0.000} & 0.000 \err{0.026}{0.000} & 0.000 \err{0.007}{0.000}\\*[4pt]
-5.0  -- 0.0 & 0.504 \err{0.046}{0.043} & 0.338 \err{0.084}{0.068} & 0.333 \err{0.040}{0.035}\\*[4pt]
0.0  -- 5.0 & 0.399 \err{0.041}{0.038} & 0.451 \err{0.094}{0.079} & 0.441 \err{0.044}{0.041}\\*[4pt]
5.0  -- 10.0 & 0.069 \err{0.019}{0.016} & 0.070 \err{0.048}{0.030} & 0.149 \err{0.028}{0.023}\\*[4pt]
$>$10.0 & 0.029 \err{0.014}{0.010} & 0.141 \err{0.060}{0.044} & 0.077 \err{0.021}{0.017}\\*[4pt]
\tableline\\*[0pt]
\multicolumn{4}{c}{OII Equivalent Width} \\
\tableline\\*[0pt]
$<$-5.0 & 0.000 \err{0.027}{0.000} & 0.000 \err{0.183}{0.000} & 0.000 \err{0.056}{0.000}\\*[4pt]
-5.0  -- 0.0 & 0.103 \err{0.055}{0.038} & 0.000 \err{0.183}{0.000} & 0.152 \err{0.101}{0.066}\\*[4pt]
0.0  -- 5.0 & 0.676 \err{0.115}{0.098} & 0.400 \err{0.315}{0.191} & 0.545 \err{0.161}{0.126}\\*[4pt]
5.0  -- 10.0 & 0.132 \err{0.060}{0.043} & 0.300 \err{0.290}{0.163} & 0.182 \err{0.108}{0.072}\\*[4pt]
$>$10.0 & 0.074 \err{0.049}{0.032} & 0.300 \err{0.290}{0.163} & 0.121 \err{0.096}{0.058}\\*[4pt]
\tableline\\*[0pt]
\multicolumn{4}{c}{OIII Equivalent Width} \\
\tableline\\*[0pt]
$<$-5.0 & 0.000 \err{0.007}{0.000} & 0.000 \err{0.027}{0.000} & 0.000 \err{0.007}{0.000}\\*[4pt]
-5.0  -- 0.0 & 0.215 \err{0.032}{0.028} & 0.101 \err{0.055}{0.037} & 0.079 \err{0.021}{0.018}\\*[4pt]
0.0  -- 5.0 & 0.674 \err{0.054}{0.050} & 0.739 \err{0.118}{0.103} & 0.728 \err{0.058}{0.053}\\*[4pt]
5.0  -- 10.0 & 0.067 \err{0.019}{0.016} & 0.058 \err{0.046}{0.028} & 0.059 \err{0.019}{0.015}\\*[4pt]
$>$10.0 & 0.041 \err{0.016}{0.012} & 0.101 \err{0.055}{0.037} & 0.134 \err{0.027}{0.023}\\*[4pt]
\enddata
\label{table:galsnid}
\end{deluxetable}

\bibliographystyle{apj}

\begin{thebibliography}{93}
\expandafter\ifx\csname natexlab\endcsname\relax\def\natexlab#1{#1}\fi

\bibitem[{{Alam} {et~al.}(2015){Alam}, {Albareti}, {Allende Prieto}, {Anders},
  {Anderson}, {Anderton}, {Andrews}, {Armengaud}, {Aubourg}, {Bailey}, \&
  et~al.}]{Alam15}
{Alam}, S., {Albareti}, F.~D., {Allende Prieto}, C., {et~al.} 2015, \apjs, 219,
  12

\bibitem[{{Arcavi} {et~al.}(2011){Arcavi}, {Gal-Yam}, {Yaron}, {Sternberg},
  {Rabinak}, {Waxman}, {Kasliwal}, {Quimby}, {Ofek}, {Horesh}, {Kulkarni},
  {Filippenko}, {Silverman}, {Cenko}, {Li}, {Bloom}, {Sullivan}, {Nugent},
  {Poznanski}, {Gorbikov}, {Fulton}, {Howell}, {Bersier}, {Riou},
  {Lamotte-Bailey}, {Griga}, {Cohen}, {Hachinger}, {Polishook}, {Xu},
  {Ben-Ami}, {Manulis}, {Walker}, {Maguire}, {Pan}, {Matheson}, {Mazzali},
  {Pian}, {Fox}, {Gehrels}, {Law}, {James}, {Marchant}, {Smith}, {Mottram},
  {Barnsley}, {Kandrashoff}, \& {Clubb}}]{Arcavi11}
{Arcavi}, I., {Gal-Yam}, A., {Yaron}, O., {et~al.} 2011, \apjl, 742, L18

\bibitem[{{Barbon} {et~al.}(1995){Barbon}, {Benetti}, {Cappellaro}, {Patat},
  {Turatto}, \& {Iijima}}]{Barbon95}
{Barbon}, R., {Benetti}, S., {Cappellaro}, E., {et~al.} 1995, \aaps, 110, 513

\bibitem[{{Bernstein} {et~al.}(2012){Bernstein}, {Kessler}, {Kuhlmann},
  {Biswas}, {Kovacs}, {Aldering}, {Crane}, {D'Andrea}, {Finley}, {Frieman},
  {Hufford}, {Jarvis}, {Kim}, {Marriner}, {Mukherjee}, {Nichol}, {Nugent},
  {Parkinson}, {Reis}, {Sako}, {Spinka}, \& {Sullivan}}]{Bernstein12}
{Bernstein}, J.~P., {Kessler}, R., {Kuhlmann}, S., {et~al.} 2012, \apj, 753,
  152

\bibitem[{{Betoule} {et~al.}(2014){Betoule}, {Kessler}, {Guy}, {Mosher},
  {Hardin}, {Biswas}, {Astier}, {El-Hage}, {Konig}, {Kuhlmann}, {Marriner},
  {Pain}, {Regnault}, {Balland}, {Bassett}, {Brown}, {Campbell}, {Carlberg},
  {Cellier-Holzem}, {Cinabro}, {Conley}, {D'Andrea}, {DePoy}, {Doi}, {Ellis},
  {Fabbro}, {Filippenko}, {Foley}, {Frieman}, {Fouchez}, {Galbany}, {Goobar},
  {Gupta}, {Hill}, {Hlozek}, {Hogan}, {Hook}, {Howell}, {Jha}, {Le Guillou},
  {Leloudas}, {Lidman}, {Marshall}, {M{\"o}ller}, {Mour{\~a}o}, {Neveu},
  {Nichol}, {Olmstead}, {Palanque-Delabrouille}, {Perlmutter}, {Prieto},
  {Pritchet}, {Richmond}, {Riess}, {Ruhlmann-Kleider}, {Sako}, {Schahmaneche},
  {Schneider}, {Smith}, {Sollerman}, {Sullivan}, {Walton}, \&
  {Wheeler}}]{Betoule14}
{Betoule}, M., {Kessler}, R., {Guy}, J., {et~al.} 2014, \aap, 568, A22

\bibitem[{{Bianco} {et~al.}(2014){Bianco}, {Modjaz}, {Hicken}, {Friedman},
  {Kirshner}, {Bloom}, {Challis}, {Marion}, {Wood-Vasey}, \& {Rest}}]{Bianco14}
{Bianco}, F.~B., {Modjaz}, M., {Hicken}, M., {et~al.} 2014, \apjs, 213, 19

\bibitem[{{Blake} {et~al.}(2008){Blake}, {Brough}, {Couch}, {Glazebrook},
  {Poole}, {Davis}, {Drinkwater}, {Jurek}, {Pimbblet}, {Colless}, {Sharp},
  {Croom}, {Pracy}, {Woods}, {Madore}, {Martin}, \& {Wyder}}]{Blake08}
{Blake}, C., {Brough}, S., {Couch}, W., {et~al.} 2008, Astronomy and
  Geophysics, 49, 5.19

\bibitem[{{Brown} {et~al.}(2014){Brown}, {Breeveld}, {Holland}, {Kuin}, \&
  {Pritchard}}]{Brown14}
{Brown}, P.~J., {Breeveld}, A.~A., {Holland}, S., {Kuin}, P., \& {Pritchard},
  T. 2014, \apss, 354, 89

\bibitem[{{Campbell} {et~al.}(2013){Campbell}, {D'Andrea}, {Nichol}, {Sako},
  {Smith}, {Lampeitl}, {Olmstead}, {Bassett}, {Biswas}, {Brown}, {Cinabro},
  {Dawson}, {Dilday}, {Foley}, {Frieman}, {Garnavich}, {Hlozek}, {Jha},
  {Kuhlmann}, {Kunz}, {Marriner}, {Miquel}, {Richmond}, {Riess}, {Schneider},
  {Sollerman}, {Taylor}, \& {Zhao}}]{Campbell13}
{Campbell}, H., {D'Andrea}, C.~B., {Nichol}, R.~C., {et~al.} 2013, \apj, 763,
  88

\bibitem[{{Childress} {et~al.}(2013){Childress}, {Aldering}, {Antilogus},
  {Aragon}, {Bailey}, {Baltay}, {Bongard}, {Buton}, {Canto}, {Cellier-Holzem},
  {Chotard}, {Copin}, {Fakhouri}, {Gangler}, {Guy}, {Hsiao}, {Kerschhaggl},
  {Kim}, {Kowalski}, {Loken}, {Nugent}, {Paech}, {Pain}, {Pecontal}, {Pereira},
  {Perlmutter}, {Rabinowitz}, {Rigault}, {Runge}, {Scalzo}, {Smadja}, {Tao},
  {Thomas}, {Weaver}, \& {Wu}}]{Childress13}
{Childress}, M., {Aldering}, G., {Antilogus}, P., {et~al.} 2013, \apj, 770, 108

\bibitem[{{Colless} {et~al.}(2003){Colless}, {Peterson}, {Jackson}, {Peacock},
  {Cole}, {Norberg}, {Baldry}, {Baugh}, {Bland-Hawthorn}, {Bridges}, {Cannon},
  {Collins}, {Couch}, {Cross}, {Dalton}, {De Propris}, {Driver}, {Efstathiou},
  {Ellis}, {Frenk}, {Glazebrook}, {Lahav}, {Lewis}, {Lumsden}, {Maddox},
  {Madgwick}, {Sutherland}, \& {Taylor}}]{Colless03}
{Colless}, M., {Peterson}, B.~A., {Jackson}, C., {et~al.} 2003, ArXiv
  Astrophysics e-prints

\bibitem[{{Conley} {et~al.}(2011){Conley}, {Guy}, {Sullivan}, {Regnault},
  {Astier}, {Balland}, {Basa}, {Carlberg}, {Fouchez}, {Hardin}, {Hook},
  {Howell}, {Pain}, {Palanque-Delabrouille}, {Perrett}, {Pritchet}, {Rich},
  {Ruhlmann-Kleider}, {Balam}, {Baumont}, {Ellis}, {Fabbro}, {Fakhouri},
  {Fourmanoit}, {Gonz{\'a}lez-Gait{\'a}n}, {Graham}, {Hudson}, {Hsiao},
  {Kronborg}, {Lidman}, {Mourao}, {Neill}, {Perlmutter}, {Ripoche}, {Suzuki},
  \& {Walker}}]{Conley11}
{Conley}, A., {Guy}, J., {Sullivan}, M., {et~al.} 2011, \apjs, 192, 1

\bibitem[{{Contreras} {et~al.}(2010){Contreras}, {Hamuy}, {Phillips},
  {Folatelli}, {Suntzeff}, {Persson}, {Stritzinger}, {Boldt}, {Gonz{\'a}lez},
  {Krzeminski}, {Morrell}, {Roth}, {Salgado}, {Jos{\'e} Maureira}, {Burns},
  {Freedman}, {Madore}, {Murphy}, {Wyatt}, {Li}, \& {Filippenko}}]{Contreras10}
{Contreras}, C., {Hamuy}, M., {Phillips}, M.~M., {et~al.} 2010, \aj, 139, 519

\bibitem[{{Ergon} {et~al.}(2014){Ergon}, {Sollerman}, {Fraser}, {Pastorello},
  {Taubenberger}, {Elias-Rosa}, {Bersten}, {Jerkstrand}, {Benetti},
  {Botticella}, {Fransson}, {Harutyunyan}, {Kotak}, {Smartt}, {Valenti},
  {Bufano}, {Cappellaro}, {Fiaschi}, {Howell}, {Kankare}, {Magill}, {Mattila},
  {Maund}, {Naves}, {Ochner}, {Ruiz}, {Smith}, {Tomasella}, \&
  {Turatto}}]{Ergon14}
{Ergon}, M., {Sollerman}, J., {Fraser}, M., {et~al.} 2014, \aap, 562, A17

\bibitem[{{Ergon} {et~al.}(2015){Ergon}, {Jerkstrand}, {Sollerman},
  {Elias-Rosa}, {Fransson}, {Fraser}, {Pastorello}, {Kotak}, {Taubenberger},
  {Tomasella}, {Valenti}, {Benetti}, {Helou}, {Kasliwal}, {Maund}, {Smartt}, \&
  {Spyromilio}}]{Ergon15}
{Ergon}, M., {Jerkstrand}, A., {Sollerman}, J., {et~al.} 2015, \aap, 580, A142

\bibitem[{{Fabricant} {et~al.}(2005){Fabricant}, {Fata}, {Roll}, {Hertz},
  {Caldwell}, {Gauron}, {Geary}, {McLeod}, {Szentgyorgyi}, {Zajac}, {Kurtz},
  {Barberis}, {Bergner}, {Brown}, {Conroy}, {Eng}, {Geller}, {Goddard},
  {Honsa}, {Mueller}, {Mink}, {Ordway}, {Tokarz}, {Woods}, {Wyatt}, {Epps}, \&
  {Dell'Antonio}}]{Fabricant05}
{Fabricant}, D., {Fata}, R., {Roll}, J., {et~al.} 2005, \pasp, 117, 1411

\bibitem[{{Falck} {et~al.}(2010){Falck}, {Riess}, \& {Hlozek}}]{Falck10}
{Falck}, B.~L., {Riess}, A.~G., \& {Hlozek}, R. 2010, \apj, 723, 398

\bibitem[{{Foley} \& {Mandel}(2013)}]{Foley13}
{Foley}, R.~J., \& {Mandel}, K. 2013, \apj, 778, 167

\bibitem[{{Foley} {et~al.}(2013){Foley}, {Challis}, {Chornock},
  {Ganeshalingam}, {Li}, {Marion}, {Morrell}, {Pignata}, {Stritzinger},
  {Silverman}, {Wang}, {Anderson}, {Filippenko}, {Freedman}, {Hamuy}, {Jha},
  {Kirshner}, {McCully}, {Persson}, {Phillips}, {Reichart}, \&
  {Soderberg}}]{Foley13b}
{Foley}, R.~J., {Challis}, P.~J., {Chornock}, R., {et~al.} 2013, \apj, 767, 57

\bibitem[{{Foreman-Mackey} {et~al.}(2013){Foreman-Mackey}, {Hogg}, {Lang}, \&
  {Goodman}}]{Foreman13}
{Foreman-Mackey}, D., {Hogg}, D.~W., {Lang}, D., \& {Goodman}, J. 2013, \pasp,
  125, 306

\bibitem[{{Frieman} {et~al.}(2008){Frieman}, {Bassett}, {Becker}, {Choi},
  {Cinabro}, {DeJongh}, {Depoy}, {Dilday}, {Doi}, {Garnavich}, {Hogan},
  {Holtzman}, {Im}, {Jha}, {Kessler}, {Konishi}, {Lampeitl}, {Marriner},
  {Marshall}, {McGinnis}, {Miknaitis}, {Nichol}, {Prieto}, {Riess}, {Richmond},
  {Romani}, {Sako}, {Schneider}, {Smith}, {Takanashi}, {Tokita}, {van der
  Heyden}, {Yasuda}, {Zheng}, {Adelman-McCarthy}, {Annis}, {Assef},
  {Barentine}, {Bender}, {Blandford}, {Boroski}, {Bremer}, {Brewington},
  {Collins}, {Crotts}, {Dembicky}, {Eastman}, {Edge}, {Edmondson}, {Elson},
  {Eyler}, {Filippenko}, {Foley}, {Frank}, {Goobar}, {Gueth}, {Gunn},
  {Harvanek}, {Hopp}, {Ihara}, {Ivezi{\'c}}, {Kahn}, {Kaplan}, {Kent},
  {Ketzeback}, {Kleinman}, {Kollatschny}, {Kron}, {Krzesi{\'n}ski}, {Lamenti},
  {Leloudas}, {Lin}, {Long}, {Lucey}, {Lupton}, {Malanushenko}, {Malanushenko},
  {McMillan}, {Mendez}, {Morgan}, {Morokuma}, {Nitta}, {Ostman}, {Pan},
  {Rockosi}, {Romer}, {Ruiz-Lapuente}, {Saurage}, {Schlesinger}, {Snedden},
  {Sollerman}, {Stoughton}, {Stritzinger}, {Subba Rao}, {Tucker}, {Vaisanen},
  {Watson}, {Watters}, {Wheeler}, {Yanny}, \& {York}}]{Frieman08}
{Frieman}, J.~A., {Bassett}, B., {Becker}, A., {et~al.} 2008, \aj, 135, 338

\bibitem[{{Ganeshalingam} {et~al.}(2010){Ganeshalingam}, {Li}, {Filippenko},
  {Anderson}, {Foster}, {Gates}, {Griffith}, {Grigsby}, {Joubert}, {Leja},
  {Lowe}, {Macomber}, {Pritchard}, {Thrasher}, \& {Winslow}}]{Ganeshalingam10}
{Ganeshalingam}, M., {Li}, W., {Filippenko}, A.~V., {et~al.} 2010, \apjs, 190,
  418

\bibitem[{{Garnavich} {et~al.}(2004){Garnavich}, {Bonanos}, {Krisciunas},
  {Jha}, {Kirshner}, {Schlegel}, {Challis}, {Macri}, {Hatano}, {Branch},
  {Bothun}, \& {Freedman}}]{Garnavich04}
{Garnavich}, P.~M., {Bonanos}, A.~Z., {Krisciunas}, K., {et~al.} 2004, \apj,
  613, 1120

\bibitem[{{Guillochon} {et~al.}(2017){Guillochon}, {Parrent}, {Kelley}, \&
  {Margutti}}]{Guillochon17}
{Guillochon}, J., {Parrent}, J., {Kelley}, L.~Z., \& {Margutti}, R. 2017, \apj,
  835, 64

\bibitem[{{Gupta} {et~al.}(2016){Gupta}, {Kuhlmann}, {Kovacs}, {Spinka},
  {Kessler}, {Goldstein}, {Liotine}, {Pomian}, {D'Andrea}, {Sullivan},
  {Carretero}, {Castander}, {Nichol}, {Finley}, {Fischer}, {Foley}, {Kim},
  {Papadopoulos}, {Sako}, {Scolnic}, {Smith}, {Tucker}, {Uddin}, {Wolf},
  {Yuan}, {Abbott}, {Abdalla}, {Benoit-L{\'e}vy}, {Bertin}, {Brooks}, {Carnero
  Rosell}, {Carrasco Kind}, {Cunha}, {da Costa}, {Desai}, {Doel}, {Eifler},
  {Evrard}, {Flaugher}, {Fosalba}, {Gazta{\~n}aga}, {Gruen}, {Gruendl},
  {James}, {Kuehn}, {Kuropatkin}, {Maia}, {Marshall}, {Miquel}, {Plazas},
  {Romer}, {S{\'a}nchez}, {Schubnell}, {Sevilla-Noarbe}, {Sobreira}, {Suchyta},
  {Swanson}, {Tarle}, {Walker}, \& {Wester}}]{Gupta16}
{Gupta}, R.~R., {Kuhlmann}, S., {Kovacs}, E., {et~al.} 2016, \aj, 152, 154

\bibitem[{{Guy} {et~al.}(2007){Guy}, {Astier}, {Baumont}, {Hardin}, {Pain},
  {Regnault}, {Basa}, {Carlberg}, {Conley}, {Fabbro}, {Fouchez}, {Hook},
  {Howell}, {Perrett}, {Pritchet}, {Rich}, {Sullivan}, {Antilogus}, {Aubourg},
  {Bazin}, {Bronder}, {Filiol}, {Palanque-Delabrouille}, {Ripoche}, \&
  {Ruhlmann-Kleider}}]{Guy07}
{Guy}, J., {Astier}, P., {Baumont}, S., {et~al.} 2007, \aap, 466, 11

\bibitem[{{Guy} {et~al.}(2010){Guy}, {Sullivan}, {Conley}, {Regnault},
  {Astier}, {Balland}, {Basa}, {Carlberg}, {Fouchez}, {Hardin}, {Hook},
  {Howell}, {Pain}, {Palanque-Delabrouille}, {Perrett}, {Pritchet}, {Rich},
  {Ruhlmann-Kleider}, {Balam}, {Baumont}, {Ellis}, {Fabbro}, {Fakhouri},
  {Fourmanoit}, {Gonz{\'a}lez-Gait{\'a}n}, {Graham}, {Hsiao}, {Kronborg},
  {Lidman}, {Mourao}, {Perlmutter}, {Ripoche}, {Suzuki}, \& {Walker}}]{Guy10}
{Guy}, J., {Sullivan}, M., {Conley}, A., {et~al.} 2010, \aap, 523, A7

\bibitem[{{Hamuy} {et~al.}(2006){Hamuy}, {Folatelli}, {Morrell}, {Phillips},
  {Suntzeff}, {Persson}, {Roth}, {Gonzalez}, {Krzeminski}, {Contreras},
  {Freedman}, {Murphy}, {Madore}, {Wyatt}, {Maza}, {Filippenko}, {Li}, \&
  {Pinto}}]{Hamuy06}
{Hamuy}, M., {Folatelli}, G., {Morrell}, N.~I., {et~al.} 2006, \pasp, 118, 2

\bibitem[{{Hlozek} {et~al.}(2012){Hlozek}, {Kunz}, {Bassett}, {Smith},
  {Newling}, {Varughese}, {Kessler}, {Bernstein}, {Campbell}, {Dilday},
  {Falck}, {Frieman}, {Kuhlmann}, {Lampeitl}, {Marriner}, {Nichol}, {Riess},
  {Sako}, \& {Schneider}}]{Hlozek12}
{Hlozek}, R., {Kunz}, M., {Bassett}, B., {et~al.} 2012, \apj, 752, 79

\bibitem[{{Hsiao} {et~al.}(2007){Hsiao}, {Conley}, {Howell}, {Sullivan},
  {Pritchet}, {Carlberg}, {Nugent}, \& {Phillips}}]{Hsiao07}
{Hsiao}, E.~Y., {Conley}, A., {Howell}, D.~A., {et~al.} 2007, \apj, 663, 1187

\bibitem[{{Jerkstrand} {et~al.}(2015){Jerkstrand}, {Ergon}, {Smartt},
  {Fransson}, {Sollerman}, {Taubenberger}, {Bersten}, \&
  {Spyromilio}}]{Jerkstrand15}
{Jerkstrand}, A., {Ergon}, M., {Smartt}, S.~J., {et~al.} 2015, \aap, 573, A12

\bibitem[{{Jones} {et~al.}(2009){Jones}, {Read}, {Saunders}, {Colless},
  {Jarrett}, {Parker}, {Fairall}, {Mauch}, {Sadler}, {Watson}, {Burton},
  {Campbell}, {Cass}, {Croom}, {Dawe}, {Fiegert}, {Frankcombe}, {Hartley},
  {Huchra}, {James}, {Kirby}, {Lahav}, {Lucey}, {Mamon}, {Moore}, {Peterson},
  {Prior}, {Proust}, {Russell}, {Safouris}, {Wakamatsu}, {Westra}, \&
  {Williams}}]{Jones09}
{Jones}, D.~H., {Read}, M.~A., {Saunders}, W., {et~al.} 2009, \mnras, 399, 683

\bibitem[{Jones(2017)}]{jonesbeams17}
Jones, D.~O. 2017, {Measuring Dark Energy With Photometrically Classified
  Pan-STARRS Supernova: Bayesian Estimation Applied to Multiple Species
  Algorithm}

\bibitem[{{Kaiser} {et~al.}(2010){Kaiser}, {Burgett}, {Chambers}, {Denneau},
  {Heasley}, {Jedicke}, {Magnier}, {Morgan}, {Onaka}, \& {Tonry}}]{Kaiser10}
{Kaiser}, N., {Burgett}, W., {Chambers}, K., {et~al.} 2010, in Society of
  Photo-Optical Instrumentation Engineers (SPIE) Conference Series, Vol. 7733,
  Society of Photo-Optical Instrumentation Engineers (SPIE) Conference Series,
  0

\bibitem[{{Kelly}(2007)}]{Kelly07}
{Kelly}, B.~C. 2007, \apj, 665, 1489

\bibitem[{{Kelly} {et~al.}(2010){Kelly}, {Hicken}, {Burke}, {Mandel}, \&
  {Kirshner}}]{Kelly10}
{Kelly}, P.~L., {Hicken}, M., {Burke}, D.~L., {Mandel}, K.~S., \& {Kirshner},
  R.~P. 2010, \apj, 715, 743

\bibitem[{{Kessler} \& {Scolnic}(2017)}]{Kessler17}
{Kessler}, R., \& {Scolnic}, D. 2017, \apj, 836, 56

\bibitem[{{Kessler} {et~al.}(2009{\natexlab{a}}){Kessler}, {Becker}, {Cinabro},
  {Vanderplas}, {Frieman}, {Marriner}, {Davis}, {Dilday}, {Holtzman}, {Jha},
  {Lampeitl}, {Sako}, {Smith}, {Zheng}, {Nichol}, {Bassett}, {Bender}, {Depoy},
  {Doi}, {Elson}, {Filippenko}, {Foley}, {Garnavich}, {Hopp}, {Ihara},
  {Ketzeback}, {Kollatschny}, {Konishi}, {Marshall}, {McMillan}, {Miknaitis},
  {Morokuma}, {M{\"o}rtsell}, {Pan}, {Prieto}, {Richmond}, {Riess}, {Romani},
  {Schneider}, {Sollerman}, {Takanashi}, {Tokita}, {van der Heyden}, {Wheeler},
  {Yasuda}, \& {York}}]{Kessler09}
{Kessler}, R., {Becker}, A.~C., {Cinabro}, D., {et~al.} 2009{\natexlab{a}},
  \apjs, 185, 32

\bibitem[{{Kessler} {et~al.}(2009{\natexlab{b}}){Kessler}, {Bernstein},
  {Cinabro}, {Dilday}, {Frieman}, {Jha}, {Kuhlmann}, {Miknaitis}, {Sako},
  {Taylor}, \& {Vanderplas}}]{Kessler09b}
{Kessler}, R., {Bernstein}, J.~P., {Cinabro}, D., {et~al.} 2009{\natexlab{b}},
  \pasp, 121, 1028

\bibitem[{{Kessler} {et~al.}(2010){Kessler}, {Bassett}, {Belov}, {Bhatnagar},
  {Campbell}, {Conley}, {Frieman}, {Glazov}, {Gonz{\'a}lez-Gait{\'a}n},
  {Hlozek}, {Jha}, {Kuhlmann}, {Kunz}, {Lampeitl}, {Mahabal}, {Newling},
  {Nichol}, {Parkinson}, {Sajeeth Philip}, {Poznanski}, {Richards}, {Rodney},
  {Sako}, {Schneider}, {Smith}, {Stritzinger}, \& {Varughese}}]{Kessler10}
{Kessler}, R., {Bassett}, B., {Belov}, P., {et~al.} 2010, \pasp, 122, 1415

\bibitem[{{Kessler} {et~al.}(2015){Kessler}, {Marriner}, {Childress},
  {Covarrubias}, {D'Andrea}, {Finley}, {Fischer}, {Foley}, {Goldstein},
  {Gupta}, {Kuehn}, {Marcha}, {Nichol}, {Papadopoulos}, {Sako}, {Scolnic},
  {Smith}, {Sullivan}, {Wester}, {Yuan}, {Abbott}, {Abdalla}, {Allam},
  {Benoit-L{\'e}vy}, {Bernstein}, {Bertin}, {Brooks}, {Carnero Rosell},
  {Carrasco Kind}, {Castander}, {Crocce}, {da Costa}, {Desai}, {Diehl},
  {Eifler}, {Fausti Neto}, {Flaugher}, {Frieman}, {Gerdes}, {Gruen}, {Gruendl},
  {Honscheid}, {James}, {Kuropatkin}, {Li}, {Maia}, {Marshall}, {Martini},
  {Miller}, {Miquel}, {Nord}, {Ogando}, {Plazas}, {Reil}, {Romer}, {Roodman},
  {Sanchez}, {Sevilla-Noarbe}, {Smith}, {Soares-Santos}, {Sobreira}, {Tarle},
  {Thaler}, {Thomas}, {Tucker}, {Walker}, \& {DES Collaboration}}]{Kessler15}
{Kessler}, R., {Marriner}, J., {Childress}, M., {et~al.} 2015, \aj, 150, 172

\bibitem[{{Kunz} {et~al.}(2007){Kunz}, {Bassett}, \& {Hlozek}}]{Kunz07}
{Kunz}, M., {Bassett}, B.~A., \& {Hlozek}, R.~A. 2007, \prd, 75, 103508

\bibitem[{{Kurtz} \& {Mink}(1998)}]{Kurtz98}
{Kurtz}, M.~J., \& {Mink}, D.~J. 1998, \pasp, 110, 934

\bibitem[{{Lampeitl} {et~al.}(2010){Lampeitl}, {Smith}, {Nichol}, {Bassett},
  {Cinabro}, {Dilday}, {Foley}, {Frieman}, {Garnavich}, {Goobar}, {Im}, {Jha},
  {Marriner}, {Miquel}, {Nordin}, {{\"O}stman}, {Riess}, {Sako}, {Schneider},
  {Sollerman}, \& {Stritzinger}}]{Lampeitl10}
{Lampeitl}, H., {Smith}, M., {Nichol}, R.~C., {et~al.} 2010, \apj, 722, 566

\bibitem[{{Le F{\`e}vre} {et~al.}(2005){Le F{\`e}vre}, {Vettolani}, {Garilli},
  {Tresse}, {Bottini}, {Le Brun}, {Maccagni}, {Picat}, {Scaramella},
  {Scodeggio}, {Zanichelli}, {Adami}, {Arnaboldi}, {Arnouts}, {Bardelli},
  {Bolzonella}, {Cappi}, {Charlot}, {Ciliegi}, {Contini}, {Foucaud},
  {Franzetti}, {Gavignaud}, {Guzzo}, {Ilbert}, {Iovino}, {McCracken}, {Marano},
  {Marinoni}, {Mathez}, {Mazure}, {Meneux}, {Merighi}, {Paltani}, {Pell{\`o}},
  {Pollo}, {Pozzetti}, {Radovich}, {Zamorani}, {Zucca}, {Bondi}, {Bongiorno},
  {Busarello}, {Lamareille}, {Mellier}, {Merluzzi}, {Ripepi}, \&
  {Rizzo}}]{LeFevre05}
{Le F{\`e}vre}, O., {Vettolani}, G., {Garilli}, B., {et~al.} 2005, \aap, 439,
  845

\bibitem[{{Leaman} {et~al.}(2011){Leaman}, {Li}, {Chornock}, \&
  {Filippenko}}]{Leaman11}
{Leaman}, J., {Li}, W., {Chornock}, R., \& {Filippenko}, A.~V. 2011, \mnras,
  412, 1419

\bibitem[{{Lewis} \& {Bridle}(2002)}]{Lewis02}
{Lewis}, A., \& {Bridle}, S. 2002, \prd, 66, 103511

\bibitem[{{Li} {et~al.}(2011){Li}, {Leaman}, {Chornock}, {Filippenko},
  {Poznanski}, {Ganeshalingam}, {Wang}, {Modjaz}, {Jha}, {Foley}, \&
  {Smith}}]{Li11}
{Li}, W., {Leaman}, J., {Chornock}, R., {et~al.} 2011, \mnras, 412, 1441

\bibitem[{{Lilly} {et~al.}(2007){Lilly}, {Le F{\`e}vre}, {Renzini}, {Zamorani},
  {Scodeggio}, {Contini}, {Carollo}, {Hasinger}, {Kneib}, {Iovino}, {Le Brun},
  {Maier}, {Mainieri}, {Mignoli}, {Silverman}, {Tasca}, {Bolzonella},
  {Bongiorno}, {Bottini}, {Capak}, {Caputi}, {Cimatti}, {Cucciati}, {Daddi},
  {Feldmann}, {Franzetti}, {Garilli}, {Guzzo}, {Ilbert}, {Kampczyk}, {Kovac},
  {Lamareille}, {Leauthaud}, {Borgne}, {McCracken}, {Marinoni}, {Pello},
  {Ricciardelli}, {Scarlata}, {Vergani}, {Sanders}, {Schinnerer}, {Scoville},
  {Taniguchi}, {Arnouts}, {Aussel}, {Bardelli}, {Brusa}, {Cappi}, {Ciliegi},
  {Finoguenov}, {Foucaud}, {Franceschini}, {Halliday}, {Impey}, {Knobel},
  {Koekemoer}, {Kurk}, {Maccagni}, {Maddox}, {Marano}, {Marconi}, {Meneux},
  {Mobasher}, {Moreau}, {Peacock}, {Porciani}, {Pozzetti}, {Scaramella},
  {Schiminovich}, {Shopbell}, {Smail}, {Thompson}, {Tresse}, {Vettolani},
  {Zanichelli}, \& {Zucca}}]{Lilly07}
{Lilly}, S.~J., {Le F{\`e}vre}, O., {Renzini}, A., {et~al.} 2007, \apjs, 172,
  70

\bibitem[{{Lochner} {et~al.}(2016){Lochner}, {McEwen}, {Peiris}, {Lahav}, \&
  {Winter}}]{Lochner16}
{Lochner}, M., {McEwen}, J.~D., {Peiris}, H.~V., {Lahav}, O., \& {Winter},
  M.~K. 2016, \apjs, 225, 31

\bibitem[{{Lunnan} {et~al.}(2015){Lunnan}, {Chornock}, {Berger}, {Rest},
  {Fong}, {Scolnic}, {Jones}, {Soderberg}, {Challis}, {Drout}, {Foley},
  {Huber}, {Kirshner}, {Leibler}, {Marion}, {McCrum}, {Milisavljevic},
  {Narayan}, {Sanders}, {Smartt}, {Smith}, {Tonry}, {Burgett}, {Chambers},
  {Flewelling}, {Kudritzki}, {Wainscoat}, \& {Waters}}]{Lunnan15}
{Lunnan}, R., {Chornock}, R., {Berger}, E., {et~al.} 2015, \apj, 804, 90

\bibitem[{{March} {et~al.}(2011){March}, {Trotta}, {Berkes}, {Starkman}, \&
  {Vaudrevange}}]{March11}
{March}, M.~C., {Trotta}, R., {Berkes}, P., {Starkman}, G.~D., \&
  {Vaudrevange}, P.~M. 2011, \mnras, 418, 2308

\bibitem[{{Marriner} {et~al.}(2011){Marriner}, {Bernstein}, {Kessler},
  {Lampeitl}, {Miquel}, {Mosher}, {Nichol}, {Sako}, {Schneider}, \&
  {Smith}}]{Marriner11}
{Marriner}, J., {Bernstein}, J.~P., {Kessler}, R., {et~al.} 2011, \apj, 740, 72

\bibitem[{{Metlova} {et~al.}(1995){Metlova}, {Tsvetkov}, {Shugarov}, {Esipov},
  \& {Pavlyuk}}]{Metlova95}
{Metlova}, N.~V., {Tsvetkov}, D.~Y., {Shugarov}, S.~Y., {Esipov}, V.~F., \&
  {Pavlyuk}, N.~N. 1995, Astronomy Letters, 21, 598

\bibitem[{{Mink} {et~al.}(2007){Mink}, {Wyatt}, {Caldwell}, {Conroy}, {Furesz},
  \& {Tokarz}}]{Mink07}
{Mink}, D.~J., {Wyatt}, W.~F., {Caldwell}, N., {et~al.} 2007, in Astronomical
  Society of the Pacific Conference Series, Vol. 376, Astronomical Data
  Analysis Software and Systems XVI, ed. R.~A. {Shaw}, F.~{Hill}, \& D.~J.
  {Bell}, 249

\bibitem[{{Modjaz} {et~al.}(2001){Modjaz}, {Li}, {Filippenko}, {King},
  {Leonard}, {Matheson}, {Treffers}, \& {Riess}}]{Modjaz01}
{Modjaz}, M., {Li}, W., {Filippenko}, A.~V., {et~al.} 2001, \pasp, 113, 308

\bibitem[{{Modjaz} {et~al.}(2014){Modjaz}, {Blondin}, {Kirshner}, {Matheson},
  {Berlind}, {Bianco}, {Calkins}, {Challis}, {Garnavich}, {Hicken}, {Jha},
  {Liu}, \& {Marion}}]{Modjaz14}
{Modjaz}, M., {Blondin}, S., {Kirshner}, R.~P., {et~al.} 2014, \aj, 147, 99

\bibitem[{{M{\"o}ller} {et~al.}(2016){M{\"o}ller}, {Ruhlmann-Kleider},
  {Leloup}, {Neveu}, {Palanque-Delabrouille}, {Rich}, {Carlberg}, {Lidman}, \&
  {Pritchet}}]{Moller16}
{M{\"o}ller}, A., {Ruhlmann-Kleider}, V., {Leloup}, C., {et~al.} 2016, JCAP,
  12, 008

\bibitem[{{Newman} {et~al.}(2013){Newman}, {Cooper}, {Davis}, {Faber}, {Coil},
  {Guhathakurta}, {Koo}, {Phillips}, {Conroy}, {Dutton}, {Finkbeiner}, {Gerke},
  {Rosario}, {Weiner}, {Willmer}, {Yan}, {Harker}, {Kassin}, {Konidaris},
  {Lai}, {Madgwick}, {Noeske}, {Wirth}, {Connolly}, {Kaiser}, {Kirby},
  {Lemaux}, {Lin}, {Lotz}, {Luppino}, {Marinoni}, {Matthews}, {Metevier}, \&
  {Schiavon}}]{Newman13}
{Newman}, J.~A., {Cooper}, M.~C., {Davis}, M., {et~al.} 2013, \apjs, 208, 5

\bibitem[{{Nugent} {et~al.}(2002){Nugent}, {Kim}, \& {Perlmutter}}]{Nugent02}
{Nugent}, P., {Kim}, A., \& {Perlmutter}, S. 2002, \pasp, 114, 803

\bibitem[{{Oguri} \& {Marshall}(2010)}]{Oguri10}
{Oguri}, M., \& {Marshall}, P.~J. 2010, \mnras, 405, 2579

\bibitem[{{Pastorello} {et~al.}(2008){Pastorello}, {Kasliwal}, {Crockett},
  {Valenti}, {Arbour}, {Itagaki}, {Kaspi}, {Gal-Yam}, {Smartt}, {Griffith},
  {Maguire}, {Ofek}, {Seymour}, {Stern}, \& {Wiethoff}}]{Pastorello08}
{Pastorello}, A., {Kasliwal}, M.~M., {Crockett}, R.~M., {et~al.} 2008, \mnras,
  389, 955

\bibitem[{{Perlmutter} {et~al.}(1999){Perlmutter}, {Aldering}, {Goldhaber},
  {Knop}, {Nugent}, {Castro}, {Deustua}, {Fabbro}, {Goobar}, {Groom}, {Hook},
  {Kim}, {Kim}, {Lee}, {Nunes}, {Pain}, {Pennypacker}, {Quimby}, {Lidman},
  {Ellis}, {Irwin}, {McMahon}, {Ruiz-Lapuente}, {Walton}, {Schaefer}, {Boyle},
  {Filippenko}, {Matheson}, {Fruchter}, {Panagia}, {Newberg}, {Couch}, \&
  {Project}}]{Perlmutter99}
{Perlmutter}, S., {Aldering}, G., {Goldhaber}, G., {et~al.} 1999, \apj, 517,
  565

\bibitem[{{Press}(1997)}]{Press97}
{Press}, W.~H. 1997, in Unsolved Problems in Astrophysics, ed. J.~N. {Bahcall}
  \& J.~P. {Ostriker}, 49--60

\bibitem[{{Rest} {et~al.}(2005){Rest}, {Stubbs}, {Becker}, {Miknaitis},
  {Miceli}, {Covarrubias}, {Hawley}, {Smith}, {Suntzeff}, {Olsen}, {Prieto},
  {Hiriart}, {Welch}, {Cook}, {Nikolaev}, {Huber}, {Prochtor}, {Clocchiatti},
  {Minniti}, {Garg}, {Challis}, {Keller}, \& {Schmidt}}]{Rest05}
{Rest}, A., {Stubbs}, C., {Becker}, A.~C., {et~al.} 2005, \apj, 634, 1103

\bibitem[{{Rest} {et~al.}(2014){Rest}, {Scolnic}, {Foley}, {Huber}, {Chornock},
  {Narayan}, {Tonry}, {Berger}, {Soderberg}, {Stubbs}, {Riess}, {Kirshner},
  {Smartt}, {Schlafly}, {Rodney}, {Botticella}, {Brout}, {Challis}, {Czekala},
  {Drout}, {Hudson}, {Kotak}, {Leibler}, {Lunnan}, {Marion}, {McCrum},
  {Milisavljevic}, {Pastorello}, {Sanders}, {Smith}, {Stafford}, {Thilker},
  {Valenti}, {Wood-Vasey}, {Zheng}, {Burgett}, {Chambers}, {Denneau}, {Draper},
  {Flewelling}, {Hodapp}, {Kaiser}, {Kudritzki}, {Magnier}, {Metcalfe},
  {Price}, {Sweeney}, {Wainscoat}, \& {Waters}}]{Rest14}
{Rest}, A., {Scolnic}, D., {Foley}, R.~J., {et~al.} 2014, \apj, 795, 44

\bibitem[{{Richardson} {et~al.}(2014){Richardson}, {Jenkins}, {Wright}, \&
  {Maddox}}]{Richardson14}
{Richardson}, D., {Jenkins}, III, R.~L., {Wright}, J., \& {Maddox}, L. 2014,
  \aj, 147, 118

\bibitem[{{Richardson} {et~al.}(2001){Richardson}, {Thomas}, {Casebeer},
  {Blankenship}, {Ratowt}, {Baron}, \& {Branch}}]{Richardson01}
{Richardson}, D., {Thomas}, R.~C., {Casebeer}, D., {et~al.} 2001, in Bulletin
  of the American Astronomical Society, Vol.~33, American Astronomical Society
  Meeting Abstracts, 1428

\bibitem[{{Richmond} {et~al.}(1996){Richmond}, {Treffers}, {Filippenko}, \&
  {Paik}}]{Richmond96}
{Richmond}, M.~W., {Treffers}, R.~R., {Filippenko}, A.~V., \& {Paik}, Y. 1996,
  \aj, 112, 732

\bibitem[{{Riess} {et~al.}(1998){Riess}, {Filippenko}, {Challis},
  {Clocchiatti}, {Diercks}, {Garnavich}, {Gilliland}, {Hogan}, {Jha},
  {Kirshner}, {Leibundgut}, {Phillips}, {Reiss}, {Schmidt}, {Schommer},
  {Smith}, {Spyromilio}, {Stubbs}, {Suntzeff}, \& {Tonry}}]{Riess98}
{Riess}, A.~G., {Filippenko}, A.~V., {Challis}, P., {et~al.} 1998, \aj, 116,
  1009

\bibitem[{{Rodney} {et~al.}(2014){Rodney}, {Riess}, {Strolger}, {Dahlen},
  {Graur}, {Casertano}, {Dickinson}, {Ferguson}, {Garnavich}, {Hayden}, {Jha},
  {Jones}, {Kirshner}, {Koekemoer}, {McCully}, {Mobasher}, {Patel}, {Weiner},
  {Cenko}, {Clubb}, {Cooper}, {Filippenko}, {Frederiksen}, {Hjorth},
  {Leibundgut}, {Matheson}, {Nayyeri}, {Penner}, {Trump}, {Silverman}, {U},
  {Azalee Bostroem}, {Challis}, {Rajan}, {Wolff}, {Faber}, {Grogin}, \&
  {Kocevski}}]{Rodney14}
{Rodney}, S.~A., {Riess}, A.~G., {Strolger}, L.-G., {et~al.} 2014, \aj, 148, 13

\bibitem[{{Rubin} {et~al.}(2015){Rubin}, {Aldering}, {Barbary}, {Boone},
  {Chappell}, {Currie}, {Deustua}, {Fagrelius}, {Fruchter}, {Hayden}, {Lidman},
  {Nordin}, {Perlmutter}, {Saunders}, {Sofiatti}, \& {Supernova Cosmology
  Project}}]{Rubin15}
{Rubin}, D., {Aldering}, G., {Barbary}, K., {et~al.} 2015, \apj, 813, 137

\bibitem[{{Sako} {et~al.}(2011){Sako}, {Bassett}, {Connolly}, {Dilday},
  {Cambell}, {Frieman}, {Gladney}, {Kessler}, {Lampeitl}, {Marriner}, {Miquel},
  {Nichol}, {Schneider}, {Smith}, \& {Sollerman}}]{Sako11}
{Sako}, M., {Bassett}, B., {Connolly}, B., {et~al.} 2011, \apj, 738, 162

\bibitem[{{Sako} {et~al.}(2014){Sako}, {Bassett}, {Becker}, {Brown},
  {Campbell}, {Cane}, {Cinabro}, {D'Andrea}, {Dawson}, {DeJongh}, {Depoy},
  {Dilday}, {Doi}, {Filippenko}, {Fischer}, {Foley}, {Frieman}, {Galbany},
  {Garnavich}, {Goobar}, {Gupta}, {Hill}, {Hayden}, {Hlozek}, {Holtzman},
  {Hopp}, {Jha}, {Kessler}, {Kollatschny}, {Leloudas}, {Marriner}, {Marshall},
  {Miquel}, {Morokuma}, {Mosher}, {Nichol}, {Nordin}, {Olmstead}, {Ostman},
  {Prieto}, {Richmond}, {Romani}, {Sollerman}, {Stritzinger}, {Schneider},
  {Smith}, {Wheeler}, {Yasuda}, \& {Zheng}}]{Sako14}
{Sako}, M., {Bassett}, B., {Becker}, A.~C., {et~al.} 2014, ArXiv e-prints

\bibitem[{{Sanders} {et~al.}(2015){Sanders}, {Soderberg}, {Gezari},
  {Betancourt}, {Chornock}, {Berger}, {Foley}, {Challis}, {Drout}, {Kirshner},
  {Lunnan}, {Marion}, {Margutti}, {McKinnon}, {Milisavljevic}, {Narayan},
  {Rest}, {Kankare}, {Mattila}, {Smartt}, {Huber}, {Burgett}, {Draper},
  {Hodapp}, {Kaiser}, {Kudritzki}, {Magnier}, {Metcalfe}, {Morgan}, {Price},
  {Tonry}, {Wainscoat}, \& {Waters}}]{Sanders15}
{Sanders}, N.~E., {Soderberg}, A.~M., {Gezari}, S., {et~al.} 2015, \apj, 799,
  208

\bibitem[{{Scodeggio} {et~al.}(2016){Scodeggio}, {Guzzo}, {Garilli}, {Granett},
  {Bolzonella}, {de la Torre}, {Abbas}, {Adami}, {Arnouts}, {Bottini}, {Cappi},
  {Coupon}, {Cucciati}, {Davidzon}, {Franzetti}, {Fritz}, {Iovino}, {Krywult},
  {Le Brun}, {Le F{\'e}vre}, {Maccagni}, {Malek}, {Marchetti}, {Marulli},
  {Polletta}, {Pollo}, {Tasca}, {Tojeiro}, {Vergani}, {Zanichelli}, {Bel},
  {Branchini}, {De Lucia}, {Ilbert}, {McCracken}, {Moutard}, {Peacock},
  {Zamorani}, {Burden}, {Fumana}, {Jullo}, {Marinoni}, {Mellier}, {Moscardini},
  \& {Percival}}]{Scodeggio16}
{Scodeggio}, M., {Guzzo}, L., {Garilli}, B., {et~al.} 2016, ArXiv e-prints

\bibitem[{{Scolnic} \& {Kessler}(2016)}]{Scolnic16}
{Scolnic}, D., \& {Kessler}, R. 2016, \apjl, 822, L35

\bibitem[{{Scolnic} {et~al.}(2014{\natexlab{a}}){Scolnic}, {Rest}, {Riess},
  {Huber}, {Foley}, {Brout}, {Chornock}, {Narayan}, {Tonry}, {Berger},
  {Soderberg}, {Stubbs}, {Kirshner}, {Rodney}, {Smartt}, {Schlafly},
  {Botticella}, {Challis}, {Czekala}, {Drout}, {Hudson}, {Kotak}, {Leibler},
  {Lunnan}, {Marion}, {McCrum}, {Milisavljevic}, {Pastorello}, {Sanders},
  {Smith}, {Stafford}, {Thilker}, {Valenti}, {Wood-Vasey}, {Zheng}, {Burgett},
  {Chambers}, {Denneau}, {Draper}, {Flewelling}, {Hodapp}, {Kaiser},
  {Kudritzki}, {Magnier}, {Metcalfe}, {Price}, {Sweeney}, {Wainscoat}, \&
  {Waters}}]{Scolnic14b}
{Scolnic}, D., {Rest}, A., {Riess}, A., {et~al.} 2014{\natexlab{a}}, \apj, 795,
  45

\bibitem[{{Scolnic} {et~al.}(2014{\natexlab{b}}){Scolnic}, {Riess}, {Foley},
  {Rest}, {Rodney}, {Brout}, \& {Jones}}]{Scolnic14}
{Scolnic}, D.~M., {Riess}, A.~G., {Foley}, R.~J., {et~al.} 2014{\natexlab{b}},
  \apj, 780, 37

\bibitem[{{Shivvers} {et~al.}(2013){Shivvers}, {Mazzali}, {Silverman},
  {Boty{\'a}nszki}, {Cenko}, {Filippenko}, {Kasen}, {Van Dyk}, \&
  {Clubb}}]{Shivvers13}
{Shivvers}, I., {Mazzali}, P., {Silverman}, J.~M., {et~al.} 2013, \mnras, 436,
  3614

\bibitem[{{Shivvers} {et~al.}(2017){Shivvers}, {Modjaz}, {Zheng}, {Liu},
  {Filippenko}, {Silverman}, {Matheson}, {Pastorello}, {Graur}, {Foley},
  {Chornock}, {Smith}, {Leaman}, \& {Benetti}}]{Shivvers17}
{Shivvers}, I., {Modjaz}, M., {Zheng}, W., {et~al.} 2017, \pasp, 129, 054201

\bibitem[{{Silverman} {et~al.}(2012){Silverman}, {Foley}, {Filippenko},
  {Ganeshalingam}, {Barth}, {Chornock}, {Griffith}, {Kong}, {Lee}, {Leonard},
  {Matheson}, {Miller}, {Steele}, {Barris}, {Bloom}, {Cobb}, {Coil},
  {Desroches}, {Gates}, {Ho}, {Jha}, {Kandrashoff}, {Li}, {Mandel}, {Modjaz},
  {Moore}, {Mostardi}, {Papenkova}, {Park}, {Perley}, {Poznanski}, {Reuter},
  {Scala}, {Serduke}, {Shields}, {Swift}, {Tonry}, {Van Dyk}, {Wang}, \&
  {Wong}}]{Silverman12}
{Silverman}, J.~M., {Foley}, R.~J., {Filippenko}, A.~V., {et~al.} 2012, \mnras,
  425, 1789

\bibitem[{{Smee} {et~al.}(2013){Smee}, {Gunn}, {Uomoto}, {Roe}, {Schlegel},
  {Rockosi}, {Carr}, {Leger}, {Dawson}, {Olmstead}, {Brinkmann}, {Owen},
  {Barkhouser}, {Honscheid}, {Harding}, {Long}, {Lupton}, {Loomis}, {Anderson},
  {Annis}, {Bernardi}, {Bhardwaj}, {Bizyaev}, {Bolton}, {Brewington}, {Briggs},
  {Burles}, {Burns}, {Castander}, {Connolly}, {Davenport}, {Ebelke}, {Epps},
  {Feldman}, {Friedman}, {Frieman}, {Heckman}, {Hull}, {Knapp}, {Lawrence},
  {Loveday}, {Mannery}, {Malanushenko}, {Malanushenko}, {Merrelli}, {Muna},
  {Newman}, {Nichol}, {Oravetz}, {Pan}, {Pope}, {Ricketts}, {Shelden},
  {Sandford}, {Siegmund}, {Simmons}, {Smith}, {Snedden}, {Schneider},
  {SubbaRao}, {Tremonti}, {Waddell}, \& {York}}]{Smee13}
{Smee}, S.~A., {Gunn}, J.~E., {Uomoto}, A., {et~al.} 2013, \aj, 146, 32

\bibitem[{{Smith} {et~al.}(2014){Smith}, {Bacon}, {Nichol}, {Campbell},
  {Clarkson}, {Maartens}, {D'Andrea}, {Bassett}, {Cinabro}, {Finley},
  {Frieman}, {Galbany}, {Garnavich}, {Olmstead}, {Schneider}, {Shapiro}, \&
  {Sollerman}}]{Smith14}
{Smith}, M., {Bacon}, D.~J., {Nichol}, R.~C., {et~al.} 2014, \apj, 780, 24

\bibitem[{{Stritzinger} {et~al.}(2011){Stritzinger}, {Phillips}, {Boldt},
  {Burns}, {Campillay}, {Contreras}, {Gonzalez}, {Folatelli}, {Morrell},
  {Krzeminski}, {Roth}, {Salgado}, {DePoy}, {Hamuy}, {Freedman}, {Madore},
  {Marshall}, {Persson}, {Rheault}, {Suntzeff}, {Villanueva}, {Li}, \&
  {Filippenko}}]{Stritzinger11}
{Stritzinger}, M.~D., {Phillips}, M.~M., {Boldt}, L.~N., {et~al.} 2011, \aj,
  142, 156

\bibitem[{{Suh} {et~al.}(2011){Suh}, {Yoon}, {Jeong}, \& {Yi}}]{Suh11}
{Suh}, H., {Yoon}, S.-c., {Jeong}, H., \& {Yi}, S.~K. 2011, \apj, 730, 110

\bibitem[{{Sullivan} {et~al.}(2006){Sullivan}, {Le Borgne}, {Pritchet},
  {Hodsman}, {Neill}, {Howell}, {Carlberg}, {Astier}, {Aubourg}, {Balam},
  {Basa}, {Conley}, {Fabbro}, {Fouchez}, {Guy}, {Hook}, {Pain},
  {Palanque-Delabrouille}, {Perrett}, {Regnault}, {Rich}, {Taillet}, {Baumont},
  {Bronder}, {Ellis}, {Filiol}, {Lusset}, {Perlmutter}, {Ripoche}, \&
  {Tao}}]{Sullivan06}
{Sullivan}, M., {Le Borgne}, D., {Pritchet}, C.~J., {et~al.} 2006, \apj, 648,
  868

\bibitem[{{Sullivan} {et~al.}(2011){Sullivan}, {Guy}, {Conley}, {Regnault},
  {Astier}, {Balland}, {Basa}, {Carlberg}, {Fouchez}, {Hardin}, {Hook},
  {Howell}, {Pain}, {Palanque-Delabrouille}, {Perrett}, {Pritchet}, {Rich},
  {Ruhlmann-Kleider}, {Balam}, {Baumont}, {Ellis}, {Fabbro}, {Fakhouri},
  {Fourmanoit}, {Gonz{\'a}lez-Gait{\'a}n}, {Graham}, {Hudson}, {Hsiao},
  {Kronborg}, {Lidman}, {Mourao}, {Neill}, {Perlmutter}, {Ripoche}, {Suzuki},
  \& {Walker}}]{Sullivan11}
{Sullivan}, M., {Guy}, J., {Conley}, A., {et~al.} 2011, \apj, 737, 102

\bibitem[{{Taubenberger} {et~al.}(2008){Taubenberger}, {Hachinger}, {Pignata},
  {Mazzali}, {Contreras}, {Valenti}, {Pastorello}, {Elias-Rosa},
  {B{\"a}rnbantner}, {Barwig}, {Benetti}, {Dolci}, {Fliri}, {Folatelli},
  {Freedman}, {Gonzalez}, {Hamuy}, {Krzeminski}, {Morrell}, {Navasardyan},
  {Persson}, {Phillips}, {Ries}, {Roth}, {Suntzeff}, {Turatto}, \&
  {Hillebrandt}}]{Taubenberger08}
{Taubenberger}, S., {Hachinger}, S., {Pignata}, G., {et~al.} 2008, \mnras, 385,
  75

\bibitem[{{Taubenberger} {et~al.}(2011){Taubenberger}, {Navasardyan}, {Maurer},
  {Zampieri}, {Chugai}, {Benetti}, {Agnoletto}, {Bufano}, {Elias-Rosa},
  {Turatto}, {Patat}, {Cappellaro}, {Mazzali}, {Iijima}, {Valenti},
  {Harutyunyan}, {Claudi}, \& {Dolci}}]{Taubenberger11}
{Taubenberger}, S., {Navasardyan}, H., {Maurer}, J.~I., {et~al.} 2011, \mnras,
  413, 2140

\bibitem[{{Tonry} \& {Davis}(1979)}]{Tonry79}
{Tonry}, J., \& {Davis}, M. 1979, \aj, 84, 1511

\bibitem[{{Tripp}(1998)}]{Tripp98}
{Tripp}, R. 1998, \aap, 331, 815

\bibitem[{{Tsvetkov} {et~al.}(2009){Tsvetkov}, {Volkov}, {Baklanov},
  {Blinnikov}, \& {Tuchin}}]{Tsvetkov09}
{Tsvetkov}, D.~Y., {Volkov}, I.~M., {Baklanov}, P., {Blinnikov}, S., \&
  {Tuchin}, O. 2009, Peremennye Zvezdy, 29

\bibitem[{{Yaron} \& {Gal-Yam}(2012)}]{Yaron12}
{Yaron}, O., \& {Gal-Yam}, A. 2012, \pasp, 124, 668

\end{thebibliography}

\end{document}